\documentclass[12pt]{article}
\usepackage{amsmath}
\usepackage{algorithmic}
\usepackage{algorithm}
\usepackage{amssymb,multirow,array,float,bm,bbm}
\usepackage{textcomp,subfigure}
\usepackage{stfloats}
\usepackage{url,multicol}
\usepackage[numbers]{natbib}
\usepackage{graphicx,enumerate}
\usepackage{appendix,epstopdf}
\usepackage[colorlinks,citecolor=blue,urlcolor=blue,linkcolor=red]{hyperref}
\usepackage[capitalize]{cleveref}

\newtheorem{definition}{Definition}
\newtheorem{proposition}{Proposition}
\newtheorem{assumption}{Assumption}

\newtheorem{theorem}{Theorem}
\newtheorem{corollary}{Corollary}
\newtheorem{remark}{Remark}
\newtheorem{example}{Example}
\newcommand{\bigO}{\mathcal{O}}
\newcommand{\dee}{\mathrm{d}}
\renewcommand{\Pr}{\mathbb{P}}
\newcommand{\I}{\mathrm{I}}
\def\EE{\mathbb{E}}

\begin{document}
	\title{Optimal Short-Term Forecast for Locally Stationary Functional Time Series}
	\author{Yan Cui and Zhou Zhou \\
		Department of Statistical Sciences, University of Toronto}
	\date{}
	\maketitle
	
	
\begin{abstract}
	Accurate curve forecasting is of vital importance for policy planning, decision making and resource allocation in many engineering and industrial applications. In this paper we establish a theoretical foundation for the optimal short-term linear prediction of non-stationary functional or curve time series with smoothly time-varying data generating mechanisms. The core of this work is to establish a unified functional auto-regressive approximation result for a general class of locally stationary functional time series. A double sieve expansion method is proposed and theoretically verified for the asymptotic optimal forecasting. A telecommunication traffic data set is used to illustrate the usefulness of the proposed theory and methodology.
\end{abstract}


\section{Introduction}\label{intro}
One of the most essential goals in time series analysis is to provide reliable predictions for future observations given a stretch of previous data. There is a large number of studies for prediction in the univariate and multivariate time series framework, see for examples, \cite{Schafer02,Nobel03,GO07,BP11,ZXW10}. Recently, forecasting functional time series whose observation at each time stamp is a continuous curve has gained much attention in various applications, such as energy systems or electricity markets (\cite{CCK18,SL15,VAP18,CL2017}), demography (\cite{HS09,HSH12,GS2017}), environment (\cite{SH2011,BS2020}), economics and finance (\cite{KZ2012,HSH12,KMZ15,Fu2020}), among others. Most of the aforementioned works assume that the functional time series is stationary, that is, the data generating mechanism does not change over time.

\begin{figure}[t!]
	\centering
	\subfigure[]{
		\centering
		\includegraphics[height=7cm,width=8cm]{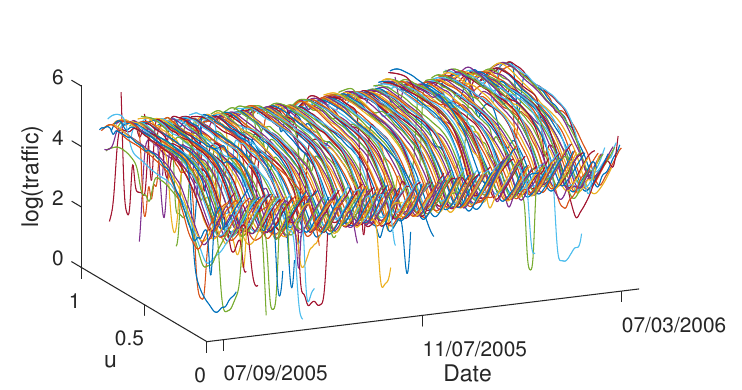}
	}\subfigure[]{
		\centering
		\includegraphics[height=6cm,width=7cm]{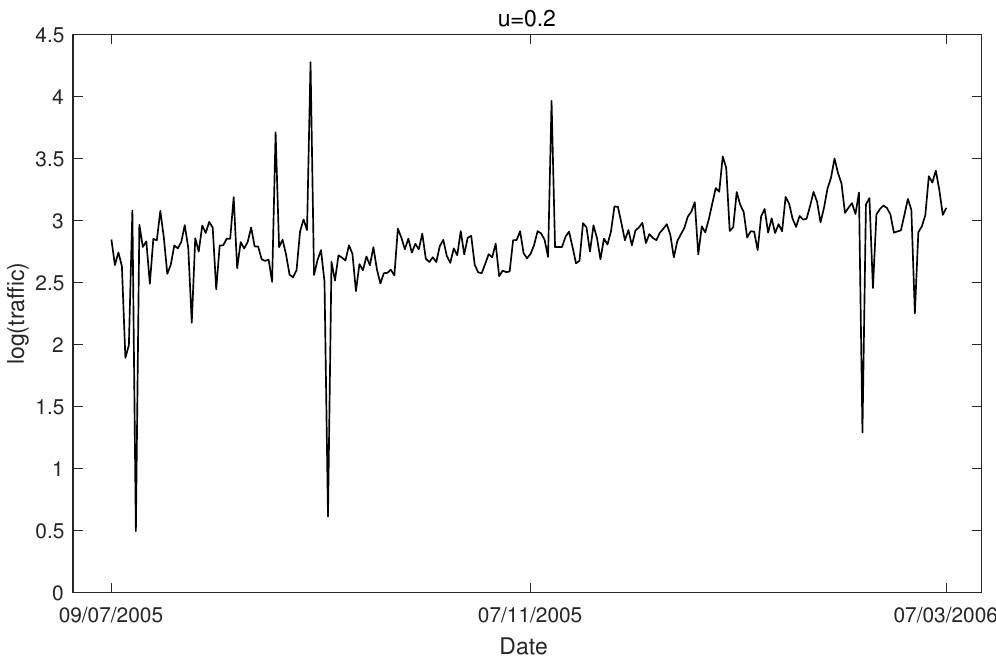}
	}
	\subfigure[]{
		\centering
		\includegraphics[height=6cm,width=7cm]{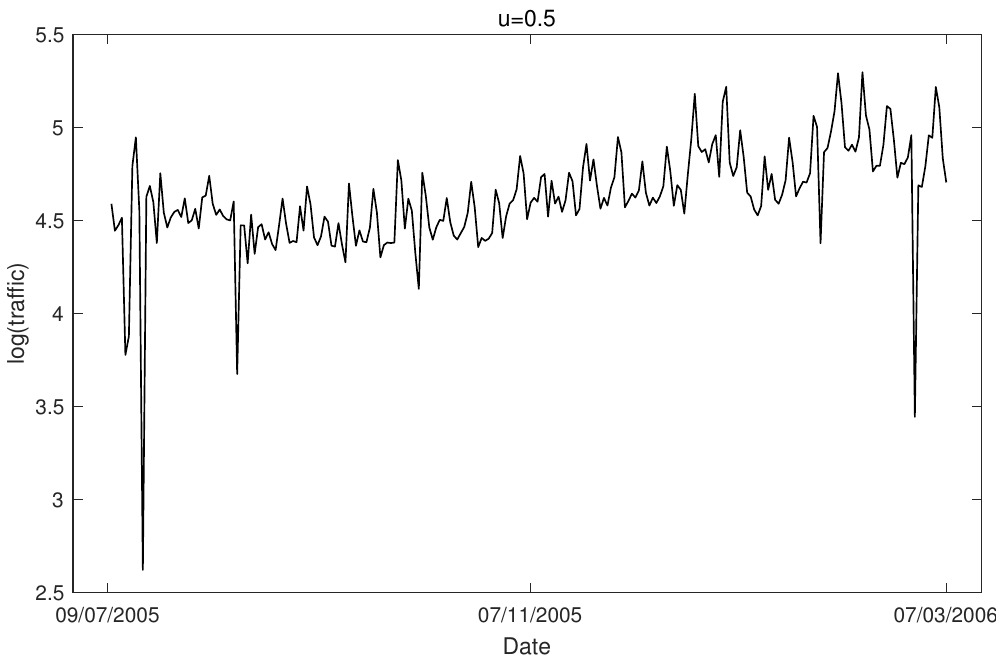}
	}~~\qquad
	\subfigure[]{
		\centering
		\includegraphics[height=6cm,width=7cm]{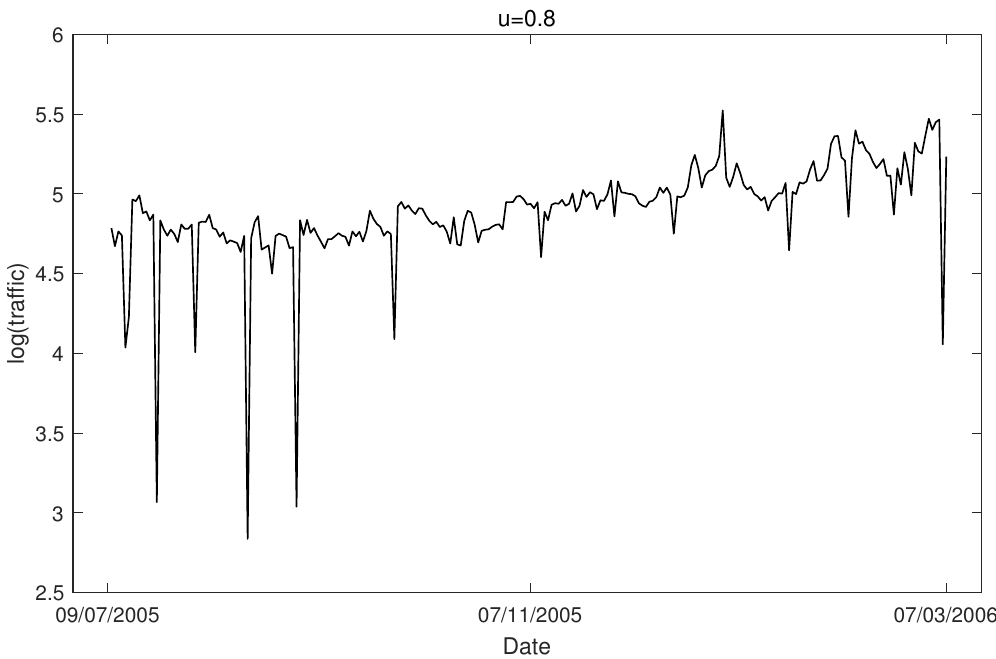}
	}
	\centering
	\caption{(a): 3D Transformed functional time series plot. (b): Transformed time series plot for fixed $u=0.2$. (c): Transformed time series plot for fixed $u=0.5$. (d): Transformed time series plot for fixed $u=0.8$.}
	\label{F1}
\end{figure}

The aim of this article is to build a theoretical foundation as well as to provide an efficient methodology for the optimal short-term linear forecasting of locally stationary functional time series. Here local stationarity refers to a smoothly or slowly time-varying data generating mechanism. Our work is motivated by a curve forecasting problem for telecommunication network traffic data. Specifically, the data set consists of user download data for a mobile infrastructure network deployed in Asia and the United States recorded minutely for a period of roughly 8 months. Though the data are recorded at a high frequency, engineers and administrators are interested in forecasting the download pattern of a future day or several days in hope of promoting efficient operation of the network system. To this end and due to the strong daily periodicity of the data, a typical way is to transform the observed data on day $i$, $Y_i(u_j)$, $i=1,...,n$ into smooth daily curves $Y_i(u)$ for $u\in [0,1]$ (refer to \cref{F1}(a) for the logarithm transformed daily curves), where $u_j=j/1440$, $j=1,2,\cdots, 1440$ denotes the $j$th minute of the day. Then one seeks to predict the future curves of downloads $Y_{n+k}(u),~k\ge 1$.

One of the most significant characteristics of the telecommunication network traffic time series lies in its non-stationarity. For instance, take a look at the log-transformed time series at $u=0.2,0.5$ and $0.8$ respectively in \cref{F1}(b)--(d). It is clear that there exists an upward trend and obvious changes of variability over time, contributing to the non-stationarity of the functional data. 

Building a unified theoretical foundation for locally stationary functional time series prediction is difficult due to the lack of insights into the structure of the series. For a univariate and weakly stationary time series, the Wiener-Kolmogorov prediction theory (\cite{Ko41,Wiener49}) elucidates that it can be represented as a white-noise-driven auto-regressive (AR) process of infinite order under some mild conditions. 
Recently, Ding and Zhou \cite{Ding21} established a unified AR approximation theory for a wide class of univariate non-stationary time series under some mild conditions. Nevertheless, it has been a difficult and open problem to build structural representations or approximations for functional time series since the intrinsic infinite-dimensional nature of such processes brings great technical difficulty to studying the structure of such complex dynamic systems. In particular, the covariance operator of a smooth functional time series is not invertible which makes it difficult to extend the existing linear approximation theory of univariate and fixed-dimensional multivariate time series directly to the functional setting.  

Our major theoretical contribution in this paper lies in establishing a functional AR approximation theory for a rich class of locally stationary functional time series. To be more specific, we prove that a wide class of short memory locally stationary functional time series can be well approximated by a locally stationary white-noise-driven functional AR process of slowly diverging order, see \cref{approx_ar} for a more precise statement. The construction of this structural approximation relies on a sieve truncation technique, the modern operator spectral theory, and the classic approximation theory which transfers the infinite-dimensional problem into a high-dimensional one and subsequently controls the decay rates of the inverse of high-dimensional banded matrices. To our best knowledge, there is no such structural approximation result in the field of functional data analysis, even under stationary scenarios. As a fundamental theory, our functional AR approximation result sheds light on the underlying linear structure of a wide class of functional time series and hence serves as a unified foundation for an optimal linear forecasting theory of such processes. Furthermore, the functional AR approximation theory could have a much wider range of applications in various fundamental problems in functional time series analysis such as covariance inference, adaptive resampling, efficient estimation, and dependence quantification.

The functional AR approximation theory is nonparametric in nature and it provides a more flexible and robust way to forecast a rich class of locally stationary functional time series without resorting to restrictive parametric modeling of the covariance operator compared to existing methods built on  parametric linear time series models. Methodologically, we propose a nonparametric double-sieve method for the estimation of the AR coefficient functions where sieve expansions are conducted and then truncated over both the function and time domains. Unlike most non-stationary time series forecasting methods in the literature where only data near the end of the sequence are utilized for the forecast (\cite{DW2022,RS2018}), the nonparametric sieve regression used in our prediction is global in the sense that it utilizes all available functional curves to determine the optimal forecast coefficients and hence is expected to be more efficient. Due to the adaptivity of the double-sieve expansion, we also claim that the prediction errors are adaptive to the smoothness of the functional time series and the strength of the temporal dependence (c.f. \cref{thm_error2}). 

There is substantial literature on prediction techniques and theory for stationary functional time series, most of which were essentially built on linear functional time series assumptions but without investigating whether the functional time series of interest can be represented or approximated by a linear model. Bosq \cite{Bosq00} suggested a one-step ahead prediction based on the functional AR process. Hyndman and Ullah \cite{hyndman2007robust} introduced a robust forecasting approach where principal component scores were predicted via a univariate time series forecasting method. As an extension of the latter, Aue et al. \cite{Aue2015} proposed a forecasting method based on vector auto-regressive forecasts of principal component scores. Later, Aue et al. \cite{AK17} considered the functional moving average (FMA) process and introduced an innovation algorithm to obtain the best linear predictor. The vector auto-regressive moving average (VARMA) model was investigated by \cite{KKW2017} for modeling and forecasting principal component scores. Other available methods for forecasting include functional kernel regression (\cite{FV03}), functional partial least squares regression (\cite{PS05}), dynamic updating approaches for incomplete trajectories (\cite{SH2011}), and robust forecasting method via dynamic functional principal component regression for data contaminated by outliers (\cite{Shang19}).

 Meanwhile, the last two decades have witnessed some developments in prediction for locally stationary time series; see for instance \cite{Das2021,DW2022,Ding21,Fryzlewicz2003,RS2018}. However, studies on locally stationary functional time series remain scarce. Recently, Van Delft and Eichler \cite{vandelft18} discussed inference and forecasting methods for a class of time-varying functional processes based on auto-regressive fitting. Kurisu \cite{Kurisu22} investigated the estimation of locally stationary functional time series and applied it to $k$-step ahead prediction using a kernel-based method.

The remainder of this paper is organized as follows. In \cref{sec_2}, we establish the functional AR approximation result under some mild assumptions. \cref{sec_3} provides one application of our theory in optimal forecasting of locally stationary functional time series. Practical implementation including the selection for tuning parameters and optimal prediction algorithm are discussed in \cref{prac_imple}. \cref{sec_simu} reports some supporting Monte Carlo simulation experiments. A real data application for the prediction of daily telecommunication downloads is carried out in \cref{real_data}. Additional  results and technical proofs are deferred to the Appendix.

\section{Functional AR approximation to locally stationary functional time series}
\label{sec_2}
Throughout this paper, let $\mathcal{L}^2([0,1])$ be a separable Hilbert space of all square integrable functions on $[0,1]$ with inner product $\langle x,y \rangle=\int_0^1 x(u)y(u)\dee u$. A square integrable random function $Y(u)\in\mathcal{L}^2[0,1]$ implies $\EE|Y(u)|_{\mathcal{L}^2}^2<\infty$, where $|Y(u)|_{\mathcal{L}^2}^2=\int_0^1 Y^2(u)\dee u$ signifies its $\mathcal{L}^2$ norm. We also denote by $\mathcal{C}^d([0,1])$ the collection of functions that are $d$-times continuously differentiable with absolutely continuous $d$-th derivative on $[0,1]$. For a random variable $Z$ and some constant $q>1$, denote by $\Vert Z\Vert_q=(\EE|Z|^q)^{1/q}$ its $L^q$ norm. The notation $\Vert\cdot\Vert$ signifies the operator norm (i.e., largest singular value) when applied to matrices and the Euclidean norm when applied to vectors. If $f(x)\asymp g(x)$, we say that functions $f(x)$ and $g(x)$ have the same order of magnitude. Also, we use $\lambda_{\min}(\cdot)$ and $\lambda_{\max}(\cdot)$ to signify the largest and smallest eigenvalues of matrices. Throughout this paper, the
symbol $C$ denotes a generic finite constant that is independent of $n$ and may vary from place to place.

\subsection{Locally stationary functional time series}\label{sec_21}
In this subsection, we will first introduce the definition of locally stationary functional time series as follows.

\begin{definition}[Locally stationary functional time series]\label{def_local}
	A non-stationary functional time series $\{Y_i(u)\}$ is a locally stationary functional time series (in covariance) if there exists a function $\gamma(t,u,v,k): [0,1]^3\times \mathbb{Z}\to \mathbb{R}$ such that
	\begin{equation}\label{def_cov}
	{\rm Cov}(Y_i(u),Y_j(v))=\gamma(t_i,u,v,i-j)+
	\bigO\left(\frac{|i-j|+1}{n}\right),~t_i=\frac{i}{n}.
	\end{equation}
	Furthermore, we assume that $\gamma$ is Lipschitz continuous in time $t$ and for any fixed $t\in[0,1]$, $\gamma(t,u,v,\cdot)$ is the autocovariance function of a stationary functional time series. 
\end{definition}

This definition only imposes a smoothness condition on the covariance structure of $\{Y_i(u)\}$ with respect to time. From \cref{def_cov}, we find that the underlying data generating mechanism evolves smoothly over time, which implies that the covariance structure of $\{Y_i(u)\}$ in any small time segment can be well approximated by that of a stationary functional process. \cref{def_local} covers a wide class of frequently used locally stationary functional time series models, and we shall provide an example in the following.

\begin{example}\label{model_example}
	Consider the following locally stationary functional time series  
	\begin{equation}\label{physi_repre_func}
	Y_i(u)=H(\frac{i}{n},u,\mathcal{F}_i),
	\end{equation} 
	where $\mathcal{F}_i=(\cdots,\eta_{i-1},\eta_i)$ with $\eta_i$ being i.i.d. random elements and $H:[0,1]^2\times \mathbb{R}^\infty\to \mathbb{R}$ is a measurable function such that $\xi_i(t,u):=H(t,u,\mathcal{F}_i)$ is a properly defined random function in $\mathcal{L}^2$. Furthermore, the following assumption is needed to ensure local stationarity.
 \begin{assumption}\label{ass_local_func}
	$H(t,\cdot,\cdot)$ defined in \eqref{physi_repre_func} satisfies the stochastic Lipschitz continuous condition across $t$, that is for some $q>2$ and any $u\in[0,1]$,
	\begin{equation}\label{lipschitz}
	\Vert H(t_1,u,\mathcal{F}_i)-H(t_2,u,\mathcal{F}_i)\Vert_q\le C|t_1-t_2|,	
	\end{equation}
	where $C>0$ and $t_1,t_2\in[0,1]$. Moreover, we assume
	\begin{equation}\label{x_bounded}
	\sup_{t,u\in[0,1]}\Vert H(t,u,\mathcal{F}_i)\Vert_q<\infty.
	\end{equation}
\end{assumption}
	 In this context, the autocovariance function $\gamma(t,u,v,j)$, $j\in\mathbb{Z}$ in \cref{def_local} can be represented as
	\begin{equation}\label{gamma}
	\gamma(t,u,v,j)={\rm Cov}(H(t,u,\mathcal{F}_0),H(t,v,\mathcal{F}_j)).	
	\end{equation}
 Under \cref{ass_local_func}, this type of locally stationary functional process in \eqref{physi_repre_func} satisfies \cref{def_local}. See Lemma 1 in Section B of the Supplementary Material for detailed proof.
\end{example}
 
Till the end of the paper, we consider a locally stationary functional time series  $\{Y_i(u)\}_{i=1}^n\in \mathcal{L}^2([0,1])$ satisfying \cref{def_local} and $\EE|Y_i(u)|_{\mathcal{L}^2}^2<\infty$, then it can always be decomposed as $$Y_i(u)=\mu_i(u)+Z_i(u),~i=1,...,n,$$ where $\mu_i(u)=\EE(Y_i(u)),~u\in[0,1]$ is the mean function and $Z_i(u)$ is the centered locally stationary functional process in $\mathcal{L}^2([0,1])$. 
For simplicity, we assume that $\mu_i(u)=0$. Let $\{\alpha_k(u)\}_{k=1}^\infty$ be a set of pre-determined orthonormal basis functions on $\mathcal{L}^2([0,1])$, the functional time series $Y_i(u)$ admits the following Karhunen-Lo\`{e}ve type expansion
\begin{equation}\label{infinite}
Y_i(u)=\sum_{k=1}^\infty r_{i,k}\alpha_k(u)=\sum_{k=1}^\infty x_{i,k}f_k\alpha_k(u),
\end{equation}
where $r_{i,k}=\int_0^1 Y_i(u)\alpha_k(u)\dee u$ is the $k$th  (random) basis expansion coefficient of $Y_i(u)$ with respect to $\{\alpha_k(u)\}_{k=1}^\infty$. For examples of commonly used basis functions, we refer readers to Section A in the Supplementary Material. In \eqref{infinite}, $f_k^2$ is the asymptotically average variance of $r_{i,k}$ over $i$, denoted by $f_k^2:=\int_{T^3} \gamma(t,u,v,0)\alpha_k(u)\alpha_k(v)\dee u\dee v\dee t$, $T=[0,1]$. $f_k$ captures the average magnitude of $r_{i,k}$ and it decays as $k$ increases. If $f_k\neq 0$, then $\{x_{i,k}:=r_{i,k}/f_k\}_{i=1}^n$ is a locally stationary (scalar) time series for any $k\ge 1$. Observe that the magnitude of $x_{i,k}$ is expected to be stable as $k$ increases.

It is worth noting that in the stationary context, one often uses $f_k:={\rm Std}(r_{i,k})$ to describe the decay speed of $r_{i,k}$ as $k$ increases and the latter representation is frequently used in the functional data analysis literature; see for instance \cite{Shang14} and \cite{CYC14}. Our definition of $f_k$ can be viewed as the corresponding extension to the locally stationary setting. 
The following assumption restricts the decay speed of the basis expansion coefficient $r_{i,k}$.

\begin{assumption}\label{ass_conti_u}
	We assume that the functional time series $Y_i(u)\in \mathcal{C}^{d_1}([0,1])$ a.s., where $d_1> 0$ is some integer. Furthermore, suppose the random coefficient $r_{i,k}=\bigO_{\Pr}(k^{-(d_1+1)})$ for $i=1,...,n$.
\end{assumption}

It is well-known that for a general $\mathcal{C}^d([0,1])$ function where $d$ is a non-negative integer, the fastest decay rate for its $k$th basis expansion coefficient is $\bigO(k^{-(d+1)})$ for a wide class of basis functions (\cite{Chen07}). For example, the Fourier basis (for periodic functions), the weighted Chebyshev polynomials (\cite{Trefethen2008}) and the orthogonal wavelets with degree $m\ge d$ (\cite{Meyer90}) admit the latter decay rate under some extra mild assumptions on the behavior of the function's $d$th derivative. On the other hand, the basis expansion coefficients may decay at slower speeds for some orthonormal bases. An example is the normalized Legendre polynomials basis function where the coefficients decay at an $\bigO(k^{-(d+1/2)})$ speed (\cite{WX11}). We remark that our functional AR approximation result can be achieved for basis functions whose corresponding coefficients decay at slower rates with the bound error slightly larger but still converging to zero under mild conditions. For the sake of brevity, we shall stick to the fastest decay \cref{ass_conti_u} for our theoretical investigations throughout this paper.

\subsection{Functional AR approximation theory}\label{sec_22}
Here, we will establish a functional AR approximation theory for locally stationary functional time series. Let $b=b(n)$ be a generic value which specifies the order of functional AR approximation. For theoretical and practical purposes, $b$ is required to be much smaller than the sample size $n$ to achieve a parsimonious approximating model. To explore the theoretical results of the functional AR approximation, we will truncate the infinite representation \eqref{infinite} to finite (but diverging) dimensional spans of basis functions as follows
\begin{equation}\label{truncation}
	Y_i(u)=\sum_{k=1}^{p} x_{i,k}f_{k}\alpha_k(u)+\bigO_ \Pr(p^{-d_1}):=Y_i^{(p)}(u)+\bigO_\Pr(p^{-d_1}),
\end{equation}
where $p=p(n)$ is the truncation number. This truncated expansion in \eqref{truncation} serves as the first dimension reduction for our theoretical investigation, which is a common technique in functional time series analysis. For example, with this approach, one could apply the initial dimension reduction by functional principal component analysis (\cite{Shang14,KR17}), or explore properties of linear regression estimators (\cite{Hall06,LiHsing07}). Some existing work suggests projecting infinite dimensional objects onto a fixed dimensional subspace to facilitate statistical calculations (\cite{Aue2015}), that is, the truncation number is a fixed constant. However, there is a growing interest in allowing the truncation number to grow to infinity with the sample size $n$ in order to make the truncation adaptive to the smoothness of the functional observations, see \cite{Hall06,LiHsing07}. Throughout this paper, we assume that the truncation number diverges to infinity at a relatively slow speed, i.e., $p\asymp n^{\beta_1}, \beta_1\in(0,1)$. We will discuss how to select it in \cref{para}.

Since the functional time series is centered, we have $\EE(x_{i,k})=0$ for any $i=1,...,n,~k\ge 1$. When $i > b$, the best linear prediction (in terms of the mean squared prediction error) of $\bm{x}_i:=(x_{i,1},...,x_{i,p})^\top$ which utilizes all its predecessors $\bm{x}_1,...,\bm{x}_{i-1}$ can be expressed as
$$\widehat{\bm{x}}_i=\sum_{j=1}^{i-1}\bm{\Phi}_{i,j}\bm{x}_{i-j},$$
where $\{\bm{\Phi}_{i,j}\}$ are the prediction coefficient matrices. By construction, $\bm{\epsilon}_i:=\bm{x}_i-\widehat{\bm{x}}_i$ is a white noise process with mean $\bm{0}$ and covariance matrix denoted by $\bm{\Sigma}_i$. Furthermore, let $\bm{\Gamma}(t,j)\in \mathbb{R}^{p\times p}$ be the autocovariance matrix of $\bm{x}_i$ at some rescaled time $t\in[0,1]$ and lag $j\in\mathbb{Z}$ with $\gamma_{k,l}(t,j)$ being its $(k,l)$th element for $k,l=1,...,p$.  Note that $\gamma_{k,l}(t,j)=\int_{T^2}\gamma(t,u,v,j)
\alpha_k(u)\alpha_l(v)\dee u\dee v/(f_kf_l)$ where $\gamma(t,u,v,j)$ is defined in \cref{def_local}. Together with Eq. \eqref{def_cov}, it also indicates that the covariance structure of the scaled multivariate time series $\{\bm{x}_i\}$ can be determined by the covariance of the functional time series $\{Y_i(u)\}$. In order to provide a theoretical foundation for the functional AR approximation, certain assumptions are required.
\begin{assumption}\label{ass_conti_t}
	For any $j\in \mathbb{Z}$, we assume that $\bm{\Gamma}(t,j)\in \mathcal{C}^{d_2}([0,1])$, where $d_2>0$ is some integer. In other words for any integers $k,l\ge 1$, each component $\gamma_{k,l}(t,j)$ is $d_2$ times continuously differentiable with respect to $t$ over $[0,1]$. 
\end{assumption} 

\begin{assumption}\label{weak_depen_components}
	For $j\in \mathbb{Z}$, suppose that 
	$\sup_{t\in[0,1]}\Vert \bm{\Gamma}(t,j)\Vert\le C(|j|+1)^{-\tau}$ for some constant $\tau>1$.
\end{assumption}

\Cref{ass_conti_t} is a local stationarity assumption and it imposes a smoothness requirement on the autocovariance matrix $\bm{\Gamma}(t,j)$. Simple calculations show that \cref{weak_depen_components} implies that $\max_{k,l}|{\rm Cov}(x_{i,k}, x_{i+j,l})| \le C(|j|+1)^{-\tau}$, which provides a polynomial decay rate of the covariance structures of random variables. In particular, \cref{weak_depen_components} states that the correlation among components of the random vector $\bm{x}_i$ is relatively weak. This condition is generally mild and can be fulfilled in most cases, as the random components $x_{i,k}$ typically exhibit weak dependence between different $k$ under appropriate basis expansions. 

Now, we will provide an example of locally stationary multivariate time series.
\begin{example}\label{multi_model_example}
	Let $\{\bm{\eta}_i\}$ be zero-mean i.i.d. $\mathbb{R}^p$ random vectors with its covariance matrix
$\Vert\bm{\Sigma}_{\eta}\Vert<\infty$. We consider the locally stationary linear process as
 $$\bm{x}_i=\sum_{m=0}^\infty \bm{A}_m(t_i)\bm{\eta}_{i-m},~~t_i=\frac{i}{n},$$
 where $\bm{A}_m(t)\in \mathbb{R}^{p\times p}$ is assumed to be a $\mathcal{C}^{d_2}([0,1])$ function with respect to $t$. Then Assumptions \ref{ass_conti_t} and \ref{weak_depen_components} will be satisfied if $$\sup_{t\in[0,1]} \left\Vert\frac{\partial^{d_2}\bm{A}_m(t)}{\partial t^{d_2}}\right\Vert\le C(m+1)^{-\tau},~~\sup_{t\in[0,1]}\Vert \bm{A}_m(t)\Vert\le C(m+1)^{-\tau}$$ hold, respectively. We refer readers to Lemma 3 in Section B of the Supplementary Material for detailed proof.
\end{example}

Furthermore, to avoid erratic behavior of the functional AR approximation, the smallest eigenvalue of the covariance matrix of multivariate time series $\{\bm{x}_i\}_{i=1}^n$ should be bounded away from zero. Similar to the uniformly-positive-definite-in-covariance (UPDC) condition for univariate time series discussed in \cite{Ding21}, we put forth an assumption for the multivariate version as follows.
\begin{assumption}[UPDC condition for multivariate time series]\label{updc}
	Denote $\bm{{\rm x}}= (\bm{x}_1^\top,\\...,\bm{x}_n^\top)^\top\in\mathbb{R}^{np}$. For all sufficiently large $n\in\mathbb{N}$, there exists a universal constant $\kappa_1>0$ such that the smallest eigenvalue of ${\rm Cov}(\bm{{\rm x}})$ is bounded away by $\kappa_1$, where ${\rm Cov}(\cdot)$ is the covariance matrix of the given vector.
\end{assumption}

This condition is necessary to avoid ill-conditioned ${\rm Cov}(\bm{{\rm x}})$ and hence makes the construction of functional AR approximation feasible. Note that it is a mild requirement and has been widely used in the statistical literature for covariance and precision matrix estimation; see for instance, \cite{CXW13}, \cite{CLZ16} and references therein. When it comes to stationary multivariate time series with short memory, \citep[Theorem 11.8.1]{BD91} states that the UPDC condition holds if its spectral density matrix is uniformly bounded below by a positive constant. To practically verify the UPDC assumption in the case of locally stationary multivariate time series, we provide a necessary and sufficient condition below.

\begin{proposition}\label{prop_updc}
	  Suppose that $\{\bm{x}_i\}_{i=1}^n$ is locally stationary multivariate time series satisfying \cref{weak_depen_components}. If there exists some constant $\kappa_1>0$ such that the smallest eigenvalue of the spectral density matrix, i.e., $\lambda_{\min}(\bm{f}(t,\omega))\ge \kappa_1$ 
	  for all $t$ and $\omega\in[-\pi,\pi]$, where 
	  \begin{equation}\label{spectral_matrix}
	  \bm{f}(t,\omega)=\frac{1}{2\pi}\sum_{h\in\mathbb{Z}}{\rm e}^{-{\rm i}h\omega}
	  \bm{\Gamma}(t,h),~~{\rm i}=\sqrt{-1},
	  \end{equation}
	  then $\{\bm{x}_i\}_{i=1}^n$ satisfies \cref{updc}. Conversely, if $\{\bm{x}_i\}_{i=1}^n$ satisfies Assumptions \ref{weak_depen_components} and \ref{updc}, then there exists some constant $\kappa_1>0$ such that $\lambda_{\min}(\bm{f}(t,\omega))\ge \kappa_1$ for all $t$ and $\omega\in[\pi,\pi]$.
	\end{proposition}
\cref{prop_updc} demonstrates that the verification of \cref{updc} boils down to checking whether the smallest eigenvalue of the local spectral density matrix $\bm{f}(t,\omega)$ is uniformly bounded from below by some positive constant. Here, we provide an example to check the UPDC condition via \cref{prop_updc}.

\begin{example}\label{exam_updc}
 Rewrite the linear process in \cref{multi_model_example} as 
	$$\bm{x}_i=\bm{\mathcal{A}}
 (\frac{i}{n},B)\bm{\eta}_i,$$
	 where $\bm{\mathcal{A}}(\cdot,B)=\sum_{m=0}^\infty 
 \bm{A}_m(\cdot)B^m$ with the backshift operator $B$, and $\{\bm{\eta}_i\}$ are zero-mean i.i.d. random vectors with non-degenerate covariance matrix $\bm{\Sigma}_{\eta}$. Using the property of linear filters for the spectral density matrix, we have 
	$$\bm{f}(t,\omega)=\frac{1}{2\pi}
 \bm{\mathcal{A}}(t,{\rm e}^{-{\rm i} \omega})\bm{\Sigma}_{\eta}\bm{\mathcal{A}}^\top(t,{\rm e}^{{\rm i}\omega}),\quad 
	-\pi\le \omega\le \pi.$$
 Therefore, by \cref{prop_updc}, we can obtain that the UPDC condition is satisfied if $$\lambda_{\min}(\bm{\mathcal{A}}(t,{\rm e}^{-{\rm i}\omega})\bm{\mathcal{A}}^\top(t,{\rm e}^{{\rm i}\omega}))\ge \kappa_1>0~\text{for~all~} t~\text{and}~\omega.$$
	\end{example}  
The next theorem is our main theoretical result and it provides a functional AR approximation theory under the short-range dependence and local stationarity conditions.
\begin{theorem}\label{approx_ar}
	Consider the locally stationary functional time series from \cref{def_local}. Under Assumptions \ref{ass_conti_u}--\ref{updc} and suppose  $\sup_{t\in[0,1]}\left\Vert \partial \bm{\Gamma}(t,j)/\partial t\right\Vert\le C$ for all $j\in\mathbb{Z}$. 
	Then we obtain that for $i\ge 2$,
	\begin{equation}\label{approx_far}
	Y_i(u)=\sum_{j=1}^{\min\{i-1,b\}}\int_0^1 
	\psi_j^{(p)}(\frac{i}{n},u,v)Y_{i-j}(v)\dee v+\varepsilon_i(u)
	+\bigO_\Pr\left(p^{1/2}b^{-\tau+2}(\log b)^{\tau-1}+\frac{p^{1/2}b^{3}}{n}+p^{-d_1}\right),
	\end{equation}
	where $\psi_j^{(p)}(\cdot,u,v)\in\mathcal{L}^2([0,1]^2)$ admits the basis expansion $\psi_j^{(p)}(t,u,v):=\sum_{k,l=1}^p  \psi_{j,kl}(t)\alpha_k(u)\alpha_l(v)$ with the coefficient $\psi_{j,kl}(t)\in \mathcal{C}^{d_2}([0,1])$ with respect to $t$ for all $j$, and the error process $\epsilon_i(u):=\bm{\alpha}_f^\top(u)\bm{\epsilon}_i$ is a functional white noise process, where $\bm{\alpha}_f(u)=(\alpha_1(u)f_1,...,\alpha_p(u)f_p)^\top$.
 \end{theorem}
\cref{approx_ar} states that a wide class of locally stationary functional time series can be efficiently approximated by a locally stationary functional autoregressive process with smoothly time-varying operators (kernels) and a slowly diverging order $b$. Notice that the functional AR coefficient function $\psi_j^{(p)}(t,u,v)$ has the same degree of smoothness over time $t$ as the time-varying covariance functions $\bm{\Gamma}(t,j)$ specified in \cref{ass_conti_t}. In addition, the first and second error terms on the right-hand side of \eqref{approx_far} describe the functional AR approximation errors based on $Y_i^{(p)}(u)$, and the third term reflects the truncation error due to \eqref{truncation}. The approximation result in \eqref{approx_far} also reveals that the error bound is adaptive to the smoothness of the functional observations ($d_1$), as well as the temporal dependence structure of $\{Y_i(u)\}$ ($\tau$). In particular, the optimal choice of the AR order $b$ can be obtained by balancing the first two error terms in \eqref{approx_far}. Simple calculations yield that the optimal $b \asymp n^{\frac{1}{\tau+1}}(\log n)^\theta$ with $\theta=\frac{\tau-1}{\tau+1}$. Similarly, the optimal choice for the truncation number $p \asymp n^{\frac{\tau-2}{(\tau+1)(d_1+1/2)}}(\log n)^{\frac{-3\theta}{d_1+1/2}}$, and minimum AR approximation error turns out to be $\bigO\Big(n^{-\frac{d_1(\tau-2)}{(\tau+1)(d_1+1/2)}}(\log n)^{\frac{3d_1\theta}{d_1+1/2}}\Big)$. For example, when the functions are infinite many times differentiable, that is $d_1\rightarrow\infty$, we have that the minimum approximation error in \eqref{approx_far} becomes $\bigO\Big(n^{-\frac{\tau-2}{\tau+1}}(\log n)^{3\theta}\Big)$. 

We will conclude this subsection by extending the functional AR approximation result to the case when the temporal dependence is of exponential decay in the following statement.
\begin{remark}\label{remark}
	When the temporal dependence of the covariance structure in \cref{weak_depen_components} is changed to exponential decay, i.e., $\sup_{t\in[0,1]}\Vert\bm{\Gamma}(t,j)\Vert \le C\rho^{|j|}$ with $\rho\in (0,1)$, then the optimal choice of $b\asymp \log n$. Consequently, the functional AR approximation result \eqref{approx_far} in \cref{approx_ar} will be updated to
	\begin{equation*}
	Y_i(u)=\sum_{j=1}^{\min\{i-1,b\}}\int_0^1 \psi_j^{(p)}(\frac{i}{n},u,v)Y_{i-j}(v)\dee v+\varepsilon_i(u)+\bigO_\Pr\left(\frac{p^{1/2}
 \log^3 n}{n}+p^{-d_1}\right).
	\end{equation*}
\end{remark}
In this case, the optimal choice for the truncation number $p\asymp n^{\frac{2}{2d_1+1}}(\log n)^{-\frac{6}{2d_1+1}}$ and the error term turns out to be $\bigO_\Pr\Big(
n^{-\frac{2d_1}{2d_1+1}}(\log n)^{\frac{6d_1}{2d_1+1}}\Big)$.

\section{Applications to optimal short-term forecast for locally stationary functional time series}\label{sec_3}
In this section, we will discuss the application of our functional AR approximation theory to optimal short-term forecasting for locally stationary functional time series. Generally speaking, \cref{approx_ar} provides a theoretical guarantee for the optimal short-term linear forecasting of a short-memory locally stationary functional time series by a locally stationary functional AR process of slowly diverging order. Section \ref{sec_31} will discuss the details. Provided that the underlying data generating mechanism is sufficiently smooth and the temporal dependence is sufficiently weak, the unknown coefficients $\psi_{j,kl}(t)$ in the basis expansion of $\psi_j^{(p)}(t,u,v)$ can be consistently estimated via a Vector Auto-Regressive (VAR) approximation and the method of sieves, which will be implemented in \cref{sec32} and \cref{sec33}.

\subsection{Optimal functional time series prediction}\label{sec_31}
In this paper, we shall focus on the best continuous linear prediction of functional time series; that is, given $i\ge 2$ and $Y_1(u),\cdots, Y_{i-1}(u)$, we try to find a linear predictor $\widehat{Y}_i(u)$ of $Y_i(u)$ in the form 
\begin{equation}\label{best_ori}
\widehat{Y}_i(u)=\sum_{j=1}^{i-1}\int_0^1
g_{i,j}(u,v)Y_{i-j}(v)\dee v
\end{equation}
such that $\EE|Y_{i}(u)-\widehat{Y}_{i}(u)|_{\mathcal{L}^2}^2$ is minimized, where the kernel function $g_{i,j}(u,v)\in \mathcal{L}([0,1]^2)$ is continuous over $u$ and $v$ for all $i$ and $j$. 
The goal of this subsection is to investigate the optimal short-term continuous prediction of locally stationary functional time series $\{Y_i(u)\}_{i=1}^n$. 

To begin with, we consider the truncated process $\{Y_i^{(p)}(u)\}_{i=1}^n$ defined in \eqref{truncation} and let the best linear predictor of $Y_i^{(p)}(u)$ in terms of $Y_1^{(p)}(u),\cdots, Y_{i-1}^{(p)}(u)$ be $\widehat{Y}_i^{(p)}(u)$. The next theorem shows the asymptotic equivalence of the best continuous linear predictor $\widehat{Y}_{n+1}(u)$  and $\widehat{Y}_{n+1}^{(p)}(u)$. 

\begin{theorem}\label{thm_error1}
	Define prediction errors as ${\rm PE}_{n+1}=Y_{n+1}(u)-\widehat{Y}_{n+1}(u)$ and ${\rm PE}_{n+1}^{(p)}=Y_{n+1}^{(p)}(u)-\widehat{Y}_{n+1}^{(p)}(u)$. Suppose Assumptions \ref{ass_conti_u} and \ref{weak_depen_components} hold, then we obtain 
	$$\EE| {\rm PE}_{n+1}|_{\mathcal{L}^2}^2-\EE| {\rm {PE}}_{n+1}^{(p)}|_{\mathcal{L}^2}^2=\bigO\left(p^{-(d_1+1)}\right).$$
\end{theorem}

This theorem illustrates that, by the fact that $Y_{n+1}(u)=Y_{n+1}^{(p)}(u)+\bigO_\Pr(p^{-d_1})$ in \eqref{truncation}, the best linear predictor $\widehat{Y}_{n+1}^{(p)}(u)$ and the best continuous linear predictor $\widehat{Y}_{n+1}(u)$ are asymptotically equivalent as $p\to \infty$. Next, denote the Auto Regressive (AR) predictor \begin{equation}\label{best_conti}
\widetilde{Y}_{i}^{(b)}(u):=\sum_{j=1}^{\min\{i-1,b\}} \int_0^1\psi_j^{(p)}(\frac{i}{n},u,v)Y_{i-j}(v)\dee v,
\end{equation}
which is the dominating term on the right hand side of \eqref{approx_far} of our AR approximation theory. The following theorem states that the best continuous linear predictor $\widehat{Y}_{n+1}(u)$ can be well approximated by the AR predictor $\widetilde{Y}_{n+1}^{(b)}(u)$. 


\begin{theorem}\label{thm_error2}
	Denote the prediction error ${\rm PE}_{n+1}^{(b)}:=Y_{n+1}^{(p)}(u)-\widetilde{Y}_{n+1}^{(b)}(u)$. Suppose that Assumptions \ref{ass_conti_u}--\ref{updc} hold and $\sup_{t\in[0,1]}\Vert \partial\bm{\Gamma}(t,j)/\partial t\Vert \le C$ for all $j\in \mathbb{Z}$, we have
	\begin{equation}\label{rate}
	\EE|{\rm PE}_{n+1}|_{\mathcal{L}^2}^2-
	\EE|{\rm PE}_{n+1}^{(b)}|_{\mathcal{L}^2}^2
	=\bigO\left(p^{-(d_1+1)}+pb^{-2\tau+3}(\log b)^{2\tau-3}+pb^5/n^2\right).
	\end{equation}
\end{theorem}

\cref{thm_error2} implies that the error bound on the right hand side of \cref{rate} converges to $0$ as $p, n\to \infty$. Specifically, with the optimal choice of $b\asymp n^{\frac{1}{\tau+1}}(\log n)^{\theta}$ where $\theta=\frac{\tau-3/2}{\tau+1}$, the optimal MSE rate \eqref{rate} turns out to be $\bigO\Big(n^{-\frac{(d_1+1)(2\tau-3)}{(d_1+2)(\tau+1)}}(\log n)^{\frac{5\theta(d_1+1)}{d_1+2}}\Big)$ by choosing $p\asymp n^{\frac{2\tau-3}{(d_1+2)(\tau+1)}}(\log n)^{-\frac{5\theta}{d_1+2}}$. Furthermore, if the smoothness parameter $d_1\to \infty$, the rate becomes $\bigO\left(n^{-2+5/(\tau+1)}(\log n)^{5\theta}\right)$. It consequently indicates that the AR predictor $\widetilde{Y}_{n+1}^{(b)}(u)$ is asymptotically equivalent to the best continuous linear predictor $\widehat{Y}_{n+1}(u)$ when $\tau>3/2$.

\begin{remark}
    Alternatively, if the covariance structure in \cref{weak_depen_components} decays exponentially fast, then one could choose $b\asymp \log n$ and
    the result in Theorem \ref{thm_error2} can be updated as 
    \begin{equation}\label{rate_exp}
        \EE|{\rm PE}_{n+1}|_{\mathcal{L}^2}^2-
	\EE|{\rm PE}_{n+1}^{(b)}|_{\mathcal{L}^2}^2
	=\bigO\left(p^{-(d_1+1)}+p(\log n)^5 /n^2\right).
    \end{equation}
    Hence, \eqref{rate_exp} equals $\bigO\left(n^{-\frac{2(d_1+1)}{d_1+2}}(\log n)^{\frac{5(d_1+1)}{d_1+2}}\right)$ by choosing $p \asymp n^{\frac{2}{d_1+2}}(\log n)^{-\frac{5}{d_1+2}}$. Similarly, if in addition $d_1\to\infty$, then the right hand side of \eqref{rate_exp} becomes $\bigO(n^{-2}(\log n)^{5})$.
\end{remark}

\subsection{Vector Auto-Regressive approximation}
\label{sec32}
With the theoretical results in \cref{sec_31}, it is clear that the short-term forecasting for locally stationary functional time series is equivalent to exploring the optimal short-term continuous linear prediction by a locally stationary functional AR process. On the other hand, we will demonstrate that the unknown coefficients $\psi_{j,kl}(t)$ in the basis expansion of $\psi_j^{(p)}(t,u,v)$ in \eqref{best_ori} can be determined by the coefficient matrix in a smoothly-varying VAR approximation. Then it turns out that the optimal short-term forecasting problem boils down to that of efficiently estimating the smoothly-varying VAR coefficient matrices at the right boundary. To this end, in this subsection we start with the prediction coefficient matrix $\bm{\Phi}_{i,j}$ defined in \cref{sec_22} and investigate its estimation. 

Consider the time series $\{\bm{x}_i\}$ of diverging dimension $p_n$ and we establish its VAR approximation. Consider the following best linear predictions:
\begin{equation}\label{var}
\bm{x}_i=\sum_{j=1}^{i-1}\bm{\Phi}_{i,j}\bm{x}_{i-j}+\bm{\epsilon}_i,
~i=2,...,n,
\end{equation}
where $\bm{\Phi}_{i,j}$ and $\bm{\epsilon}_i$ have been defined in \cref{sec_22}. Let $\bm{x}_{i-1}^{(i)}=
(\bm{x}_{i-1}^\top,...,\bm{x}_1^\top)^\top\in 
\mathbb{R}^{(i-1)p}$ be a block vector and $\bm{\Gamma}_i={\rm Cov}(\bm{x}_{i-1}^{(i)},\bm{x}_{i-1}^{(i)})\in \mathbb{R}^{(i-1)p\times(i-1)p}$ be the covariance matrix of $\bm{x}_{i-1}^{(i)}$. Similar to the univariate AR approximation result established in \cite{Ding21}, we will demonstrate that a rich class of locally stationary multivariate time series $\bm{x}_i$ can be well approximated by a VAR($b$) process under some mild conditions.

Now, denote $\bm{\Phi}_i=(\bm{\Phi}_{i,1}^\top,...,\bm{\Phi}_{i,i-1}^\top)^\top\in \mathbb{R}^{(i-1)p\times p}$, then we have the Yule-Walker equation 
$$\bm{\Phi}_i=\bm{\Omega}_i\bm{\gamma}_i,$$
where $\bm{\Omega}_i=\bm{\Gamma}_i^{-1}$ and $\bm{\gamma}_i={\rm Cov}(\bm{x}_{i-1}^{(i)},\bm{x}_i)\in \mathbb{R}^{(i-1)p\times p}$. See \citep[Section 11.3]{BD91} for more details on Yule-Walker equation for multivariate time series. We shall first state the following results regarding the coefficient matrices $\bm{\Phi}_{i,j}$.

\begin{proposition}\label{prop1}
	Under Assumptions \ref{weak_depen_components} and \ref{updc}, then for VAR process \eqref{var}, there exists some constant $C>0$ such that
	\begin{equation}\label{phi_order}
	\max_i\Vert\bm{\Phi}_{i,j}\Vert \le C\left(\frac{j}{\log j+1}\right)^{-\tau+1},~~\text{for}~~j\ge 1.
	\end{equation}
\end{proposition}
\cref{prop1} provides a polynomial decay rate of the coefficient matrices $\bm{\Phi}_{i,j}$ in \eqref{phi_order} when $\tau>1$. 

Next, define $\bm{\Phi}^{(b)}(t)=(\bm{\Phi}_1^\top(t),...,\bm{\Phi}_b^\top(t))^\top\in\mathbb{R}^{bp\times p}$ via the Yule-Walker equation $$\bm{\Phi}^{(b)}(t)=\bm{\Gamma}_n^{-1}(t)\bm{\gamma}_n(t),$$
where $\bm{\Gamma}_n(t) \in\mathbb{R}^{bp\times bp}$ with its $(i,i+j)$th block matrix as $\bm{\Gamma}(t,j)$ for $j=0,\pm 1,...,\pm(b-1)$ and $\bm{\gamma}_n(t)=(\bm{\Gamma}^\top(t,1),\cdots,\bm{\Gamma}^\top(t,b))^\top\in
\mathbb{R}^{bp\times p}$. It is worth mentioning that there exists a one-to-one mapping from the coefficient matrix $\bm{\Phi}_j(t)$ to the coefficient function $\psi_j^{(p)}(t,u,v)$ for $j=1,...,b$ in light of the fact that $\psi_j^{(p)}(t,u,v)=\bm{\alpha}_f^\top(u){\rm diag}(f_1,\cdots,f_p)\bm{\Phi}_{j}(t)\\{\rm diag}(1/f_1,\cdots,1/f_p)
\bm{\alpha}_f(v)$ where $\bm{\alpha}_f(\cdot)$ is defined in \cref{approx_ar}, and we refer readers to find out more details in the proof of \cref{approx_ar} in Section C.1 of the Supplementary Material. Next proposition implies that the coefficient matrix $\bm{\Phi}_{i,j}$ can be well approximated by the smooth function $\bm{\Phi}_j(\frac{i}{n})$ when $i>b$ and $1\le j\le b$.

\begin{proposition}\label{prop2}
	Under Assumptions \ref{ass_conti_t}--\ref{updc}, for any $j=1,...,b$, we have $\bm{\Phi}_j(t)=\{\Phi_{j,kl}(t)\}_{k,l=1}^p \\ \in \mathcal{C}^{d_2}([0,1])$, that is each entry of $\bm{\Phi}_j(t)$ is $d_2$ times continuously differentiable over $[0,1]$. Furthermore suppose $\sup_{t\in[0,1]}\Vert \partial\bm{\Gamma}(t,j)/\partial t\Vert \le C$ for all $j\in \mathbb{Z}$ holds true, then there exists some constant $C>0$, such that for all $j=1,...,b$,
	\begin{equation}\label{prop2_error}
	\max_{i>b}\left\Vert\bm{\Phi}_{i,j}-\bm{\Phi}_j\left(\frac{i}{n}\right)
	\right\Vert \le C\left(b^{-\tau+1}(\log b)^{\tau-1}+
	\frac{b^2}{n}\right).
	\end{equation}
\end{proposition}

The first term on the right hand side of \eqref{prop2_error} is the truncation error by using VAR$(b)$ process to approximate the VAR$(i-1)$, which is also the error rate in Lemma 6 in Section C.2 of the Supplementary Material. The second part is the error caused by using the smooth VAR coefficient matrices $\bm{\Phi}_j(\cdot)$ to approximate $\bm{\Phi}_{i,j}$ for all $j=1,...,b$. 

Combing Propositions \ref{prop1} and \ref{prop2}, we can rewrite \eqref{var} as
\begin{equation}\label{approx2}
\left\Vert \bm{x}_i-\sum_{j=1}^{\min\{i-1,b\}}\bm{\Phi}_j(\frac{i}{n})
\bm{x}_{i-j}-\bm{\epsilon}_i\right\Vert
=\bigO_\Pr\left(p^{1/2}b^{-\tau+2}(\log b)^{\tau-1}+\frac{p^{1/2}b^{3}}{n}\right),~i\ge 2.
\end{equation}
This formula indicates that the multivariate time series $\bm{x}_i$ can be approximated by a locally stationary VAR$(b)$ process with smoothly time-varying coefficients as long as the error terms on the right-hand side of \eqref{approx2} vanish as $n\to \infty$.

\subsection{Sieve estimation for coefficient matrices}\label{sec33}
Based on our discussions in Sections \ref{sec_31} and \ref{sec32}, optimal prediction of locally stationary functional time series boils down to efficient estimation of the matrix functions $\bm{\Phi}_j(\cdot)$, $j=1,2,\cdots b$. This subsection is devoted to the estimation of such matrix coefficient functions. Observe that the smoothness of $\bm{\Phi}_j(t)$ over $t$ (c.f. \cref{prop2}) allows us to conduct another basis expansion and thus estimate them via the method of sieves. Employing the method of sieves on the time-varying coefficient matrices can effectively reduce the dimension of the parameter space in the sense that one only needs to perform a multiple linear regression with a slowly diverging number of predictors. Prior to utilizing this method, we need an assumption concerning the derivatives of $\bm{\Phi}_j(t)$.
\begin{assumption}\label{ass_deriv}
	The derivatives of $\bm{\Gamma}(t,j)$ over $t$ decay with $|j|$ as follows
	$$\sup_{t\in[0,1]}
 \sum_{j\in\mathbb{Z}}\left\Vert \frac{\partial^{d_2} \bm{\Gamma}(t,j)}{\partial t^{d_2}}\right\Vert<\infty.$$ 
	\end{assumption}
We mention that \cref{ass_deriv} entails that all derivatives of $\bm{\Gamma}(t,j)$ over $t$ up to the order $d_2+1$ decay as $|j|$ increases. For the linear process in \cref{multi_model_example}, this condition will be satisfied if $\sup_{t\in[0,1]}\Vert \partial^{d_2} \bm{A}_m(t)\big/ \partial t^{d_2}\Vert \le C(m+1)^{-\tau}$. See Lemma 3 of the Supplementary Material for detailed proof. Armed with the relation $\bm{\Phi}_j(t)=\bm{E}_j^\top\bm{\Omega}_n(t)\bm{\gamma}_n(t)$ where $\bm{E}_j^\top\in \mathbb{R}^{p\times bp}$ has $\bm{I}_p$ at its $j$th block and $\bm{0}_p$ at others, the general Leibniz rule as well as the implicit differentiation, it also guarantees that the $d_2$th derivative of $\bm{\Phi}_j(t)$ over $t$ is bounded (\citep[Section 2.3.1]{Chen07}), so that $\bm{\Phi}_j(\cdot)$ can be approximated by linear sieves. 

Let $\Phi_{j}^{(lm)}(\frac{i}{n})$ be the $(l,m)$th element of coefficient matrix $\bm{\Phi}_j(\frac{i}{n})$ for $l,m=1,...,p$. 
By \citep[Section 2.3]{Chen07} and \cref{ass_deriv}, we have that for any $j=1,...,b$, 
\begin{equation}\label{truncation_coef}
\Phi_j^{(lm)}\left(\frac{i}{n}\right)=\sum_{k=1}^c \phi_{jk}^{(lm)}v_k(\frac{i}{n})+
\bigO(c^{-d_2}),~i>b,
	\end{equation}
	where $\phi_{jk}^{(lm)}$ for $j=1,...,b$ and $k=1,...,c$ is the $(l,m)$th element in the coefficient matrix $\bm{\phi}_{jk}$, $\{v_k(\cdot)\}$ is also a set of pre-chosen orthogonal basis functions on $[0,1]$ and $c=c(n)$ is the truncation number of basis functions. Notice that the first basis expansion in \eqref{truncation} and the second basis expansion in \eqref{truncation_coef} constitute our methodology of double-sieve expansion. Furthermore, let  $\bm{z}_{kj}(\frac{i}{n}):=v_k(\frac{i}{n})\bm{x}_{i-j}$ and similar to \eqref{approx2}, we have for $i=b+1,...,n$,
	\begin{equation}\label{approx3}
	\left\Vert\bm{x}_i-\sum_{j=1}^b\sum_{k=1}^c\bm{\phi}_{jk}\bm{z}_{kj}
	\left(\frac{i}{n}\right)-\bm{\epsilon}_i\right\Vert
	=\bigO_\Pr\left(p^{1/2}b^{-\tau+2}(\log b)^{\tau-1}+\frac{p^{1/2}b^{3}}{n}+b^{1/2}pc^{-d_2}\right).
	\end{equation}
	
 As a result, the estimation of unknown parameter $\bm{\phi}_{jk}$ boils down to dealing with the above multiple linear regression problem \eqref{approx3}. Equipped with the method of double-sieve expansion and the VAR approximation \eqref{approx3}, one can  consequently estimate the functional AR coefficient as described in \eqref{approx_far}. In order to facilitate the estimation of coefficient matrix $\bm{\phi}_{jk}$, we impose a regularity condition.
\begin{assumption}\label{ass5}
	For any $j=1,...,b$, denote $\bm{W}^{(j)}(t)\in\mathbb{R}^{jp\times jp}$ with its $(k,l)$th block entry $\bm{W}_{kl}^{(j)}(t)=\bm{\Gamma}(t,k-l)
	\in\mathbb{R}^{p\times p}$ for $k,l=1,...,j$. We assume that the eigenvalues of $$\int_0^1 \bm{W}^{(j)}(t)\otimes(\bm{v}(t) \bm{v}^\top(t))~\dee t$$ are bounded above and below from zero by a constant $\kappa_2>0$, where $\bm{v}(t)=(v_1(t),...,v_c(t))^\top\in \mathbb{R}^c$. 
\end{assumption}
Since $\bm{W}^{(j)}(t)\otimes (\bm{v}(t)\bm{v}^\top(t))$ is positive semi-definite for all $t\in [0,1]$, the above integral is always positive semi-definite. This assumption is mild and it is easy to check that when $\bm{x}_i$ is a stationary process with weak inter-element dependence, the above assumption will hold immediately by UPDC condition and the orthonormality of the basis functions. Moreover, this condition guarantees the invertibility of the design matrix $\bm{Y}$ in the following equation \eqref{multi_linear} and the existence of the least squares solution.
 
In this context, denote by $\bm{\beta}$ the $bcp\times p$ block matrix  with block rectangular element $\{\bm{\beta}_j\}_{j=1}^b\in \mathbb{R}^{cp\times p}$, where $\bm{\beta}_j=(\bm{\phi}_{j1},...,\bm{\phi}_{jc})^\top$. Let $s=1,...,bc,~j_s=\lfloor{\frac{s-1}{c}}\rfloor+1$ and $k_s=s-\lfloor{\frac{s-1}{c}}\rfloor\times c$, then we can define $\bm{y}_i\in \mathbb{R}^{bcp}$ by letting its block vector  $\bm{y}_{is}=\bm{z}_{k_s,j_s}(\frac{i}{n})$. Moreover, let $\bm{Y}^\top$ be the $bcp\times(n-b)$ rectangular matrix whose columns are $\{\bm{y}_i\}_{i=b+1}^n$ and we also denote $\bm{x}=(\bm{x}_{b+1},...,\bm{x}_n)^\top\in\mathbb{R}^{(n-b)\times p},~\bm{\epsilon}=
(\bm{\epsilon}_{b+1},...,\bm{\epsilon}_n)^\top\in\mathbb{R}^{(n-b)\times p}$. Then the matrix form of the multiple linear regression for \eqref{approx3} can be constructed as
\begin{equation}\label{multi_linear}
\bm{x}=\bm{Y}\bm{\beta}+\bm{\epsilon}+\bm{Q}_1+\bm{Q}_2,
\end{equation}
where 
{\footnotesize
\begin{align*}
\bm{Q}_1&=\left(\sum_{j=1}^b\left[\bm{\Phi}_{b+1,j}-\bm{\Phi}_j(\frac{b+1}{n})\right]
\bm{x}_{b+1-j},\cdots,\sum_{j=1}^b\left[\bm{\Phi}_{n,j}-\bm{\Phi}_j(1)\right]
\bm{x}_{n-j}\right)^\top,\\
\bm{Q}_2&=\left(\sum_{j=1}^b
\sum_{k=c+1}^\infty\bm{\phi}_{jk}
\bm{z}_{kj}(\frac{b+1}{n}),
\sum_{j=1}^b\sum_{k=c+1}^\infty
\bm{\phi}_{jk}\bm{z}_{kj}(\frac{b+2}{n})+\bm{\Phi}_{b+2,b+1}\bm{x}_1,\cdots,
\sum_{j=1}^b\sum_{k=c+1}^\infty
\bm{\phi}_{jk}\bm{z}_{kj}(1)+
\sum_{j=b+1}^{n-1}
\bm{\Phi}_{n,j}\bm{x}_{n-j}\right)^\top.
\end{align*}}
\noindent
According to the proof of \cref{prop3} in Section C.2 of the Supplementary Material, we find that the error terms $\bm{Q}_1, \bm{Q}_2$ are negligible in the regression  \eqref{multi_linear}, then by the multiple least squares method, we have $$\widehat{\bm{\beta}}=(\bm{Y}^\top\bm{Y})^{-1}\bm{Y}^\top\bm{x} \approx \bm{\beta}+(\bm{Y}^\top\bm{Y})^{-1}\bm{Y}^\top\bm{\epsilon}.
$$ 
Similarly, one can decompose the estimator $\widehat{\bm{\beta}}$ into its block elements, denoted by $\{\widehat{\bm{\beta}}_j\}_{j=1}^b\in\mathbb{R}^{cp\times p}$. Hence, the estimate of the time-varying coefficient matrix in \eqref{approx2} can be represented as \begin{equation}\label{phi_est}
\widehat{\bm{\Phi}}_j(\frac{i}{n})=\widehat{\bm{\beta}}_j^\top\bm{A}(\frac{i}{n})
\quad \text{with}~\bm{A}(\cdot)=(v_1(\cdot)\bm{\I}_p, ...,v_c(\cdot)\bm{\I}_p)^\top\in \mathbb{R}^{cp\times p}.
\end{equation} 
The rest of this subsection is devoted to the investigation of the convergence rate of $\widehat{\bm{\Phi}}_j(\cdot)$. We consider the difference
\begin{equation}\label{coef_matrix_diff}
\widehat{\bm{\Phi}}_j\left(\frac{i}{n}\right)-\bm{\Phi}_j\left(\frac{i}{n}\right)
=(\widehat{\bm{\beta}}_j-\bm{\beta}_j)^\top \bm{A}\left(\frac{i}{n}\right)-\bm{\Delta}_c,
\end{equation}
where $\bm{\Delta}_c$ is a $p\times p$ matrix with its $(l,m)$th entry being $\sum_{k=c+1}^\infty \phi_{j,k}^{(lm)}v_k(\frac{i}{n})$. Till the end of this paper, we assume $c=\bigO(n^{\nu_1})$ and denote $\zeta_c:=\sup_{t}\Vert\bm{v}(t)\Vert$ where $\bm{v}(\cdot)$ is defined in \cref{ass5}. Several additional assumptions are needed. First from \cref{physi_repre_func} and the basis expansion in \cref{infinite}, we assume that 
$\{\bm{x}_i\}$ admits a physical representation 
	\begin{equation}\label{physi_vector}
		 \bm{x}_i=\bm{G}(\frac{i}{n},\mathcal{F}_i),
	\end{equation}
 where $\bm{G}=(G_1,...,G_p)^\top$ is a measurable function similar to $H$ defined in \eqref{physi_repre_func}. The representation form \eqref{physi_vector} includes many commonly used locally stationary time series models, see for \cite{WZ2011,Zhang18} for examples. Consequently, the the $k$th entrywise of $\bm{x}_{i}$ can be written as $x_{i,k}=G_k(\frac{i}{n},\mathcal{F}_i)$. Under the above physical representation, we define the physical dependence measure for the functional time series $\{Y_i(u)\}$ with respect to the basis $\{\alpha_k(u)\}_{k=1}^\infty$ as
	\begin{equation}\label{phy_dep_x}
	\delta_x(l,q)=\sup_{t\in [0,1]}\max_{1\le k\le\infty}\Vert G_k(t,\mathcal{F}_i)-G_{k}(t,\mathcal{F}_{i,l})\Vert_q,~l\ge 0,
	\end{equation}
	where $\mathcal{F}_{i,l}=(\mathcal{F}_{i-l-1},\eta_{i-l}^\ast,\eta_{i-l+1},\cdots,
	\eta_i)$ with $\eta_{i-l}^\ast$ 
 being an i.i.d. copy of $\eta_{i-l}$.

\begin{assumption}\label{depen_decay}
	There exists some constant $\tau>1$ such that for some constant $C>0$, the physical dependence measure in \eqref{phy_dep_x} satisfies $\delta_x(l,q)\le C(l+1)^{-\tau}$ for $l\ge 0$.
	\end{assumption}
	
    
\begin{assumption}\label{ass6}
	For some constant $C>0$,\\
	(i) there exist $\omega_1,~\omega_2\ge 0$ such that $\sup_t\Vert\nabla\bm{v}^\top(t)\Vert \le Cn^{\omega_1}c^{\omega_2}$ where 
	$\nabla\bm{v}^\top(t)$ is the first derivative with respect to $t$.\\ 
	(ii) there exist $\bar{\omega}_1\ge 0,~\bar{\omega}_2>0$ such that $\zeta_c\le Cn^{\bar{\omega}_1}c^{\bar{\omega}_2}$.
\end{assumption}

\begin{assumption}\label{ass7}
	We assume that the smoothness order $d_2$ defined in \cref{ass_conti_t}, the order $\tau$ for the temporal dependence of the locally stationary process, the order $\beta_1$ for the truncation number $p$ of the first sieve expansion and the order $\nu_1$ for the truncation number $c$ of the second sieve expansion satisfy 
	\begin{equation}\label{order_constraint}
	\frac{C}{\tau+1}+2\beta_1+4\nu_1<1,~~2(d_2-1)\nu_1>\beta_1+\frac{1}{\tau+1}, 
	\end{equation}
	where $C>2$ is some finite constant. 
\end{assumption}
We comment on the above conditions. \cref{depen_decay} imposes a polynomial decay speed on the physical dependence measure, which implies a short-range dependence property of the functional time series. We refer readers to Examples 1 and 2 in Section A.2 of the Supplementary Material on how to calculate $\delta_x(l,q)$ for a class of functional MA$(\infty)$ and functional AR$(1)$ processes, respectively. \cref{ass6} is a mild condition for basis functions. For example, $\omega_1=0, \omega_2=1/2, \bar{\omega}_1=0, \bar{\omega}_2=1/2$ for tensor-products of univariate polynomial splines and orthogonal wavelets, while $\omega_1=0, \omega_2=3, \bar{\omega}_1=0, \bar{\omega}_2=1$ when we use tensor-products of orthonormal Legendre polynomial bases (see, e.g., \cite{NEWEY1997147}, \cite{Huang98} and \cite{Chen07}). \cref{ass7} puts some mild constraints to control the error bound in the technical proof. Notice that if we choose the optimal $b\asymp n^{\frac{1}{\tau+1}}(\log n)^{\theta}$ with $\theta=\frac{\tau-3/2}{\tau+1}$, the truncation number $p\asymp n^{\frac{2\tau-3}{(\tau+1)(d_1+2)}}(\log n)^{\frac{-5\theta}{d_1+2}}$ studied in \cref{sec_22} and the optimal $c$ discussed in \cref{coro2}, then \eqref{order_constraint} can be easily satisfied by properly choosing smoothing parameters $d_1$ and $d_2$. When the physical dependence is of exponential decay, the constraint \eqref{order_constraint} will be reduced to $2\beta_1+4\nu_1<1$ and $2(d_2-1)\nu_1>\beta_1$. In the following, we will show the estimation consistency of the coefficient matrix.

\begin{proposition}\label{prop3}
	With Assumptions \ref{ass_conti_t}, \ref{weak_depen_components}, \ref{ass_deriv}--\ref{ass7}, we have 
	\begin{equation}\label{consis_mat}
	\max_{i>b,j\le b}\left\Vert\widehat{\bm{\Phi}}_j
	\left(\frac{i}{n}\right)-\bm{\Phi}_j\left(\frac{i}
	{n}\right)\right\Vert\le C\left(\zeta_c^2
 \sqrt{\frac{bp\log n}{n}}
 +\zeta_c\sqrt{bp}c^{-d_2}\right).
	\end{equation}
\end{proposition}

As we can see from \cref{prop3}, the convergence rate on the right hand side of \eqref{consis_mat} comprises of the standard deviation and bias term, respectively. Furthermore, the above proposition indicates that $\widehat{\bm{\Phi}}_j(\frac{i}{n})$ are consistent estimators for $\bm{\Phi}_j(\frac{i}{n})$ uniformly in $i>b$ for all $j=1,...,b$. The \cref{coro2} below provides the optimal convergence rate by balancing the aforementioned two types of errors.

\begin{corollary}\label{coro2}
	Under conditions in \cref{prop3}, when one uses the orthonormal bases with the fastest decay rates for its basis expansion coefficients and chooses $c\asymp (n/\log n)^{\frac{1}{2d_2+1}}$ by balancing the standard deviation term and the bias term in \eqref{consis_mat}, we have
 $$\max_{i>b,j\le b}\left\Vert\widehat{\bm{\Phi}}_j
	\left(\frac{i}{n}\right)-\bm{\Phi}_j\left(\frac{i}
	{n}\right)\right\Vert\le C
\sqrt{bp}\left(\frac{n}{\log n}\right)^{\frac{-d_2+1/2}{2d_2+1}}.$$	
 On the other hand, if we employ the basis functions with a slower decay rate and selects the optimal truncation number $c\asymp (n/\log n)^{\frac{1}{2(d_2+1)}}$, then \eqref{consis_mat} becomes $\bigO\Big(\sqrt{bp}\big(\frac{n}
 {\log n}\big)^{\frac{-d_2+1}{2(d_2+1)}}\Big)$. 
\end{corollary}

\begin{remark}
The basis functions with the fastest decay speed at its basis expansion coefficient includes trigonometric polynomials, spline series, orthogonal wavelets and weighted orthogonal Chebyshev polynomials; On the other hand, normalized Legendre polynomials are an example where the basis expansion coefficients decay at slower speeds. Additionally, when $d_1, d_2=\infty$ and $\tau$ is sufficiently large, the convergence rate in \cref{coro2} reduces to $\bigO(\sqrt{\log n/n})$.
\end{remark}

\subsection{Asymptotically optimality of empirical predictors}\label{sec34}
This subsection concludes the consistency of our empirical (estimated) optimal linear predictor with the original best linear continuous predictor. 
Let $$\widehat{\bm{x}}_{n+1}^{(b)}:=\sum_{j=1}^b
\widehat{\bm{\Phi}}_j(1)\bm{x}_{n+1-j}$$ 
be the estimated linear forecast of $\bm{x}_{n+1}$ using our double-sieve method. From the VAR process \eqref{approx3} and discussions in \cref{sec33}, we can write the estimated functional predictor as
\begin{equation}\label{opt_fun}
\widehat{Y}_{n+1}^{(b)}(u)=\bm{\alpha}_f^\top(u)
\sum_{j=1}^b\widehat{\bm{\Phi}}_j(1)\bm{x}_{n+1-j}.
\end{equation}
In the next theorem, we demonstrate that the best linear continuous predictor $\widehat{Y}_{n+1}(u)$ can be well approximated by the empirical functional predictor $\widehat{Y}_{n+1}^{(b)}(u)$. 

\begin{theorem}\label{thm_error3}
	Under conditions in \cref{thm_error2} and further suppose Assumptions \ref{ass_conti_t}--\ref{ass7} hold. Denote the prediction error as $\widehat{\rm PE}_{n+1}^{(b)}:=Y_{n+1}(u)-\widehat{Y}_{n+1}^{(b)}(u)$, then we have 
	\begin{align}\label{final_error}
	&\EE| {\rm {PE}}_{n+1}|_{\mathcal{L}^2}^2-\EE|\widehat{\rm PE}_{n+1}^{(b)}|_{\mathcal{L}^2}^2\notag \\
	=&\bigO\left(p^{-(d_1+1)}+
	pb^{-2\tau+3}(\log b)^{2\tau-3}
	+pb^5/n^2+
 \frac{bp\zeta_c^4\log(n)}{n}+bp\zeta_c^2c^{-2d_2}\right).
	\end{align}
\end{theorem}

The error terms on the right hand side of \eqref{final_error} contain four types of error rates, sequentially from left to right including the prediction error by our first sieve expansion from \cref{thm_error1}, the error from the VAR approximation, the error by smoothing approximation of VAR coefficient matrices as well as the last two terms as estimation errors of the smoothed VAR coefficients, respectively. In the following, we will discuss the optimal error rate in \eqref{final_error} with some typical cases, and ultimately conclude the asymptotic optimality of our empirical functional linear predictor. 

\begin{corollary}\label{coro3}
By choosing the optimal truncation number $p\asymp n^{\frac{2\tau-3}{(\tau+1)(d_1+2)}}
(\log n)^{\frac{-5\theta}{d_1+2}}$ with $\theta=\frac{\tau-3/2}{\tau+1}$ in the first basis expansion, the functional AR order $b\asymp n^{\frac{1}{\tau+1}}(\log n)^{\theta}$, together with the optimal truncation number $c\asymp (n/\log n)^{\frac{1}{2d_2+1}}$ for the fastest decay rate of the basis expansion coefficient, then the above result turns out to be
	\begin{align}\label{optimal_rate}
	\EE| {\rm PE}_{n+1}|_{\mathcal{L}^2}^2-\EE| \widehat{\rm {PE}}_{n+1}^{(b)}|
 _{\mathcal{L}^2}^2
=\bigO\left(n^{\frac{1}{\tau+1}+\frac{
2\tau-3}{(\tau+1)(d_1+2)}+
\frac{-2d_2+1}{2d_2+1}}
 (\log n)^{\theta+
 \frac{2d_2-1}{2d_2+1}-\frac{5\theta}{d_1+2}}\right),
	\end{align} 
 where $\theta=\frac{\tau-3/2}{\tau+1}$.
\end{corollary}

In particular, when $\tau$ is sufficiently large, then the error bound in \eqref{optimal_rate} converges to $0$ if $d_1(1-2d_2)+4<0$. Further suppose that the functional time series $Y_i(u)$ is infinitely many differentiable ($d_1=\infty$) over $u\in [0,1]$, \eqref{optimal_rate} reduces to $\bigO\left(\big(n/\log n\big)^{-
\frac{2d_2-1}{2d_2+1}}\right)$. Specifically, if $d_2=\infty$, then the MSE turns out to be $\bigO\big(\log n/n\big)$.

 \section{Practical implementation}\label{prac_imple}
\subsection{Choices of tuning parameters}\label{para}

In this subsection, we will discuss how to choose the tuning parameters in the functional time series forecasting procedure. From Eqs. \eqref{truncation} and \eqref{approx3}, one needs to choose three parameters in order to get an accurate prediction: the truncation number $p$ for the first sieve expansion of the locally stationary functional time series, the lag order $b$ for the functional AR approximation and the truncation number $c$ for the second sieve expansion.

First, we employ the cumulative percentage of total variance (CPV) method to choose the truncated number $p$. For any $p\in \mathbb{N}$, consider the largest $p$ empirical eigenvalues $\lambda_1,\lambda_2,...,\lambda_p$ of $\widehat{\rm Cov}(X(u),X(v))$ for any $X\in \mathcal{L}^2$. The CPV$(p)$ is defined as 
$${\rm CPV}(p):=\sum_{i=1}^p\lambda_i \Big/ \sum_{i=1}^\infty \lambda_i.$$
In the simulation studies, we choose $p$ such that the CPV$(p)$ exceeds a predetermined high percentage value (say 95\% used in the simulation), which means the first $d$ functional principal component scores explain at least 95\% of the variability of the data. 

For the rest of parameters $b$ and $c$, we use Akaike information criterion (AIC) to choose them simultaneously. More specifically, we first propose a sequence of candidate pairs $(b_j,c_k)$ ranging from the initial pair $(1,1)$ to $(w,v)$, where $w,v$ are some given integers. For each pair of the choices $(b_j,c_k)$, we fit a time-varying VAR$(b_j)$ model for the sieve expansion of order $c_k$ and calculate its corresponding AIC as $2b_jc_kp^2-2\log(L)$ where $L$
is the pseudo-Gaussian likelihood function of $\{\bm{x}_i\}_{i=1}^n$. Then we choose the optimal pair $(b_{j^\ast},c_{k^\ast})$ by selecting the minimum AIC.

\subsection{Prediction algorithm by the method of double sieve expansions}\label{procedure}
Here, we describe our prediction algorithm as follows.

\begin{enumerate}[Step 1.]
	\item Choose the truncation number $p$ for the centered functional time series $\{Y_i(u)\}_{i=1}^n$ by CPV criterion in Section \ref{para}.
	\item Decompose the functional time series via the first sieve expansion, calculate $\{f_k\}_{k=1}^p$ and find the scaled sequence $\{x_{i,k}\}_{i=1}^n$ using \eqref{truncation}.
	\item For each pair $(b_j,c_k)$, fit a time-varying VAR($b_j$) model for the scaled multivariate time series $\{\bm{x}_i\}_{i=1}^n$ with the truncation number $c_k$ used in the second sieve expansion, then select the optimal pair $(b_{j^\ast},c_{k^\ast})$ by AIC described in Section \ref{para}.
	\item Estimate the coefficient matrix $\bm{\phi}_{jk}$ for $j=1,...,b_{j^\ast}$ and $k=1,...,c_{k^\ast}$ by multivariate least squares estimation discussed in Section \ref{sec32}, consequently calculate $\widehat{\bm{\Phi}}_j(1)$ by Eq. \eqref{phi_est}. 
	\item Obtain the optimal one-step ahead forecast $\widehat{Y}_{n+1}^{(b)}(u)$ via \eqref{opt_fun}.
\end{enumerate}

\section{Simulation studies}\label{sec_simu}
To show the finite-sample prediction performance of our optimal forecasting algorithm by the method of double sieve expansions (hereafter named sieve method for short), we conduct a comparative simulation study among several state-of-the-art functional forecasting methods. In each simulation scenario, we compare our sieve method with (1) univariate time series forecasting technique proposed by \cite{HS09}, namely, an ARIMA model; (2) Naive method which uses the last observation as a prediction ($\widehat{Y}_{n+1}(u)=Y_n(u)$); (3) Standard functional prediction proposed by \cite{Bosq00} where the multiple testing procedure of \cite{KR13} is used to determine the order $p$ of the functional auto-regressive (FAR) model to be fitted; (4) VAR model introduced by \cite{Aue2015} and (5) VARMA model considered by \cite{KKW2017}. Specifically, we use R package \textsf{forecast} for the ARIMA forecasting method, while for the VAR and VARMA forecasting methods, we employ R packages \textsf{vars} and \textsf{MTS}, respectively.

We will investigate the following four kinds of stationary functional time series models and three types of locally stationary functional time series models, which include both linear and nonlinear cases. Here, we rewrite the basis expansion of a general functional time series as
$Y_i(u)=\bm{\alpha}_\ast^\top(u)\bm{r}_i$ where $\bm{\alpha}_\ast(u)=(\alpha_1(u),\alpha_2(u),\cdots)^\top$ and $\bm{r}_i=(r_{i,1},r_{i,2},\cdots)^\top$. Different models for the random vector scores are specified below. 

\begin{enumerate}[(1)]
	\item \label{case1} Stationary MA(1) model. 
	Let $\bm{r}_i=(r_{i,1},...r_{i,\infty})^\top$, consider $$\bm{r}_i=\bm{\epsilon}_{i}+\bm{A}_1\bm{\epsilon}_{i-1},$$ where $\bm{A}_1$ is a infinite-dimensional matrix with $a$ at its diagonal and $a/3$ at its off-diagonals. In this case, we choose the dependence parameter $a=0.5$ or 1. 
	\item \label{case2} Stationary AR(2) model. Let $\bm{r}_i=(r_{i,1},r_{i,2})^\top$, consider $$\bm{r}_i=\bm{\Phi}_1\bm{r}_{i-1}+\bm{\Phi}_2\bm{r}_{i-2}+\bm{\epsilon}_i,$$ where $\bm{\Phi}_1=\begin{pmatrix}
	0.5 & 0.2\\
	-0.2 & -0.5
	\end{pmatrix}$ and $\bm{\Phi}_2=\begin{pmatrix}
	-0.3 & -0.7\\
	-0.1 & 0.3
	\end{pmatrix}$.
	\item \label{case3} Stationary bivariate bilinear BL(1,0,1,1) model. Let $\bm{r}_i=(r_{i,1},r_{i,2})^\top$, and
	$$\bm{r}_i=\bm{A}\bm{r}_{i-1}+\bm{B}{\rm vec}(\bm{r}_{i-1}\bm{\epsilon}_{i-1}^\top)+\bm{\epsilon}_i,$$
	where $\bm{A}=\begin{pmatrix}
	-0.3 & 0.3\\
	0.4 & 0.5
	\end{pmatrix}$ and $\bm{B}=\begin{pmatrix}
	0.4 & -0.5 & 0.4 & -0.5\\
	0.3 & 0.4 & 0.3 & 0.4
	\end{pmatrix}$.
	\item \label{case4} Stationary BEKK(1,0) model. Let $\bm{r}_i=(r_{i,1},r_{i,2})^\top$, consider
	\begin{align*}
	\bm{r}_i&=\bm{\Sigma}_{i}^{1/2}\bm{\epsilon}_i\\
	\bm{\Sigma}_{i}&=\bm{D}+
 \bm{C r}_{i-1}\bm{r}_{i-1}^\top\bm{C}^\top,
	\end{align*}
	where $\bm{C}=\begin{pmatrix}
	0.5 & 0.2\\
	0.2 & 0.4\end{pmatrix}$ and $\bm{D}=\begin{pmatrix}
	0.4 & 0\\
	0 & 0.3\end{pmatrix}$.
	\item \label{case5} Locally stationary MA(1) model. Similar to Case (\ref{case1}), consider $$\bm{r}_i=\bm{\epsilon}_{i}+\bm{A}_2(\frac{i}{n})\bm{\epsilon}_{i-1},$$ where $\bm{A}_2(\frac{i}{n})=a(2i/n-1)\bm{A}_1$ with $\bm{A}_1$ defined in Case (\ref{case1}). The dependence parameter is also chosen as 0.5 or 1.
	\item \label{case6} Time-varying ARMA(1,1) (TV-ARMA(1,1)) model. Let $\bm{r}_i=(r_{i,1},r_{i,2})^\top$, and
	$$\bm{r}_i=(0.5+2(i/n-0.5)^2)\bm{\Phi r}_{i-1}+\bm{\epsilon}_i+
	\cos(2\pi\frac{i}{n})\bm{\Theta}\bm{\epsilon}_{i-1},$$ where 
	$\bm{\Phi}=\begin{pmatrix}
	0.2 & 0\\
	0 & 0.5 \end{pmatrix}$ and $\bm{\Theta}=\begin{pmatrix}
	0.4 & 0.5\\
	-0.6 & 0.7
	\end{pmatrix}$.
	\item \label{case7} Time-varying threshold AR(1) (TV-TAR(1)) model. Let $\bm{r}_i=(r_{i,1},r_{i,2})^\top$, consider
	\begin{equation*}
	\bm{r}_i=\left\{
	\begin{array}{lr}
	\sin(\pi\frac{i}{n})\bm{\Psi}_1\bm{r}_{i-1}+\bm{\epsilon}_{i}, &r_{i-1,1}\ge0,\\
	-\cos(\pi\frac{i}{n})\bm{\Psi}_2\bm{r}_{i-1}+\bm{\epsilon}_{i}, &r_{i-1,1}<0,
	\end{array}\right.
	\end{equation*}
	where $\bm{\Psi}_1=\begin{pmatrix}
	0.5 & 0.2\\
	-0.2 & 0.5
	\end{pmatrix}$ and $\bm{\Psi}_2=\begin{pmatrix}
	-0.3 & -0.7\\
	-0.1 & 0.3
	\end{pmatrix}$.
\end{enumerate} 

For functional moving average models in Cases \eqref{case1} and \eqref{case5}, denoted by FMA(1), we consider the following data generating processes for the innovations. Let $\bm{e}_i=(e_{i1},e_{i2},...,e_{i\infty})^\top$ i.i.d. follows multivariate normal distribution $\mathcal{MN}(0,\bm{\Sigma}_1)$, where $\bm{\Sigma}_1$ has 1 at diagonal and 0.4 at off-diagonal. The innovation process is generated as $\epsilon_{i1}=e_{i1}, \epsilon_{i2}=0.8e_{i2}, \epsilon_{i3}=-0.5e_{i3}, \epsilon_{ik}=k^{-2}e_{ik}$ for $k\ge 4,~i=1,...,n$. For models \eqref{case2} and \eqref{case6}, let $\bm{e}_i=(e_{i1},e_{i2})^\top$ i.i.d. follow centered multivariate $t$ distribution with degree of freedom 6 and the scale parameter $\bm{\Sigma}_2=\begin{pmatrix}
1 & 0.4\\
0.4 & 1\end{pmatrix}$. The innovation process is generated by $\epsilon_{i1}=e_{i1},~\epsilon_{i2}=0.5e_{i2},~i=1,...,n$ for model \eqref{case2} and $\epsilon_{i1}=(0.4+0.5\sin(2\pi i/n))e_{i1},~\epsilon_{i2}=0.8(0.4+0.5\sin(2\pi i/n))e_{i2}$ for model \eqref{case6}. For models \eqref{case4} and \eqref{case7}, consider that $\bm{e}_i=(e_{i1},e_{i2})^\top$ i.i.d. follows $\mathcal{MN}(0,\bm{\Sigma}_2)$ and $\epsilon_{i1}=e_{i1}$ and $\epsilon_{i2}=0.8e_{i2}$ for $i=1,...,n$. Lastly in the BL(1,0,1,1) model \eqref{case3}, we let $\bm{e}_i=(e_{i1},e_{i2})^\top$ i.i.d. follow $\mathcal{MN}(0,\bm{\Sigma}_3)$ with $\bm{\Sigma}_3=\begin{pmatrix}
0.2 & 0\\
0 & 0.2
\end{pmatrix}$ and $\epsilon_{i1}=e_{i1}$ and $\epsilon_{i2}=0.8e_{i2}$ for $i=1,...,n$.  

For generating functional time series based on our sieve method, the Legendre polynomial basis functions are employed in Cases \eqref{case1}--\eqref{case3}, \eqref{case5} and \eqref{case6}, while the orthogonal Daubechies-9 wavelets based on the father wavelet representation (Eq. (2)) in Example 2 of the Supplementary Material are used in Cases \eqref{case4} and \eqref{case7}. The aim in this simulation study is evaluating the one-step ahead curve forecast accuracy. As discussed in \cref{prac_imple}, we implement one-step ahead prediction for each method and the corresponding forecast accuracy in terms of prediction errors are computed via MSE, which is defined as 
$${\rm MSE}=\frac{1}{N}\sum_{s=1}^N
[Y_{n+1}(u_s)-\widehat{Y}_{n+1}(u_s)]^2,$$ where $N$ is the total number of equally spaced grids. For both stationary and non-stationary functional models, we evaluate the percentage of relative differences (RD) between our sieve method and the optimal approach among five other existing methods. Additionally for locally stationary functional time series cases, we also consider the relative ratio (RR) on the MSE deviations from the true MSE value for our method compared to the optimal method among the aforementioned existing methods. These quantities can be defined as 
$${\rm RD}=\frac{|{\rm MSE_{sieve}}-{\rm MSE_{opt}}|}{\min\{{\rm MSE_{opt}},{\rm MSE_{sieve}}\}}\times 100\%,\quad {\rm RR}=\frac{{\rm MSE_{opt}}-{\rm MSE_{true}}}{{\rm MSE_{sieve}}-{\rm MSE_{true}}},$$ where ${\rm MSE}_{sieve}$ denotes the mean squared error under our sieve method, ${\rm MSE_{opt}}$ is the mean squared error based on the best method among the existing five methods and ${\rm MSE_{true}}$ stands for the true mean squared error of the best linear forecast.

\begin{table*}[htbp!]
	\centering
	\footnotesize
	\caption{Comparison results on forecast accuracy for stationary functional models (1)--(4).}
	\label{T1}
	\vspace*{0.1in}
	\begin{tabular}{|c|cc|cc|cc|cc|cc|cc|cc|}
		\hline
		&\multicolumn{2}{c|}{FMA(1) ($a=0.5$)}&\multicolumn{2}{c|}{FMA(1)  ($a=1$)}&\multicolumn{2}{c|}{FAR(2)}&\multicolumn{2}{c|}{BL(1,0,1,1)}
		&\multicolumn{2}{c|}{BEKK(1,0)}\\
		\hline
		Method&$n=200$&$n=400$&
		$n=200$&$n=400$&$n=200$&$n=400$&
		$n=200$&$n=400$&$n=200$&$n=400$\\
		\hline
		ARIMA&\textit{2.017}&\textit{2.002}&
		\textit{3.019}&2.955&2.681&2.425&
         0.434&0.407&1.091&\textbf{0.992}\\
		\hline
		Naive&3.063&3.015&4.750&4.571&
           4.166&3.681&1.091&1.030&
		1.998&1.984\\
		\hline
		Standard&2.561&2.383&4.486&4.127
           &3.538&3.252&0.447&
	    0.439&1.096&\textit{1.014}\\
		\hline
		VAR&2.221&2.095&3.248&\textbf{2.635}&
            2.280&2.051&\textit{0.412}&
            0.399&1.118&0.997\\
		\hline
		VARMA&\textbf{2.009}&\textbf{1.988}
		&3.493&\textit{2.667}
            &\textbf{2.225}&\textbf{2.012}&
		\textbf{0.408}&\textbf{0.394}&
            \textbf{1.014}&0.999\\
		\hline
  Sieve&2.064&2.020&\textbf{3.006}&2.762
  &\textit{2.241}&\textit{2.027}
           &0.414&\textit{0.396}&
		\textit{1.072}&\textit{0.994}\\
		\hline
  RD(\%)&2.74&1.61&0.43&4.82&0.72&0.75&
  1.47&0.51&5.72&0.20\\
  \hline
	\end{tabular}
\end{table*}

In the simulation experiments for stationary cases \eqref{case1}--\eqref{case4} , we use the sample sizes $n=200, 400$ as the training samples, while for the locally stationary models \eqref{case5}--\eqref{case7}, the training samples are chosen as $n=200, 400$ and $800$. The purpose is to investigate the one-step ahead forecasts at $n=201, 401$ or $801$ and the procedure is repeated for $m=1000$ times. In Table \ref{T1}, we show the comparison results on MSE criterion among the aforementioned methods for stationary time series models \eqref{case1}--\eqref{case4}. The MSE values typically decrease as the sample sizes grow. If no confusion arises, the smallest values of MSE under each case are imposed to be bold and the second smallest values are marked as italic. With the small values of RD in percentage, one will observe that our sieve method is comparable with other methods for all stationary cases. Furthermore, for the stationary FMA(1) scenario, as the temporal dependence of functional time series becomes stronger, the corresponding prediction errors turn out to be larger. One explanation is that variances of estimators become higher under stronger dependence, which reduces the accuracy of predictions.

\begin{table*}[htbp!]
	\centering
	\footnotesize
	\caption{Comparison results on forecast accuracy among six methods for non-stationary functional MA(1) model.}
	\label{T3}
	\vspace*{0.1in}
	\begin{tabular}{|c|ccc|ccc|ccc|}
		\hline
		FMA(1)&\multicolumn{3}{c|}{$a=0.5$}&\multicolumn{3}{c|}{$a=1$}\\
		\hline
		Method&{$n=200$}&{$n=400$}&{$n=800$}&{$n=200$}&{$n=400$}&{$n=800$}\\
		\hline
  ARIMA&2.582&2.541&2.418&4.476&4.324&4.205\\
		\hline
		Naive&3.051&3.008&2.908&
		4.702&4.548&4.621\\
		\hline
		Standard&\textit{2.561}&\textit{2.514}
		&2.374&\textit{4.355}&\textit{4.311}&
		4.141\\
		\hline
		VAR&2.702&2.593&2.387&
		4.655&4.437&4.170\\
		\hline
		VARMA&2.673&2.561&\textit{2.373}&
		4.494&4.354&\textit{4.133}\\
		\hline
		Sieve&\textbf{2.312}&\textbf{2.209}&
		\textbf{1.987}&\textbf{3.673}&
		\textbf{3.458}&\textbf{3.193}\\
		\hline
		RD(\%)&9.72&12.13&16.27&
		15.66&19.79&22.74\\
		\hline
		RR&1.60&1.98&5.29&1.39&1.55&1.73\\
		\hline
	\end{tabular}
\end{table*}

The comparison results for locally stationary functional time series generated from Cases \eqref{case5}--\eqref{case7} are shown in Tables \ref{T3}--\ref{T4}. One can find that in all three models, our sieve method performs best among other methods for sample sizes $n=200, 400$ and $800$.  In addition, we observe that with the increasing sample size, the values of MSE gradually decrease and approach to the theoretical MSE, especially under weak temporal dependence.  
\begin{table*}[htbp!]
	\centering
	\footnotesize
	\caption{Comparison results on forecast accuracy among six methods for functional TV-ARMA(1,1) and TV-TAR(1) models.}
	\label{T4}
	\vspace*{0.1in}
	\begin{tabular}{|c|ccc|ccc|ccc|ccc|}
		\hline
		&\multicolumn{3}{c|}{TV-ARMA(1,1)}&\multicolumn{3}{c|}{TV-TAR(1)}\\
		\hline
		Method&{$n=200$}&{$n=400$}&{$n=800$}&{$n=200$}&{$n=400$}&{$n=800$}\\
		\hline
		ARIMA&0.582&0.490&\textit{0.470}&
    \textit{2.114}&\textit{2.035}&2.010\\
		\hline
		Naive&0.591&\textit{0.490}&0.504&
		4.334&4.153&3.916\\
		\hline
		Standard&\textit{0.571}&0.554&0.541&
		2.125&2.106&2.004\\
		\hline
		VAR&0.777&0.553&0.492&2.143
            &2.112&1.979\\
		\hline
		VARMA&0.691&0.530&0.477&
		2.151&2.092&\textit{1.973}\\
		\hline
		Sieve&\textbf{0.562}&\textbf{0.421}&
		\textbf{0.383}&\textbf{1.965}&
            \textbf{1.891}&\textbf{1.761}\\
		\hline
	RD(\%)&1.58&14.08&18.51&7.05&7.08&10.75\\
		\hline
		RR&1.03&1.57&2.04&1.46&1.57&2.75\\
		\hline
	\end{tabular}
\end{table*}

To further compare the convergent speeds of the computed MSEs, the theoretical true MSEs of the best linear forecast for models \eqref{case5}--\eqref{case7} are listed in \cref{T5}. In light of the quantity RR displayed in Tables \ref{T3}--\ref{T4}, we find that the computed MSEs under our sieve method approximates to its theoretical MSE at a faster rate than other methods. This demonstrates that under the locally stationary framework, our sieve method provides an asymptotically optimal best continuous linear forecast under weak temporal dependence and sufficiently large sample size in view of its optimal convergence to the true MSE. Moreover, with the quantity RD, we also find that that our sieve method to some extent improve the functional forecasting accuracy for cases \eqref{case5}--\eqref{case7}. In contrast, other existing methods fail to reach the best linear forecast error even at a moderately large sample size. 

\begin{table}[htbp!]
	\centering
	\footnotesize
	\caption{Theoretical MSE of the best linear forecast for models \eqref{case5}--\eqref{case7} when $n=800$.}
	\label{T5}
	\vspace*{0.15in}
	\begin{tabular}{|c|c|c|c|c|}
		\hline
		Model&MA(1)~$(a=0.5)$&MA(1)~$(a=1)$&TV-ARMA(1,1)&TV-TAR(1)\\
		\hline
		True MSE&1.897&1.905&0.300&1.640\\
		\hline
	\end{tabular}
\end{table}

In summary, this simulation experiment verifies that our proposed forecasting strategy via the method of double-sieve expansion can be efficiently used for predicting both stationary and locally stationary functional time series with short-range temporal dependence. In particular, for the locally stationary functional time series, our double-sieve methodology will produce an asymptotically optimal short-term forecasting based on all available preceding functional time series.

\section{Empirical data example}\label{real_data}
We apply the double-sieve methodology to forecast the telecommunication network traffic dataset described in \cref{intro}. The user download data set of interest consists of voice communication and digital items including ring tones, wall paper, music, video, games, etc for mobile users on their mobile devices. It is worth noting that the wireless networks are more complex, expensive to handle both voice and digital items and they require more bandwidth than traditional networks that handle voice only. Hence, accurate predictions are crucial for telecommunication system to manage resource allocation, maintenance plan and price policy. The hourly transaction counts of this data set has been investigated in \cite{ZXW10} to construct long-term prediction intervals. 

We consider the telecommunication traffic counts per minute from 0:00 AM July 9th, 2005 to 12:00 PM March 7th, 2006. In the first step, some missing data points are left out, mainly from 0:00 AM September 5th--12:00 PM September 6th, 2006 due to the system outage. Very few zero data points possibly resulted from system maintenance or upgrade are also removed. Next, we take logarithm of the data to stabilize the variance, and transform the daily high-dimensional data to a functional time series $\{Y_i(u)\}_{i=1}^n$ by the local polynomial smoothing technique, which leads to $n=240$ daily curves in \cref{F1} of \cref{intro}. Combining the intuitive information from \cref{F1}(b)--(d), we conduct the stationarity test (\cite{HKR14}) to the centered transformed functional data and verify the non-stationarity of the dataset with a statistically significant $p$-value $0.0228$.

In order to make functional prediction for the usage curves of the week following this 8-month
period, we consider several alternative models described in \cref{sec_simu} as contrasts. The one-step out-of-sample forecasts of the transaction curves for the last 7 days (March 1st--March 7th, 2006) are computed. Notice that under our double-sieve method, we use the Legendre polynomial (Leg.) and Daubechies-9 (D-9) wavelet basis functions based on Eq. (1) in Example 2 of the Supplementary Material. The truncation number under the first basis expansion is chosen as $p=6$ to explain 85.99\% of the variability of the data. Here, we also compute the average MSE of the out-of-sample predictions over the last 7 days, and the prediction accuracy of each method is presented in \cref{T_1}. Note that the smallest value of MSE among all methods is imposed to be bold and the smallest value among the exsiting five approaches is marked as italic.

\begin{table*}[htbp!]
	\centering
	\caption{Prediction Accuracy for the last 7 observations based on different methods.}
	\label{T_1}
	\vspace*{0.15in}
	\begin{tabular}{|c|c|c|c|c|c|c|c|}
		\hline
		Method&ARIMA&Naive&Standard&VAR&VARMA&Sieve (Leg.)&Sieve (D-9)\\
		\hline
		MSE&0.0748&0.0654&0.1859&0.0796&\textit{0.0524}&0.0513&
		\textbf{0.0453}\\
		\hline
	\end{tabular}
\end{table*}

\begin{figure}[htbp!]
	\centering
	\includegraphics[height=8cm,width=12cm]{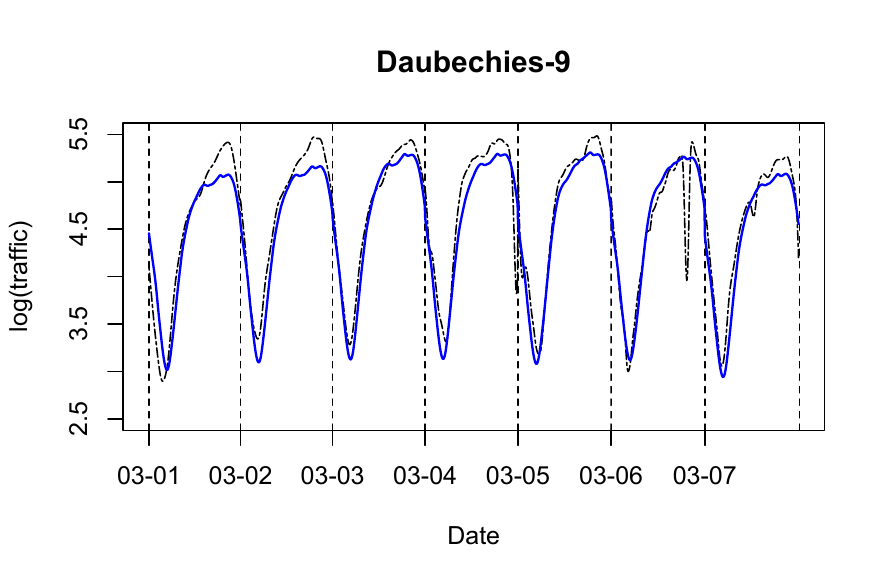}
	\caption{True functional data in black dashed curve and one-step ahead prediction based on sieve method with Daubechies-9 wavelets in blue solid curve for March 1st--7th.}
	\label{F4}
	\vspace{-0.3cm}
\end{figure}

In the forecasting procedure, we find that the Standard functional prediction method always choose the lag order of FAR model as zero in this data set, which simplifies the predictor as the mean of preceding functional time series instead. Due to the under-estimation of the FAR order, its MSE value is significantly larger than those by using other forecasting methods. From \cref{T_1}, we observe that our sieve method outperforms other prediction methods in terms of MSE. Notably compared to the best methods among the existing five state-of-art methods, our sieve prediction method with Daubechies-9 wavelet bases improve  13.55\% compared to the VARMA approach. Finally, we plot the true last seven-day curves and their one-step prediction curves based on our sieve prediction method with Daubechies-9 wavelet basis functions in \cref{F4}. One can obviously find out that the out-of-sample prediction is relatively accurate, except for the wave crests of the daily curves.

\vspace{-0.3cm}

\bibliographystyle{abbrv}
\bibliography{main}
\end{document}


\title{Supplementary Material of ``Optimal Short-Term Forecast for Locally Stationary Functional Time Series"}
	\author{Yan Cui and Zhou Zhou \\
		Department of Statistical Sciences, University of Toronto}
	\date{}
	\maketitle

\begin{abstract}
Section \ref{examples} introduces commonly used basis functions and several illustrative examples for checking Assumptions 1 and 8 of the main paper. Section \ref{add_result} provides additional results and Section \ref{add_proof} demonstrates some technical proofs for the main theoretical results of the paper. In Section \ref{add_lemma}, we collect some auxiliary lemmas used in the proofs. Throughout the Supplementary Material, the symbol $C$ denotes a generic finite constant whose value may vary from place to place.
\end{abstract}

\tableofcontents

 \section{Additional examples}\label{examples}
 \subsection{Commonly used basis functions}\label{appen_a1}
Here, we list two commonly used orthogonal basis functions which will be employed in our paper.
	\begin{example}[Normalized Legendre polynomials \cite{Bell04}]\label{example1}
		The Legendre polynomial of degree $n$ can be obtained using Rodrigue's formula 
		$$P_n(t)=\frac{1}{2^n n!}\frac{\dee^n}{\dee t^n}(t^2-1)^n,~-1\le t \le 1.$$ For $t\in [0,1]$, the normalized Legendre polynomials turn out to be
		\begin{align*}
		\alpha_k(t)=\begin{cases}
		1,&k=0,\\
		\sqrt{2k+1}P_k(2t-1),&k>0.
		\end{cases}
		\end{align*}
	\end{example}
	
	\begin{example}[Daubechies orthogonal wavelet \cite{Daubechies88,Daubechies92}]
		For $N\in\mathbb{N}$, a Daubechies (mother) wavelet of class D-N is a function $\psi\in L^2(\mathbb{R})$ defined by
		$$\psi(x):=\sqrt{2}\sum_{k=1}^{2N-1}(-1)^kh_{2N-1-k}\varphi(2x-k),$$ where $h_0,h_1,...,h_{2N-1}\in\mathbb{R}$ are the constant (high pass) filter coefficients satisfying the conditions $\sum_{k=0}^{N-1}h_{2k}=\sum_{k=0}^{N-1}h_{2k+1}=1/\sqrt{2}$, and for $l=0,1,...,N-1$, $$\sum_{k=2l}^{2N-1+2l}h_kh_{k-2l}=
		\begin{cases}
		1, &l=0,\\
		0, &l\neq 0.
		\end{cases}$$
		Moreover, $\varphi(x)$ is the scaling (father) wavelet function supported on $[0,2N-1)$ and satisfies the recursion equation $\varphi(x)=\sqrt{2}\sum_{k=1}^{2N-1}h_k\varphi(2x-k)$, as well as the normalization $\int_\mathbb{R}\varphi(x)\dee x=1$ and $\int_{\mathbb{R}}\varphi(2x-k)\varphi(2x-l)\dee x=0$ when $k\neq l$. The order $N$, which determines the support of our wavelet, provides the regularity condition in the sense that $$\int_\mathbb{R}x^j\psi(x)\dee x=0,~j=0,...,N,~\text{where}~N>d.$$
		In this paper, we will employ Daubechies wavelet with a sufficiently high order. The basis functions can be generated using the \textsf{wavefun} in the \textsf{Wavelet Toolbox} of Matlab. Now in order to construct a sequence of orthogonal wavelet, we will follow the dyadic construction of \cite{Daubechies88}. For some given integers $J_n$ and $J_0$, we consider the following periodic wavelets on $[0,1]$,
		\begin{equation}\label{wavelet}
		\left\{\varphi_{J_0 k}(x), 0\le k\le 2^{J_0}-1;
		\psi_{jk}(x), J_0\le j\le J_n-1, 0\le k\le 2^j-1. \right\}
		\end{equation}
		where
		$\varphi_{J_0 k}(x)=2^{J_0}\sum_{l\in\mathbb{Z}}\varphi(2^{J_0}x+2^{J_0}l-k), \psi_{jk}(x)=2^{j/2}\sum_{l\in\mathbb{Z}}\psi(2^j x+2^j l-k),$ or equivalently based on father wavelet functions (see \cite{Meyer90}) 
		\begin{equation}\label{father}
		\left\{\varphi_{J_n k}(x), 0\le k\le 2^{J_n}-1\right\}.
		\end{equation}	
	\end{example}
	
	\subsection{Examples of functional time series models}\label{appen_a2}
	Next, we will provide two models to verify Assumption 1 in Section 2.1 and illustrate how to calculate the physical dependence measure $\delta_x(l,q)$ in Assumption 8, Section 3.2 of the main paper.
	\begin{example}[Functional MA$(\infty)$ model] \label{exam1}
		Let $\eta_i(v)$ be i.i.d. centered and continuous Gaussian random functions with $\sup_{v\in[0,1]}\Vert\eta_i(v)\Vert_{2}<\infty$. For $i\in \mathbb{Z}$ and $m\ge 0$, let $\beta_m(t,u,v)=a_m\beta_{m}^\ast(t,u,v)$ where $\{a_m\}$ is a positive deterministic sequence with $\sum_{m=0}^\infty a_m<\infty$ and  $\beta_{m}^\ast(t_i,u,v)=\beta_{i,m}^
  \ast(u,v)$ is a $\mathcal{C}^0([0,1]^2)$ function for some fixed time $t_i=i/n$ such that $\max_{m}|\beta^*_{m}(t,u,v)|\le C$ for all $t,u,v$ and some finite constant $C$. Consider the locally stationary functional MA$(\infty)$ model,
		\begin{equation}\label{exp1}
		Y_i(u)=\sum_{m=0}^\infty \int_0^1 \beta_{m}(t_i,u,v)\eta_{i-m}(v)\dee v.
		\end{equation}
		Armed with the technique of basis expansion, it is easy to find that Eq. (3) in Assumption 1 can be satisfied if 
		$$\sum_{m=0}^\infty \sup_{t,u,v}\left|\frac{\partial
  \beta_m(t,u,v)}{\partial t}\right|<\infty.$$ 
		
		\noindent
		Let $C(u,v):=\EE(\eta_i(u)\eta_i(v))$ be the covariance function of $\eta_i(u)$. Let $\xi_{1}(u), \xi_{2}(u),\cdots$ and the corresponding $\lambda_{1}\ge \lambda_{2}\ge \cdots$ be the eigenfunctions and eigenvalues of $C(u,v)$. By the basis expansion method, we can write $Y_i(u)=\sum_{j=1}^\infty r_{i,j}\alpha_j(u),
		~\eta_i(v)=\sum_{j=1}^\infty\eta_{i,j}\xi_j(v)$ and $\beta_{m}^\ast(t_i,u,v)=
		\sum_{j=1}^\infty\sum_{k=1}^\infty
  \beta_{jk}^m(\frac{i}{n})
  \alpha_j(u)\xi_k(v)$ where $\{\alpha_j(\cdot)\}$ is the predetermined basis function. Next, by substituting the above relations into \eqref{exp1}, 
		we obtain
		$$r_{i,j}=\sum_{m=0}^\infty a_m\left(\sum_{k=1}^\infty
		\beta_{jk}^m(\frac{i}{n})\eta_{i-m,k}\right)=: f_{j}x_{i,j},$$
		where $f^2_{j}=\sum_{m=0}^\infty a_m^2\theta_{jm}$ with $\theta_{jm}:=\sup_{t\in[0,1]}\sum_{k=1}^\infty[\beta_{jk}^m(t)]^2 \lambda_k$. Similar to the discussion of \citep[Example 1]{CZ2022}, we obtain that $\delta_x(l,q)=O(a_l)$ for any given $q\ge 2$ if $\sum_{k=0}^\infty a_k^2\theta_{jk}\ge C\theta_{jm}$ for sufficiently large $j$ and $m$. Consequently, the physical dependence measure for \eqref{exp1} accords with Assumption 8 when we choose $a_l=(l+1)^{-\tau}$.		
	\end{example}
	
	The polynomial temporal dependence used in the paper (c.f. Assumption 4 of the main paper) can be easily extended to the case of exponential decay, i.e, $\sup_{t\in[0,1]}\Vert\bm{\Gamma}(t,j)\Vert \le C\rho^{|j|}$ for $j\in\mathbb{Z}$. We provide the following example to check assumptions in the exponential scenario.
 
	\begin{example}[Functional AR(1) model] \label{exam2}
		Let $\epsilon_i(u)$ be i.i.d. centered and continuous Gaussian random functions with $\sup_{u\in[0,1]}\Vert\epsilon_i(u)\Vert_{2}<\infty$. Consider the following model \begin{equation}\label{model}
		Y_i(u)=\int_0^1 \beta(t_i,u,v)Y_{i-1}(v)\dee v+\epsilon_i(u),\quad t_i=\frac{i}{n},
		\end{equation}
		where $\beta(t_i,u,v)=\beta_i(u,v): [0,1]^2\to \mathbb{R}$ is a continuous, symmetric function satisfying $\int_0^1\int_0^1\beta_i^2(u,v)\dee u\dee v<\infty$ and $\int_0^1\int_0^1\beta_i(u,v)x(u)x(v)\dee u\dee v\ge 0$ with any random function $x(u)\in \mathcal{L}^2([0,1])$ for any $i=1,...,n$. Thus for any fixed $t_i$, $\beta(t_i,u,v)$ is called a symmetric and positive-definite kernel on $[0,1]^2$. With \eqref{model}, the Lipschitz continuous condition in Assumption 1 will be verified if 
		$$\sup_{t,u,v}\left|\frac{\partial\beta(t,u,v)}
		{\partial t}\right|<\infty.$$ 
		
		\noindent
		Define $C(u,v):=\EE(\epsilon_i(u)\epsilon_i(v))$ as the covariance function of $\epsilon_i(u)$. Let $\xi_1(u), \xi_2(u),\cdots$ and the corresponding $\lambda_1\ge \lambda_2\ge \cdots$ be the eigenfunctions and eigenvalues of $C(u,v)$. By the basis expansion method, we can write
		$Y_i(u)=\sum_{k=1}^\infty r_{i,k}\xi_k(u),
		~\epsilon_i(u)=\sum_{k=1}^\infty \epsilon_{i,k}\xi_k(u)$ and $\beta(t_i,u,v)=\sum_{k=1}^\infty \beta_{i,k}\xi_k(u)\xi_k(v)$ with $\beta_{i,k}:=\beta_k(\frac{i}{n})$ changing over time. Now, we can rewrite \eqref{model} as 
		\begin{equation}\label{basis_expan}
		r_{i,k}=\beta_{i,k}r_{i-1,k}+\epsilon_{i,k}=
		\epsilon_{i,k}+\sum_{m=1}^\infty \left(\prod_{l=0}^{m-1} \beta_{i-l,k}\right)\epsilon_{i-m,k}.
		\end{equation}
		Here denote $f^2_k=\sum_{m=0}^\infty\left(\int_0^1 [\beta_k(t)]^m\dee t
		\right)^2\lambda_k$ and $x_{i,k}=r_{i,k}/f_k$ when $f_k\neq 0$. If we let $\rho:=\sup_{i,k}|\beta_{i,k}|\in (0,1)$, then $f_k^2\ge \lambda_k$ for $k\ge 1$. Further denote $x_{i,k}^\ast$ and $\epsilon_{i,k}^\ast$ as the coupling versions of $x_{i,k}$ and $\epsilon_{i,k}$ respectively, with the random element $\eta_{i-l}$ from the generating process replaced by its i.i.d. copy $\eta'_{i-l}$. Observe that $\{\epsilon_{i,k}\}$ are independent Gaussian random variables across $k$, hence $(\epsilon_{i-l,k}-\epsilon_{i-l,k}^\ast)$ is normally distributed with mean 0 and variance  $2\lambda_k$, finally for any $k\ge 1$, we have
		\begin{align*}
		\Vert x_{i,k}-x_{i,k}^\ast\Vert_q&=\left\Vert \left(\Pi_{l=0}^{m-1}\beta_{i-l,k}\right)
		(\epsilon_{i-l,k}-\epsilon_{i-l,k}^\ast)/f_k\right\Vert_q\\
		&\le C_q\sqrt{2\lambda_k}\sup_{i,k}|\beta_{i,k}|^l/
  \sqrt{\lambda_k}\le C\rho^l.
		\end{align*}
	\end{example}
 \section{Additional results}\label{add_result}	

 \begin{lemma}\label{lemma_def}
Under Assumption 1, the locally stationary functional time series represented in Eq. (2) satisfies Definition 1 of the main paper.
\end{lemma}

\noindent
	\textit{Proof}:
 Let $t_i=\frac{i}{n}$ and with Eq. (2) of the paper, we have
 \begin{align*}
     &{\rm Cov}(Y_i(u),Y_j(v))=\EE\left[H(\frac{i}{n},u,\mathcal{F}_i)H(\frac{j}{n},v,\mathcal{F}_j)\right]\\
     =&\EE\left[H(\frac{i}{n},u,\mathcal{F}_i)\bigg(H(\frac{j}{n},v,\mathcal{F}_j)-H(\frac{i}{n},v,\mathcal{F}_j)\bigg)\right]+\EE\left[H(\frac{i}{n},u,\mathcal{F}_i)H(\frac{i}{n},v,\mathcal{F}_j)\right]\\
     \le &\Vert H(\frac{i}{n},u,\mathcal{F}_i)\Vert_2\Vert H(\frac{j}{n},v,\mathcal{F}_j)-H(\frac{i}{n},v,\mathcal{F}_j)\Vert_2+\gamma(\frac{i}{n},u,v,|i-j|)\\
     = &\gamma(t_i,u,v,|i-j|)+\bigO\left(
     \frac{|i-j|+1}{n}\right),
 \end{align*}
 which is in accord with Eq. (1) in Definition 1. Furthermore, the Lipschitz continuity in time $t$ of the autocovariance function $\gamma(t,u,v,\cdot)$ can be verified as follows. For $t_1,t_2\in[0,1]$ and any $u,v \in[0,1]$,
 \begin{align*}
     &\left|\gamma(t_1,u,v,|i-j|)-
     \gamma(t_2,u,v,|i-j|)\right|\\
     \le &\big|\EE\left[\left(H(t_1,u,
     \mathcal{F}_i)-
     H(t_2,u,\mathcal{F}_i)\right)H(t_1,v,
\mathcal{F}_j)\right]\big|
+\big|\EE\left[H(t_2,u,\mathcal{F}_i)(H(t_1,v,\mathcal{F}_j)-H(t_2,v,\mathcal{F}_j))\right]\big|\\
\le &\Vert H(t_1,u,\mathcal{F}_i)-H(t_2,u,\mathcal{F}_i)\Vert_2 \Vert H(t_1,v,\mathcal{F}_j)\Vert_2+ \Vert
H(t_2,u,\mathcal{F}_i)\Vert_2
\Vert H(t_1,v,\mathcal{F}_j)-H(t_2,v,\mathcal{F}_j)\Vert_2\\
\le &C|t_1-t_2|,
 \end{align*}
  where the last inequality follows by Assumption 1.
 $\hfill \square$
 
	\begin{lemma}\label{lemma_gamma}
		Recall $\bm{\Gamma}(t,j)=\{\gamma_{k,l}(t,j)\}_{k,l=1}^p$. Under Assumption 8 of the main paper, then there exists some constant $C>0$ such that
		$$\sup_{t\in [0,1]}\max_{k,l}
		|\gamma_{k,l}(t,j)|\le C(|j|+1)^{-\tau},~j\in \mathbb{Z}.$$
		Furthermore, suppose that $x_{i,k}$ satisfies Lipschitz continuity across time $i$ and $\max_{i,k}\Vert x_{i,k}\Vert_q<\infty$ and for any $k$ and $l$, then $\gamma_{k,l}(t,j)$ satisfies the stochastic Lipschitz continuity over $t\in[0,1]$, that is,
		$$\Vert \gamma_{k,l}(t_1,j)-\gamma_{k,l}(t_2,j)\Vert_q
		\le C|t_1-t_2|,~t_1,t_2\in [0,1].$$
	\end{lemma}
	
	\noindent
	\textit{Proof}:
	Without loss of generality, we will prove the decay rate for $j\ge 0$ and the case for $j<0$ is straightforward. Let $\mathcal{P}_m(\cdot)=\EE(\cdot|\mathcal{F}_m)-\EE(\cdot|\mathcal{F}_{m-1})$, then we can write $x_{i,k}=\sum_{m=-\infty}^i\mathcal{P}_m(x_{i,k})$. For any fixed $t_i=i/n$ and $k, l$, we have
	\begin{align}
	|\bm{\gamma}_{k,l}(t_i,j)|&=\left|\EE\left(\sum_{m=-\infty}^i \mathcal{P}_m(x_{i,k})\sum_{m=-\infty}^{i+j}\mathcal{P}_m(x_{i+j,l})
	\right)\right| \notag \\ 
	&\le \left|\EE\left(\sum_{m=-\infty}^\infty\mathcal{P}_m(x_{i,k})
	\mathcal{P}_m(x_{i+j,l})\right)\right| \notag \\
	&\le \sum_{m=-\infty}^\infty\Vert\mathcal{P}_m(x_{i,k})\Vert_2\cdot
	\Vert\mathcal{P}_m(x_{i+j,l})\Vert_2 \notag \\
	&\le \sum_{m=-\infty}^i \delta_x(i-m,2)\delta_x(i+j-m,2). \label{star}
	\end{align} 
	The first inequality follows by the orthogonality of $\mathcal{P}_m(\cdot)$ and the second inequality is due to Fubini's theorem and Cauchy-Schwarz inequality. The last inequality follows from the argument in \citep[Theorem 1]{Wu05}. Therefore, we have $\sup_{t\in [0,1]}\max_{k,l}|\gamma_{k,l}(t,j)|\le C(j+1)^{-\tau}$ for $j\ge 0$. Since $\gamma_{k,l}(t,j)=\gamma_{l,k}(t,-j)$, we conclude the upper bound for $j \in \mathbb{Z}$ as $C(|j|+1)^{-\tau}$.
	
	On the other hand, with the Lipchitz continuous condition on $x_{i,k}$, we have
	\begin{align*}
	&\Vert\gamma_{k,l}(t_1,j)-\gamma_{k,l}(t_2,j)\Vert_q\\
	=&\left\Vert\EE[(G_k(t_1,\mathcal{F}_0)-G_k(t_2,\mathcal{F}_0))
	G_l(s_1,\mathcal{F}_j)]\right.\\
	&{}\left.+\EE[G_k(t_2,\mathcal{F}_0)(G_l(t_1,\mathcal{F}_j)-
	G_l(t_2,\mathcal{F}_j))]\right\Vert_q\\
	\le &\Vert G_k(t_1,\mathcal{F}_0)-G_k(t_2,\mathcal{F}_0)\Vert_{2q} 
	\Vert G_l(t_1,\mathcal{F}_j)\Vert_{2q}\\
	&{}+ \Vert G_k(t_2,\mathcal{F}_0)\Vert_{2q}
	\Vert G_l(t_1,\mathcal{F}_j)-G_l(t_2,\mathcal{F}_j)\Vert_{2q}\\
	\le & C|t_1-t_2|.
	\end{align*}
$\hfill \square$

 Next lemma illustrates how to verify Assumptions 3, 4 in Example 2 of the main paper and Assumption 6 in Section 3.2.
 \begin{lemma}\label{lemma_example2}
 With the multivariate linear model in Example 2 of the main paper, we find that Assumptions 3 and 4 will be satisfied if 
 $$\sup_{t\in[0,1]} \left\Vert\frac{\partial^{d_2}\bm{A}_m(t)}{\partial t^{d_2}}\right\Vert\le C(m+1)^{-\tau};~~\sup_{t\in[0,1]}\Vert \bm{A}_m(t)\Vert\le C(m+1)^{-\tau}.$$ As a byproduct, Assumption 6 will be met if $\sup_{t\in[0,1]} \left\Vert \partial^{d_2}\bm{A}_m(t)/\partial t^{d_2}\right\Vert\le C(m+1)^{-\tau}$ holds true.
 \end{lemma}

 \noindent
	\textit{Proof}: For any fixed $t$, we have
 \begin{equation}\label{gamma_expression}
 \bm{\Gamma}(t,j)=\sum_{m=0}^\infty\bm{A}_m(t)
 \bm{\Sigma}_{\eta}\bm{A}_{m+j}^\top(t),~~j\in \mathbb{Z}.
 \end{equation}
 For the smoothness condition in Assumption 3, with \cref{gamma_expression} and integration by parts, we need to show
 $$\sup_{t\in[0,1]}\left\Vert \frac{\partial^{d_2}\bm{\Gamma}(t,j)}{\partial t^{d_2}}\right\Vert \le \sum_{m=0}^\infty\sum_{k=0}^{d_2}
 \sup_t\left\Vert \frac{\partial^{k}\bm{A}_m(t)}{\partial t^{k}}\right\Vert \Vert\bm{\Sigma}_{\eta}\Vert \sup_t
 \left\Vert \frac{\partial^{d_2-k}\bm{A}_{m+j}(t)}{\partial t^{d_2-k}}\right\Vert <\infty.$$ Consequently we find that if $\sup_{t\in[0,1]}\Vert \partial^{d_2}\bm{A}_m(t)/\partial t^{d_2}\Vert \le C(m+1)^{-\tau}$ holds, then one can easily obtain $\bm{\Gamma}(t,j)\in\mathcal{C}^{d_2}([0,1])$ and $\sup_{t\in[0,1]}\sum_{j\in\mathbb{Z}}
 \Vert \partial^{d_2}\bm{\Gamma}(t,j)/\partial t^{d_2}\Vert<\infty$.

Similarly with the relation of \eqref{gamma_expression}, we have for $j\in\mathbb{Z}$,
$$\sup_{t\in[0,1]}\Vert\bm{\Gamma}(t,j)\Vert\le 
\sum_{m=0}^\infty \sup_t\Vert\bm{A}_m(t)\Vert
\Vert \bm{\Sigma}_{\eta}\Vert \sup_t\Vert \bm{A}_{m+j}(t)\Vert\le C(|j|+1)^{-\tau}$$ if $\sup_{t\in[0,1]}\Vert\bm{A}_m(t)\Vert \le C(m+1)^{-\tau}$. $\hfill \square$

	It is clear that the random vector $\bm{\epsilon}_i=(\epsilon_{i,1},...,
 \epsilon_{i,p})^\top$ is the best linear prediction error in the VAR model. Denote $\epsilon_{i,k}=G_k^\ast(\frac{i}{n},\mathcal{F}_i)$ where $G_k^\ast$ is some measurable function for each $k=1,...,p$, similar to $G_k$ defined in Section 3.3 of the main paper. Then we have the following lemma on its physical dependence measure.
	
	\begin{lemma}\label{eps_decay}
		Suppose Assumptions  4, 5 and 8  hold true, denote the physical dependence measure of $\{\epsilon_{i,k}\}_{i\in\mathbb{Z}}$ as $\delta_\epsilon(l,q):=\max_{1\le k\le p}\Vert G_k^\ast(\frac{i}{n},\mathcal{F}_i)-G_k^\ast(\frac{i}{n},\mathcal{F}_{i,l})\Vert_q$. Then, there exists some constant $C>0$ such that $\delta_\epsilon(l,q)\le C(l+1)^{-\tau},~l\ge 0$.
	\end{lemma}
	\noindent
	\textit{Proof}:
	Rewrite the VAR model in Eq. (14) of the main paper as $\bm{x}_i=\sum_{h=1}^{i-1}\bm{\Phi}_{i,h}\bm{x}_{i-h}+\bm{\epsilon}_i$, since $\max_{i,k}\Vert x_{i,k}\Vert_q<\infty$ and $\max_{i,k}\Vert \epsilon_{i,k}\Vert_q<\infty$, we have $$\max_{i,k}\left\Vert \sum_{j=1}^{i-1}\sum_{m=1}^p
	\bm{\Phi}_{i,h}(k,m)x_{i-h,m}\right\Vert_q<\infty,$$ where $\bm{\Phi}_{i,h}(k,m)$ represents the $(k,m)$th element of the matrix $\bm{\Phi}_{i,h}$. On this basis, we can derive that $\max_{i,k}|\sum_{m=1}^p{\bm{\Phi}}_{i,h}(k,m)|$ decays with $h$. Consequently, we have  
	\begin{align*}
	\Vert \epsilon_{i,k}-\epsilon_{i,k}^\ast \Vert_q=&\left\Vert
	x_{i,k}-x_{i,k}^\ast+\sum_{h=1}^{i-1}\sum_{m=1}^p
	\bm{\Phi}_{i,h}(k,m)(x_{i-h,m}^\ast-x_{i-h,m})\right\Vert_q\\
	\le &\Vert x_{i,k}-x_{i,k}^\ast\Vert_q+\left\Vert\sum_{h=1}^{i-1}
	\sum_{m=1}^p\bm{\Phi}_{i,h}(k,m)(x_{i-h,m}-x_{i-h,m}^\ast)\right\Vert_q\\
	\le &C(l+1)^{-\tau}.
	\end{align*} $\hfill \square$
	
	\section{Technical proofs of main results}\label{add_proof}
	This subsection is devoted to the technical proofs in Sections 2 and 3.
	\subsection{Proof of the results in Section 2}\label{sec_c1}
	\noindent
	\textbf{Proof of Proposition 1}.
	In the following, we will show that the eigenvalues of ${\rm Cov}(\bm{{\rm x}})$ can be approximated by those of a banded matrix $\bm{\Sigma}_{d_n}$, especially the smallest eigenvalue. The parameter $d_n$ controls the bandedness and will be chosen later in the proof. For the sufficiency part in Proposition 1, we will provide a statement in \cref{lemma_updc} that if the smallest eigenvalue of the spectral density matrix in Eq. (8) is bounded from below, then for a length $d_n+1$ sequence of the multivariate time series, it must satisfy the UPDC condition. For the necessity part, we will first construct a function $\bm{f}_n(t,\omega)$ following the classic theory of spectral density matrix, which is close to  $\bm{f}(t,\omega)$ when $n$ is sufficiently large. Consequently, we aim to show that the smallest eigenvalue  of $\bm{f}_n(t,\omega)$ is bounded from below by constructing another matrix $\bm{g}_n(t,\omega)$ using the banded matrix $\bm{\Sigma}_{d_n}$, whose smallest eigenvalue is bounded from below by a positive constant using UPDC assumption and the short-range dependence assumption. In addition, the smallest eigenvalues of $\bm{f}_n(t,\omega)$ and $\bm{g}_n(t,\omega)$ are sufficiently close when $n$ is sufficiently large.
	
 Denote the covariance matrix of 
 $\bm{{\rm x}}=(\bm{x}_1^\top,
 ...,\bm{x}_n^\top)^\top$ as $\bm{\Sigma}_n$ with $p\times p$ block entries $\{(\bm{\Sigma}_{n})_{kl}\}_{k,l=1}^n$. For a given truncation number $d_n$ such that $pd_n^2/n\to 0$ as $n$ goes to infinity, we define the banded matrix $\bm{\Sigma}_{d_n}$ by defining its $(k,l)$th block entry as
	\begin{equation}\label{prop_banded_def}
	(\bm{\Sigma}_{d_n})_{kl}:=\begin{cases}
	(\bm{\Sigma}_{n})_{kl}, &|k-l|\le d_n,\\
	\bm{0}, &\text{otherwise}.
	\end{cases}
	\end{equation}
		Under Assumption 4 and by \cref{lemma_w}, we have 
	\begin{equation}\label{diff_double}
	\begin{split}
	\lambda_{\min}(\bm{\Sigma}_{d_n})-\lambda_{\min}(\bm{\Sigma}_{n})&=
	\lambda_{\min}(\bm{\Sigma}_{d_n})+\lambda_{\max}(-\bm{\Sigma}_{n})\ge
	-\lambda_{\max}(\bm{\Sigma}_{n}-\bm{\Sigma}_{d_n})\ge -Cd_n^{-\tau+1},\\
	\lambda_{\min}(\bm{\Sigma}_{n})-\lambda_{\min}(\bm{\Sigma}_{d_n})&=
	\lambda_{\min}(\bm{\Sigma}_{n})+\lambda_{\max}(-\bm{\Sigma}_{d_n})\le
	\lambda_{\max}(\bm{\Sigma}_{n}-\bm{\Sigma}_{d_n})\le Cd_n^{-\tau+1}.
	\end{split}
	\end{equation}
	Thus, one can conclude that for any $u,v$, 
	\begin{equation}\label{cov_diff}
	|\lambda_{\min}(\bm{\Sigma}_{d_n}(u,v))-
	\lambda_{\min}(\bm{\Sigma}_{n}(u,v))|\le Cd_n^{-\tau+1}.
	\end{equation}
	Here, it suffices to investigate the UPDC condition for $\bm{\Sigma}_{d_n}$. Next, we consider a longer locally stationary functional time series $\{\bm{x}_i\}_{i=-d_n}^{n+d_n}$, where we use the convention $\bm{x}_i=\bm{G}(0,\mathcal{F}_i)$ for $i<0$ and $\bm{x}_i=\bm{G}(1,\mathcal{F}_i)$ for $i>n$ with $\bm{G}=(G_1,...,G_p)^\top$. We need the following lemma to prove the sufficiency.
	\begin{lemma}\label{lemma_updc}
		Let $\bm{\Sigma}_{d_n}^{(i)}$ be the covariance matrix of $(\bm{x}_i^\top,...,\bm{x}_{i+d_n}^\top)^\top$ for $i=-d_n+1,...,n$ where $d_n$ has a slow diverging speed such that $pd_n^2/n\to 0$ as $n\to \infty$. Then if the smallest eigenvalue of the spectral density matrix defined in Eq. (8) in Proposition 1 of the paper is bounded below from zero, we have that for some constant $\delta>0$,
		$$\lambda_{\min}(\bm{\Sigma}_{d_n}^{(i)})\ge \delta,~\text{for~all}~i.$$
	\end{lemma}
\textit{Proof}: Without loss of generality, we set $i=0$. Consider the stationary process $\{\bm{x}_i^0\}$ with $\bm{\Gamma}(0,\cdot)$ as its autocovariance function. Then denote 
$\bm{\Sigma}_{d_n}^0={\rm Cov}
(((\bm{x}_0^0)^\top,...,(\bm{x}_{d_n}^0)^\top)^\top)$, by \citep[Theorem 11.8.1]{BD91}
and when the smallest eigenvalue of the spectral density matrix is bounded below from zero, we have
\begin{equation}\label{updc_stationary}
\lambda_{\min}(\bm{\Sigma}_{d_n}^0)\ge \delta>0~~\text{for~any~large}~d_n.
\end{equation}
On the other hand, from Assumption 4 of the paper and the Lipschitz continuity of $\bm{\Gamma}$ in $t$, we find that 
$$\max_{k,l}|{\rm Cov}(x_{i,k},x_{j,l})-{\rm Cov}(x_{i,k}^0,x_{j,l}^0)|\le \frac{C\max(i,j)}{n}.$$
Similar to the discussion in \eqref{diff_double}, by Lemmas \ref{g_circle_block} and \ref{lemma_w}, we have  
\begin{equation*}
\left|\lambda_{\min}(\bm{\Sigma}_{d_n}^0)-
\lambda_{\min}(\bm{\Sigma}_{d_n}^{(i)})\right|\le |\lambda_{\max}(\bm{\Sigma}_{d_n}^0-\bm{\Sigma}_{d_n}^{(i)})|\le \frac{Cpd_n^2}{n}.
\end{equation*}
Combining with \cref{updc_stationary}, we find that
$$\lambda_{\min}(\bm{\Sigma}_{d_n}^{(i)})\ge \lambda_{\min}(\bm{\Sigma}_{d_n}^0)-\frac{Cpd_n^2}{n}\ge \delta.$$ 
$\hfill \square$

\noindent
Now we first prove the sufficiency part in Proposition 1. For any non-zero vector $\bm{a}=(a_{-d_n+1},...,a_0,...,a_{n+d_n})^\top\in\mathbb{R}^{n+2d_n}$, define
$$\bm{F}(\bm{a},i):=\sum_{k=1}^{d_n+1}\sum_{l=1}^{d_n+1}a_{i+k-1}
\Sigma_{d_n,kl}^{(i)}a_{i+l-1},$$ where $\Sigma_{d_n,kl}^{(i)}$ is the $(k,l)$th element of $\bm{\Sigma}_{d_n}^{(i)}$. Furthermore, let $\bm{A}_{d_n}^{(i)}=(a_i\bm{I}_p,...,a_{i+d_n}\bm{I}_p)^\top,
~\bm{\Sigma}_{d_n}^{(i)}=
\bm{P}_{d_n}^\top\bm{\Lambda}_{d_n}^{(i)}\bm{P}_{d_n}$ where $\bm{P}_{d_n}$ is the orthogonal matrix such that $\bm{\Lambda}_{d_n}^{(i)}$ is the diagonal matrix containing all eigenvalues. Denote the block matrix on the diagonal of $\bm{\Lambda}_{d_n}^{(i)}$ as $\bm{\Lambda}_k\in\mathbb{R}^{p\times p}$, $\bm{\Lambda}_{k}^0=\lambda_{\min}(\bm{\Lambda}_k)\bm{I}_p$ and $\bm{B}_{d_n}^{(i)}=\bm{P}_{d_n}\bm{A}_{d_n}^{(i)}$. Denote $\bm{B}_{d_n}^{(i)}$ as the $k$th column of $\bm{B}_{d_n}^{(i)}$, then under \cref{lemma_updc}, we have
\begin{align*}
\lambda_{\min}(\bm{F}(\bm{a},i))&=\lambda_{\min}\left(\sum_{k=1}^{d_n+1}
(\bm{B}_{d_n}^{(i)})_k^\top[\bm{\Lambda}_k-\bm{\Lambda}_k^0](\bm{B}_{d_n}^{(i)})_k+\sum_{k=1}^{d_n+1}(\bm{B}_{d_n}^{(i)})_k^\top\bm{\Lambda}_k^0
(\bm{B}_{d_n}^{(i)})_k\right)\\
&\ge \lambda_{\min}\left(\sum_{k=1}^{d_n+1}\lambda_{\min}(\bm{\Lambda}_k)
(\bm{B}_{d_n}^{(i)})_k^\top(\bm{B}_{d_n}^{(i)})_k\right)\\
&\ge \lambda_{\min}(\bm{\Sigma}_{d_n}^{(i)})
\lambda_{\min}\left(\sum_{k=1}^{d_n+1}
(\bm{B}_{d_n}^{(i)})_k^\top(\bm{B}_{d_n}^{(i)})_k\right)\ge \delta\sum_{k=i}^{i+d_n}a_k^2,
\end{align*}
where the first inequality follows by \cref{lemma_w} and the fact $(\bm{B}_{d_n}^{(i)})_k^\top[\bm{\Lambda}_k-\bm{\Lambda}_k^0](\bm{B}_{d_n}^{(i)})_k$ is positive semi-definite matrix. Next we let the first $d_n$ and the last $d_n$ entries of $\bm{a}$ be zeros. Then by elementary calculation, we find that 
\begin{equation}\label{quadratic_bound}
\frac{1}{d_n+1}\lambda_{\min}\left(
\sum_{i=-d_n+1}^n \bm{F}(\bm{a},i)\right)\ge \delta\sum_{k=1}^n a_k^2.
\end{equation}
Furthermore, by \cref{g_circle_block} again, it is obvious to find that for some constant $C>0$,
$$\lambda_{\max}\left(
\frac{1}{d_n}\sum_{i=-d_n}^n\bm{F}(\bm{a},i)-
\sum_{k=1}^n\sum_{l=1}^n a_{k}\Sigma_{d_n,kl}a_l\right)
\le \frac{C}{d_n}\sum_{k=1}^na_k^2.$$
Together with \eqref{quadratic_bound} and \cref{lemma_w}, we have
$$\lambda_{\min}\left(\sum_{k=1}^n\sum_{l=1}^na_k\Sigma_{d_n,kl}a_l\right)\ge \left(\frac{d_n+1}{d_n}\delta-\frac{C}{d_n}\right)\sum_{k=1}^na_k^2\ge \frac{\delta}{2}\sum_{k=1}^n a_k^2,$$ when $n$ is large enough. This shows that $\lambda_{\min}(\bm{\Sigma}_{d_n})$ is bounded below from zero and hence $\lambda_{\min}(\bm{\Sigma}_n)$ is bounded below by some constant via \eqref{cov_diff}.

On the other side, we will prove the necessity. For any given $t_i=\frac{i}{n}$, define 
\begin{align*}
\bm{f}_n(t_i,\omega)&=\frac{1}{2\pi n}\sum_{k,l=1}^n{\rm e}^{-{\rm i}k\omega}
\bm{\Gamma}(t_i,k-l){\rm e}^{{\rm i}l\omega}\\
&=\frac{1}{2\pi}\sum_{|h|< n}\left(1-\frac{|h|}{n}\right){\rm e}^{-{\rm i}h\omega}
\bm{\Gamma}(t_i,h).
\end{align*}
By the short-range dependence condition and $|{\rm e}^{-{\rm i}h\omega}|=1$, it is obvious to find that for sufficiently large $n$,
\begin{align}\label{approx_1}
\Vert\bm{f}_n(t,\omega)-\bm{f}(t_i,\omega)\Vert &\le
C\left\Vert\sum_{|h|<n}\frac{|h|}{n}\bm{\Gamma}(t_i,h)\right\Vert
+C\left\Vert \sum_{|h|> n}\bm{\Gamma}(t_i,h)\notag \right\Vert\\
&\le C\left(\frac{d_n}{n}+d_n^{-\tau+1}\right).
\end{align}
As a consequence, it is equivalent to show that $\lambda_{\min}(\bm{f}_n(t_i,\omega))$ is bounded from below by some positive constant for all $i,\omega$ and sufficiently large $n$. Now, we construct a function $\bm{g}_n(t_i,\omega)$ using block entries of the banded matrix $\bm{\Sigma}_{d_n}$ as
\begin{equation}\label{g_n kernel}
\bm{g}_n(t_i,\omega)=\frac{1}{2\pi}\sum_{|h|<n}\left(1-\frac{|h|}{n}\right)
{\rm e}^{-{\rm i}h\omega}(\bm{\Sigma}_{d_n})_{i,i+|h|},
\end{equation} 
where $(\bm{\Sigma}_{d_n})_{i,i+|h|}$ denotes the $(i,i+|h|)$th block entry of the banded matrix $\bm{\Sigma}_{d_n}$. By elementary calculations, we have that when $n$ is sufficiently large,

\begin{align}
&\Vert\bm{g}_n(t_i,\omega)-\bm{f}_n(t_i,\omega)\Vert \notag\\
=& \left\Vert\frac{1}{2\pi}\sum_{|h|<n}
\left(1-\frac{|h|}{n}\right){\rm e}^{-{\rm i}h\omega}\left(
\bm{\Gamma}(t_i,h)-(\bm{\Sigma}_{d_n})_{i,i+|h|}\right)\right\Vert\notag\\
\le &C\left(\sum_{|h|\le d_n}\left(1-\frac{|h|}{n}\right)\left\Vert
\bm{\Gamma}(t_i,h)-{\rm Cov}(\bm{x}_i,\bm{x}_{i+h})\right\Vert+
\left\Vert\sum_{|h|\ge d_n}\left(1-\frac{|h|}{n}\right)
\bm{\Gamma}(t_i,h)\right\Vert\right)\notag\\
\le &C\left(\frac{pd_n^2}{n}+d_n^{-\tau+1}\right)\label{approx_2}
\end{align}
for some constant $C$, where the last inequality holds by Assumptions 4 and 8 of the main paper. Notice that Combining with the assumption $\lambda_{\min}(\bm{\Sigma}_n)\ge \kappa_1$, Eq. \eqref{cov_diff} and the structure of $\bm{\Sigma}_{d_n}$ used in \eqref{g_n kernel}, we can obtain that $$\lambda_{\min}(\bm{g}_n(t_i,\omega))\ge \kappa_1.$$ In light of Eqs. \eqref{approx_1}, \eqref{approx_2}, armed with \cref{lemma_w} as well as the continuity of $\bm{f}(t,\omega)$ over $t$, we conclude that $\lambda_{\min}(\bm{f}(t,\omega))\ge \kappa_1$ for all $t$ and $\omega$.
$\hfill \square$

\noindent
	\textbf{Proof of Theorem 1}.
With the results in Propositions 2--3, Eqs. (7) and (17) of the paper, we can derive that
	\begin{align}\label{thm_left_eq}
	&Y_i(u)=Y_i^{(p)}(u)+\bigO_\Pr(p^{-{d_1}})\notag \\
	=&\bm{\alpha}_f^\top(u)\left[\sum_{j=1}^{i-1}\bm{\Phi}_{i,j}\bm{x}_{i-j}
	+\bm{\epsilon}_i\right]+\bigO_\Pr(p^{-{d_1}})\notag \\
	=&\bm{\alpha}_f^\top(u)\left[\sum_{j=1}^{\min\{i-1,b\}}
	\bm{\Phi}_{j}(\frac{i}{n})\bm{x}_{i-j}+
\bm{\epsilon}_i\right]+\bm{\alpha}_f^\top(u)\sum_{j=1}^{\min\{i-1,b\}}
	\left[\bm{\Phi}_{i,j}-
	\bm{\Phi}_j(\frac{i}{n})\right]\bm{x}_{i-j}\notag \\
	&{}+\bm{\alpha}_f^\top(u)
	\sum_{j=b+1}^{i-1}\bm{\Phi}_{i,j}\bm{x}_{i-j}+\bigO_\Pr(p^{-{d_1}})\notag \\
	=&\bm{\alpha}_f^\top(u)\sum_{j=1}^{\min\{i-1,b\}}
	\bm{\Phi}_{j}(\frac{i}{n})\bm{x}_{i-j}+\varepsilon_i(u)+\bigO_\Pr\left(p^{1/2}
	b^{-\tau+2}(\log b)^{\tau-1}+p^{1/2}b^3/n+p^{-d_1}\right).
	\end{align}
	
	Now, consider the main part at the right hand side of Eq. (9) of the paper. Write the time-varying coefficient matrix as $\bm{\Psi}_j(t)\in\mathbb{R}^{p\times p}$ with its $(k,l)$th component denoted by $\psi_{j,kl}(t)$. Using basis expansions on $\psi_j^{(p)}(\frac{i}{n},u,v)$ and $Y_{i-j}(v)$, we have
 
	\begin{align}\label{thm_right_eq}
	&\sum_{j=1}^{\min\{i-1,b\}}\int_0^1 \psi_j^{(p)}(\frac{i}{n},u,v)Y_{i-j}(v)\dee v \notag\\
	=&\sum_{j=1}^{\min\{i-1,b\}}\sum_{k=1}^p\sum_{l=1}^p\psi_{j,kl}
	(\frac{i}{n})f_lx_{i-j,l}\alpha_k(u) \notag \\
	=&\bm{\alpha}_f^\top(u)\sum_{j=1}^{\min\{i-1,b\}}{\rm diag}(1/f_1,...,1/f_p)
	\bm{\Psi}_j(\frac{i}{n}){\rm diag}(f_1,...,f_p)\bm{x}_{i-j}.
	\end{align} 
	Notice that if we choose $\psi_{j,kl}(\cdot)=\phi_{j,kl}(\cdot)f_k/f_l$ for all $j,k,l$, then it is easily to find that $\bm{\Psi}_j(\frac{i}{n})={\rm diag}(f_1,...,f_p)\bm{\Phi}_j(\frac{i}{n}){\rm diag}(f_1^{-1},...,f_p^{-1})$. Consequently, we conclude the approximation result by combining Eqs. \eqref{thm_left_eq} and \eqref{thm_right_eq}. $\hfill \square$

 \subsection{Proof of the results in Section 3}
 \noindent
	\textbf{Proof of Theorem 2}.
	First, from Assumption 2 of the main paper, we can obtain for all $i=1,...,n$, \begin{equation}\label{trun_l2}
 \EE\left| Y_i(u)-Y_i^{(p)}(u)\right|_{\mathcal{L}^2}^2=
	\EE\left(\sum_{k=p+1}^\infty r_{i,k}^2\right)=
	\bigO\left(p^{-(2d_1+1)}\right).
 \end{equation}
	Since the best linear one-step ahead predictor of $\widehat{Y}_{n+1}(u)$ in Eq. (10) belongs to $\mathcal{L}^2([0,1])$, we have
	\begin{equation}\label{g_j}
	\EE\left|\sum_{j=1}^n\int_0^1 g_{n,j}(u,v)Y_{n+1-j}(v)\dee v \right|^2_{\mathcal{L}^2}<\infty.
	\end{equation}
	By the continuity property of the kernel function $g_{n,j}(u,v)$ defined in Eq. (10), it admits a unique representation as
	\begin{equation}\label{g_infinite}
	g_{n,j}(u,v)=\sum_{k=1}^\infty\sum_{l=1}^\infty
	h_{n,kl}^{(j)}\alpha_k(u)\alpha_l(v),~j=1,...,n,
	\end{equation} 
	where $\{h_{n,kl}^{(j)}\}_{k,l=1}^\infty$ is the constant coefficient. Now denote $\{a_{n,j}\}$ as a positive deterministic sequence defined by $a_{n,j}:=\sqrt{\int_0^1\int_0^1 g_{n,j}^2(u,v)\dee u\dee v}$. Moreover, let $g_{n,j}(u,v):=a_{n,j}g_{n,j}^\ast(u,v)$ where $g_{n,j}^\ast(u,v)\in \mathcal{C}^0([0,1]^2)$ for $j=1,...,n$. Notice $g_{n,j}^\ast(u,v)$ admits the basis expansion representation as follows $$g_{n,j}^\ast(u,v)=\sum_{k=1}^\infty\sum_{l=1}^\infty h_{nj,kl}^\ast\alpha_k(u)\alpha_l(v),$$ where $\{h_{nj,kl}^\ast\}_{k,l=1}^\infty$ is also the corresponding coefficient. According to the consequences in \eqref{g_j} and \eqref{g_infinite}, we will obtain $\sum_{j=1}^n a_{n,j}<\infty$ and $h_{nj,kl}^\ast=\bigO\big((kl)^{-1}\big)$ for all $j=1,...,n$. As a result, we conclude that the coefficient $h_{n,kl}^{(j)}=a_{n,j}h_{nj,kl}^\ast$ in the basis expansion representation \eqref{g_infinite} decays with respect to the indices $k,l$ and $j$. 
	
	Now, we construct an intermediate predictor $Y_{n+1}^\dagger(u):=
	\sum_{j=1}^n\langle g_{n,j}(u,v),Y_{n+1-j}^{(p)}(v)\rangle$. Let $\widehat{\rm PE}_{n+1}:=
 Y_{n+1}(u)-Y_{n+1}^\dagger(u)$, then it yields
	
	\begin{align*} 
	\EE\left|\widehat{\rm PE}_{n+1}\right|_{\mathcal{L}^2}^2&=\EE\left| Y_{n+1}(u)-\sum_{j=1}^n\langle g_{n,j}(u,v), Y_{n+1-j}(v)\rangle+\sum_{j=1}^n\langle g_{n,j}(u,v),Y_{n+1-j}(v)-Y_{n+1-j}^{(p)}(v)\rangle\right|_{\mathcal{L}^2}^2\\
	&=\EE\left| {\rm PE}_{n+1}\right|_{\mathcal{L}^2}^2+ \EE\left|\sum_{j=1}^n\langle g_{n,j}(u,v),Y_{n+1-j}(v)-Y_{n+1-j}^{(p)}(v)\rangle\right|_{\mathcal{L}^2}^2\\
	&\le \EE\left| {\rm PE}_{n+1}\right|_{\mathcal{L}^2}^2+
	\sum_{j_1=1}^n\sum_{j_2=1}^n\sum_{k=1}^\infty
	\sum_{l_1=p+1}^\infty\sum_{l_2=p+1}^\infty\left|
	h_{n,kl_1}^{(j_1)}h_{n,kl_2}^{(j_2)}\right|\left|\EE (r_{n+1-j_1,l_1}r_{n+1-j_2,l_2})\right|\\
	&\le \EE\left| {\rm PE}_{n+1}\right|_{\mathcal{L}^2}^2+Cp^{-2(d_1+1)},
	\end{align*}
 where the last inequality holds true using the decay rate of $g_{n,j}(u,v)$ and \cref{lemma_gamma}. For the second equality, the intersection term vanishes by the orthogonality property of ${\rm PE}_{n+1}$ and $Y_n^\ast(u):=\sum_{j=1}^n\langle g_{n,j}(u,v),Y_{n+1-j}(v)-Y_{n+1-j}^{(p)}(v)\rangle$ for any $u\in[0,1]$. To be more specific, we first discretize the integral of intersection term. Let $0=u_0<u_1<\cdots<u_n=1$ be the fine enough discrete points on $[0,1]$, then for any fixed $u_i\in[0,1]$, we find that the best linear forecast of $Y_{n+1}(u_i)$ is $\sum_{j=1}^n\langle g_{n,j}(u_i,v),Y_{n+1-j}(v)\rangle$. Consequently for $j=1,...,n$, $\langle g_{n,j}(u_i,v),Y_{n+1-j}(v)\rangle$ is the projection mapping of $Y_{n+1-j}(u_i)$ onto the space of its linear combinations. 
	By Fubini's theorem and the definition of integration, we have
	$$\EE\int_0^1{\rm PE}_{n+1}(u)Y_n^\ast(u)\dee u=\int_0^1\EE \left[{\rm PE}_{n+1}(u)Y_n^\ast(u)\right]\dee u=
	\lim_{n\to \infty}\frac{1}{n}\sum_{i=1}^n\EE\left[{\rm PE}_{n+1}(u_i)Y_n^\ast(u_i)\right]=0,$$
	which implies that the integral of intersection term equals to zero. 
	
	From the above deduction and the fact $\EE\left|\widehat{{\rm PE}}_{n+1}\right|_{\mathcal{L}^2}^2 \ge \EE\left|{\rm PE}_{n+1}\right|_{\mathcal{L}^2}^2$, we conclude
	\begin{equation}\label{err1}
	\EE|\widehat{\rm PE}_{n+1}|_{\mathcal{L}^2}^2-
	\EE|{\rm PE}_{n+1}|_{\mathcal{L}^2}^2\le Cp^{-2(d_1+1)}.
	\end{equation} 
	Next, consider the best linear one-step ahead predictor using the preceding truncated functional time series $Y_{n+1-j}^{(p)}(u)$, denoted as $\widetilde{Y}_{n+1}^{(p)}(u):=\sum_{j=1}^n
 \langle \widetilde{g}_{n,j}(u,v),Y_{n+1-j}^{(p)}(v)\rangle$ where $\widetilde{g}_{n,j}(u,v)\in\mathcal{C}^0([0,1]^2)$ has similar basis expansion to \cref{g_infinite} with $h_{n,kl}^{(j)}$ replaced by $\tilde{h}_{n,kl}^{(j)}$ for $j=1,...,n$. Let $\widetilde{\rm PE}_{n+1}:=Y_{n+1}(u)-\widetilde{Y}_{n+1}^{(p)}(u)$ be its corresponding best linear forecast error based on the historical data $\{Y_i(u)\}_{i=1}^n$. With the above discussion, we have 
	$$ \EE\left|{\rm PE}_{n+1}\right|_{\mathcal{L}^2}^2\le \EE\left|\widetilde{\rm PE}_{n+1}\right|_{\mathcal{L}^2}^2 \le \EE\left|\widehat{\rm PE}_{n+1}\right|_{\mathcal{L}^2}^2.$$
	Furthermore, combining with \cref{err1}, we will obtain 
	\begin{equation}\label{err2}
	\EE\left|\widetilde{\rm PE}_{n+1}\right|_{\mathcal{L}^2}^2-\EE\left|{\rm PE}_{n+1}\right|_{\mathcal{L}^2}^2\le Cp^{-2(d_1+1)}.
	\end{equation}
	Meanwhile, consider $\widetilde{Y}_{n+1}^{(p)}(u)$ as a linear (not necessarily the best) predictor of truncated process $Y_{n+1}^{(p)}(u)$ and denote its prediction error as $\widetilde{\rm PE}_{n+1}^{(p)}:= Y_{n+1}^{(p)}(u)-\widetilde{Y}_{n+1}^{(p)}(u)$, then we can derive
	\begin{align*}
	\EE\left|\widetilde{\rm PE}_{n+1}^{(p)}\right|_{\mathcal{L}^2}^2&=\EE\left| Y_{n+1}^{(p)}(u)-Y_{n+1}(u)+Y_{n+1}(u)-\sum_{j=1}^n\langle \widetilde{g}_{n,j}(u,v), 
 Y_{n+1-j}^{(p)}(v)\rangle\right|_{\mathcal{L}^2}^2\\
	&\le \sum_{k=p+1}^\infty \EE\left(r_{n+1,k}^2\right)+
	\EE\left|\widetilde{\rm PE}_{n+1}\right|_{\mathcal{L}^2}^2+2\sum_{j=1}^n\sum_{k=p+1}^\infty\sum_{l=1}^{p}\max_{j} \left|\tilde{h}_{n,kl}^{(j)}\right|\left|
 \EE(r_{n+1,k}r_{n+1-j,l})\right|\\
	&\le \EE\left|\widetilde{\rm PE}_{n+1}\right|_{\mathcal{L}^2}^2+
	C\left(p^{-(2d_1+1)}+p^{-(d_1+1)}\right),
	\end{align*}
	where the last inequality follows by \cref{lemma_gamma} and $\max_j|\tilde{h}_{n,kl}^{(j)}|\le C(kl)^{-1}$.
	Hence it yields that
	\begin{equation}\label{err3}
	\EE|\widetilde{\rm PE}_{n+1}^{(p)}|_{\mathcal{L}^2}^2
 -\EE|\widetilde{{\rm PE}}_{n+1}|_{\mathcal{L}^2}^2\le C\left(p^{-(2d_1+1)}+p^{-(d_1+1)}\right).
	\end{equation}
	Recall ${\rm PE}_{n+1}^{(p)}=Y_{n+1}^{(p)}(u)-\widehat{Y}_{n+1}^{(p)}(u)$ defined in Theorem 2 and $\widehat{Y}_{n+1}^{(p)}(u):=\sum_{j=1}^n\langle
 g_{n,j}^{(p)}(u,v),Y_{n+1-j}^{(p)}(v)\rangle$ where $g_{n,j}^{(p)}(u,v)\in \mathcal{C}^0([0,1]^2)$, it is clear to see that $\EE|{\rm PE}_{n+1}^{(p)}|_{\mathcal{L}^2}^2\le\EE\left| \widetilde{{\rm PE}}_{n+1}^{(p)}\right|_{\mathcal{L}^2}^2$. Armed with \cref{err3}, we have  
	\begin{equation}\label{err4}
	\EE|{\rm PE}_{n+1}^{(p)}|_{\mathcal{L}^2}^2-\EE|\widetilde{\rm PE}_{n+1}|_{\mathcal{L}^2}^2\le C\left(p^{-(2d_1+1)}+p^{-(d_1+1)}\right)\le Cp^{-(d_1+1)}.
	\end{equation} 
	Lastly, we introduce another intermediate prediction error term 
	${\rm PE}_{n+1}^\ast:=Y_{n+1}(u)-\sum_{j=1}^n\langle g_{n,j}^{(p)}(u,v),Y_{n+1-j}^{(p)}(v)\rangle$, where $g_{n,j}^{(p)}(u,v)$ retains the same as that defined in $\widehat{Y}_{n+1}^{(p)}(u)$. Similar to \eqref{g_infinite}, the kernel function $g_{n,j}^{(p)}(u,v)$ admits the following representation as
	\begin{equation*}
	g_{n,j}^{(p)}(u,v)=a_{n,j}^{(p)}\sum_{k=1}^\infty\sum_{l=1}^\infty
	h_{n,kl}^{(j,p)}\alpha_k(u)
 \alpha_l(v),~j=1,...,n,
	\end{equation*}
 where $a_{n,j}^{(p)}:=\sqrt{\int_0^1\int_0^1 [g_{n,j}^{(p)}(u,v)]^2\dee u\dee v}$ and $h_{n,kl}^{(j,p)}=\bigO((kl)^{-1})$ for all $j=1,...,n$. Then with the similar routine, we can obtain
	\begin{align*}
	\EE\left|{\rm PE}_{n+1}^\ast\right|_{\mathcal{L}^2}^2&=\EE\left| Y_{n+1}(u)-Y_{n+1}^{(p)}(u)+Y_{n+1}^{(p)}(u)-\sum_{j=1}^n\langle g_{n,j}^{(p)}(u,v), Y_{n+1-j}^{(p)}(v)\rangle\right|_{\mathcal{L}^2}^2\\
	&\le \sum_{k=p+1}^\infty\EE\left(r_{n+1,k}^2\right)
	+\EE\left|{\rm PE}_{n+1}^{(p)}\right|_{\mathcal{L}^2}^2
	+2\sum_{j=1}^n\sum_{k=p+1}^\infty
 \sum_{l=1}^{p}a_{n,j}^{(p)}\max_{j} \left|h_{n,kl}^{(j,p)}\right|\left|
 \EE(r_{n+1,k}r_{n+1-j,l})\right|.
	\end{align*}
	Consequently, it yields that
	\begin{equation}\label{err5}
	\EE|{\rm PE}_{n+1}^\ast|_{\mathcal{L}^2}^2-\EE|{\rm PE}_{n+1}^{(p)}|_{\mathcal{L}^2}^2\le C\left(p^{-(2d_1+1)}+p^{-(d_1+1)}\right)\le Cp^{-(d_1+1)}.
	\end{equation}
 Since $\widetilde{\rm PE}_{n+1}$ is the best linear forecast error of $Y_{n+1}(u)$ based on $Y_{n+1}^{(p)}(u)$ while ${\rm PE}_{n+1}^\ast$ is not, we conclude that $\EE|\widetilde{\rm PE}_{n+1}|_{\mathcal{L}^2}^2\le\EE\left|{\rm PE}_{n+1}^\ast\right|_{\mathcal{L}^2}^2$. Combining with \eqref{err5}, then we have 
	\begin{equation}\label{err6}
	\EE|\widetilde{\rm PE}_{n+1}|^2-\EE|{\rm PE}_{n+1}^{(p)}|_{\mathcal{L}^2}^2\le Cp^{-(d_1+1)}.
	\end{equation}
	With \cref{err4,err6}, we conclude that $$\EE|{\rm PE}_{n+1}^{(p)}|_{\mathcal{L}^2}^2-\EE|\widetilde{\rm PE}_{n+1}|_{\mathcal{L}^2}^2=
	\bigO\left(p^{-(d_1+1)}\right).$$
	Moreover, by \cref{err2} and the triangle inequality, we complete our proof with
	$$\EE\left|{\rm PE}_{n+1}^{(p)}\right|
 _{\mathcal{L}^2}^2-\EE\left|{\rm PE}_{n+1}\right|_{\mathcal{L}^2}^2=
 \bigO\left(p^{-(d_1+1)}+p^{-(2d_1+1)}
	\right)=\bigO\left(p^{-(d_1+1)}\right).$$ $\hfill \square$
	
	\noindent
	\textbf{Proof of Theorem 3}.
	First, recall that $\widehat{\bm{x}}_{n+1}$ and $\widehat{\bm{x}}_{n+1}^b$ are the best linear forecasts of $\bm{x}_{n+1}$ based on $\bm{x}_1,...,\bm{x}_n$ and $\bm{x}_n,...\bm{x}_{n+1-b}$, respectively. According to Eqs. (7) and (14), we can rewrite the best linear one-step ahead predictor of 
 $Y_{n+1}^{(p)}(u)$ as $$\widehat{Y}_{n+1}^{(p)}(u)=\bm{\alpha}_f^\top(u)\sum_{j=1}^n
	\bm{\Phi}_{n,j}\bm{x}_{n+1-j}.$$ Then, we can obtain 
	\begin{align*}
	{\rm PE}_{n+1}^{(b)}&=Y_{n+1}^{(p)}(u)-\widehat{Y}_{n+1}^{(p)}(u)+\widehat{Y}_{n+1}^{(p)}(u)-
	\widetilde{Y}_{n+1}^{(b)}(u)\\
	&={\rm PE}_{n+1}^{(p)}+
 \bm{\alpha}_f^\top(u)\sum_{j=1}^b
	\left[\bm{\Phi}_{n,j}-\bm{\Phi}_j(1)\right]\bm{x}_{n+1-j}+
	\bm{\alpha}_f^\top(u)
 \sum_{j=b+1}^n\bm{\Phi}_{n,j}\bm{x}_{n+1-j}.
	\end{align*}
	Calculating $\mathcal{L}^2$ norm and taking expectation on both sides of the above relation, we have
	\begin{align*}
	\EE\left|{\rm PE}_{n+1}^{(b)}\right|_{\mathcal{L}^2}^2
	&=\EE\left|{\rm PE}_{n+1}^{(p)}\right|_{\mathcal{L}^2}^2+2\EE\langle
	\sum_{j=1}^b\bm{\alpha}_f^\top(u)[\bm{\Phi}_{n,j}-\bm{\Phi}_j(1)]
	\bm{x}_{n+1-j},\sum_{j=b+1}^n\bm{\alpha}_f^\top(u)\bm{\Phi}_{n,j}
	\bm{x}_{n+1-j}\rangle\\&{}~+\EE\left|\sum_{j=1}^b\bm{\alpha}_f^\top(u)
	\left[\bm{\Phi}_{n,j}-\bm{\Phi}_j(1)\right]\bm{x}_{n+1-j}\right|
	_{\mathcal{L}^2}^2+\EE\left|\bm{\alpha}_f^\top(u)
	\sum_{j=b+1}^n\bm{\Phi}_{n,j}\bm{x}_{n+1-j}\right|_{\mathcal{L}^2}^2\\
	&\le \EE\left|{\rm PE}_{n+1}^{(p)}\right|_{\mathcal{L}^2}^2+
	C\left(pb^{-2\tau+3}(\log b)^{2\tau-3}+pb^5/n^2\right),
	\end{align*}
	where the first equality holds because two intersection terms involving ${\rm PE}_{n+1}^{(p)}$ vanish by the fact that $\widehat{Y}_{n+1}^{(p)}(u)$ is the best linear forecast of $Y_{n+1}^{(p)}(u)$ and ${\rm PE}_{n+1}^{(p)}$ is uncorrelated to any linear combination of $\{\bm{x}_1,...,\bm{x}_n\}$. Similarly, we also obtain
	$$\EE\left|{\rm PE}_{n+1}^{(p)}\right|_{\mathcal{L}^2}^2-
	\EE\left|{\rm PE}_{n+1}^{(b)}\right|
 _{\mathcal{L}^2}^2\le C\left(
	pb^{-2\tau+3}(\log b)^{2\tau-3}+pb^5/n^2\right).$$ Therefore, we conclude that 
	$$\EE\left|{\rm PE}_{n+1}^{(b)}\right|_{\mathcal{L}^2}^2-
	\EE\left|{\rm PE}_{n+1}^{(p)}\right|_{\mathcal{L}^2}^2=\bigO\left(
	pb^{-2\tau+3}(\log b)^{2\tau-3}+pb^5/n^2\right).$$ 
	Together with the result in Theorem 2 and the triangular inequality, we have 
	\begin{align*}
	&\left|\EE\left|{\rm PE}_{n+1}\right|_{\mathcal{L}^2}^2-
	\EE\left|{\rm PE}_{n+1}^{(b)}\right|_{\mathcal{L}^2}^2\right|\\
	\le& \left|\EE\left|{\rm PE}_{n+1}\right|_{\mathcal{L}^2}^2-
	\EE\left|{\rm PE}_{n+1}^{(p)}\right|_{\mathcal{L}^2}^2\right|+
	\left|\EE\left|{\rm PE}_{n+1}^{(p)}\right|_{\mathcal{L}^2}^2-
	\EE\left|{\rm PE}_{n+1}^{(b)}\right|_{\mathcal{L}^2}^2\right|\\
	\le &C\left(p^{-(d_1+2)}+pb^{-2\tau+3}(\log b)^{2\tau-3}+pb^5/n^2\right).
	\end{align*}
	$\hfill \square$
 
	\noindent
	\textbf{Proof of Proposition 2}.
	For $i>b$, denote the $(i-1)p\times (i-1)p$ symmetric block banded matrix $\bm{\Gamma}_i^\ast$ by defining its $p\times p$ block element as 
	\begin{equation*}
	(\bm{\Gamma}_i^\ast)_{kl}=
	\begin{cases}
	(\bm{\Gamma}_i)_{kl},~&|k-l|\le \frac{j}{m_0\log j},\\
	~\bm{0}_{p\times p},~&\text{otherwise},
	\end{cases}
	\end{equation*}
	where $m_0$ is some large constant and $k,l=1,...,i-1$. Since $\lambda_{\min}(\bm{\Gamma}_i)\ge \kappa_1>0$ via Assumption 5, by Weyl's inequality and Assumption 4, we can obtain
	\begin{align}
	&\lambda_{\min}(\bm{\Gamma}_i^\ast)\ge \kappa_1-Cj^{-\tau+1}(m_0\log j)^{\tau-1}>0,\label{lower_bound}\\
	&\lambda_{\max}(\bm{\Gamma}_i^\ast)\le\sum_{k=1}^{\lfloor{j/m_0\log j}\rfloor+1}\Vert \bm{\Gamma}(t,k)\Vert\le C. \label{upper_bound}
	\end{align} 
	Therefore, the support of the spectrum of $\bm{\Gamma}_i^\ast$ is bounded from both above and below by some constants. Recall $\bm{\gamma}_i={\rm Cov}(\bm{x}_{i-1}^i,\bm{x}_i)$, define $\bm{\Phi}_i^\ast=\bm{\Omega}_i^\ast\bm{\gamma}_i$ where $\bm{\Omega}_i^\ast=(\bm{\Gamma}_i^\ast)^{-1}$. By the fact that any operator norm is sub-multiplicative, then we have 
	\begin{align}
	\Vert\bm{\Phi}_i-\bm{\Phi}_i^\ast\Vert&=
	\Vert(\bm{\Omega}_i-\bm{\Omega}_i^\ast)
	\bm{\gamma}_i\Vert \le
	\Vert\bm{\Gamma}_i^{-1}(\bm{\Gamma}_i^\ast-\bm{\Gamma}_i)
	(\bm{\Gamma}_i^\ast)^{-1}\Vert
	\Vert\bm{\gamma}_i\Vert \notag \\
	&=\Vert\bm{\Gamma}_i^{-1}\Vert \Vert \bm{\Gamma}_i^\ast-\bm{\Gamma}_i\Vert
	\Vert(\bm{\Gamma}_i^\ast)^{-1}\Vert
	\Vert\bm{\gamma}_i\Vert
	\le Cj^{-\tau+1}(\log j)^{\tau-1}. \label{A.1} 
	\end{align}
	where the last inequality follows by Assumption 5 of the paper, Eq. \eqref{lower_bound} and \cref{g_circle}. Thus, 
	\begin{equation}\label{diff_order}
	\left\Vert \bm{\Phi}_{i,j}-\bm{\Phi}_{i,j}^\ast\right\Vert \le Cj^{-\tau+1}(\log j)^{\tau-1},
	\end{equation} 
	where $\bm{\Phi}_{i,j}^\ast$ is the $j$-th block matrix of $\bm{\Phi}_i^\ast$. Now, it suffices to control the operator norm of $\bm{\Phi}_{i,j}^\ast$. Note that $\bm{\Gamma}_i^\ast$ is a block $2(j+1)p/m_0\log j$-banded positive definite bounded matrix by its definition and denote $\omega_i^\ast(k,l)$ as the $(k,l)$th element of $\bm{\Omega}_i^\ast$. According to \cref{lemma_band}, there exists a $\lambda_1\in (0,1)$ and some constant $C>0$, for each element $k,l=1,...,(i-1)p$ we have
	\begin{equation}
	\left|\omega_i^\ast(k,l)\right| \le C\lambda_1^{\frac{m_0\log j|k-l|}{(j+1)p}}. \label{A.2}
	\end{equation}
	Notice that $\bm{\Phi}_{i,j}^\ast$ is the product of the block matrix $\bm{\Omega}_i^\ast$'s $j$-th block row and the $(i-1)p\times p$ block matrix $\bm{\gamma}_i$. Further denote $(\bm{\Omega}_i^\ast)_{jl}$ as the $(j,l)$th $p\times p$ block matrix of $\bm{\Omega}_i^\ast$ and let $\bm{\gamma}_{i,l}$ be the $l$th $p\times p$ block matrix. Through \cref{A.2} and elementary calculations, we obtain that 
	\begin{align}
	\Vert\bm{\Phi}_{i,j}^\ast\Vert&\le \left\Vert \sum_{l=1}^{j-1} (\bm{\Omega}_i^\ast)_{jl} \bm{\gamma}_{i,l}\right\Vert+
	\left\Vert\sum_{l=j}^{i-1} (\bm{\Omega}_i^\ast)_{jl} \bm{\gamma}_{i,l}\right\Vert \notag\\
	&\le C\sum_{l=1}^{j-1} \lambda_1^{\frac{m_0\log j[(j-l-1)p+1]}{(j+1)p}} (l+1)^{-\tau}+
	C\sum_{l=j}^{i-1}(l+1)^{-\tau}. \label{summation1}
	\end{align}
	The second term in \eqref{summation1} can be easily controlled by $\bigO(j^{-\tau+1})$ via elementary calculations. For the first term in \eqref{summation1}, since $j$ diverges with $n$, we have $\lambda_1^{k\log j}\asymp j^{-k}$ for $\lambda_1\in (0,1)$. Then it suffices to control the first term in \eqref{summation1} by considering the product under summation as a function of $l$, that is, $f(l):=j^{-\frac{m_0[(j-l-1)p+1]}{(j+1)p}}(l+1)^{-\tau}$. Observe that $f(l)$ is decreasing in $[1,\lfloor \frac{\tau(j+1)}{m_0\log j}\rfloor-1]$ and increasing in $[\lfloor \frac{\tau(j+1)}{m_0\log j}\rfloor,j-1]$. Consequently, it yields that
		\begin{align*}
		\sum_{l=1}^{j-1}j^{-\frac{m_0[(j-l-1)p+1]}{(j+1)p}}(l+1)^{-\tau}&\le 
		\left(\left\lfloor \frac{\tau(j+1)}{m_0\log j}\right\rfloor-1\right)
		j^{-\frac{m_0[(j-2)p+1]}{(j+1)p}}2^{-\tau}+\sum_{l=\left\lfloor \frac{\tau(j+1)}{m_0\log j}\right\rfloor}^{j-1}(l+1)^{-\tau}\\
			&\le C(j/\log j)^{-\tau}.	
		\end{align*} 
	Combining \eqref{diff_order} and \eqref{summation1}, Eq. (15) in Proposition 2 holds true by triangle inequality. $\hfill \square$ 

 To prove the results in Proposition 3, we need an additional lemma.
 \begin{lemma}\label{addit_lemma}
 When $i>b$, let $\bm{x}_i^{(b)}=(\bm{x}_{i-1}^\top,...,\bm{x}_{i-b}^\top)^\top\in \mathbb{R}^{bp}$ with $\bm{x}_{i-k}=(x_{i-k,1},...,x_{i-k,p})^\top\in \mathbb{R}^{p}$ for $k=1,...,b$. Define $\bm{\Phi}_i^{(b)}:=\bm{\Omega}_i^{(b)}
 \bm{\gamma}_i^{(b)}:=(\bm{\Phi}_{i,1}^{(b)\top},...,
 \bm{\Phi}_{i,b}^{(b)\top})^\top\in \mathbb{R}^{bp\times p}$ as the best linear forecast coefficient matrices of $\bm{x}_i$ based on $\bm{x}_{i-1},...,\bm{x}_{i-b}$, where $\bm{\Omega}_i^{(b)}=[{\rm Cov}(\bm{x}_i^{(b)},\bm{x}_i^{(b)})]^{-1}$ and $\bm{\gamma}_i^{(b)}={\rm Cov}(\bm{x}_i^{(b)},\bm{x}_i)$. Then we can obtain
	\begin{equation}\label{prop_error2}
	\max_{i>b}\left\Vert\bm{\Phi}_{i,j}-
	\bm{\Phi}_{i,j}^{(b)}\right\Vert\le Cb^{-\tau+1}(\log b)^{\tau-1},~1\le j\le b.
	\end{equation}
 \end{lemma}
 
 \noindent
 \textit{Proof}: we rewrite the original autocovariance matrix $\bm{\Gamma}_i$ as the following block matrix 
	$$\bm{\Gamma}_i=\begin{pmatrix}
	{\rm Cov}(\bm{x}_i^{(b)},\bm{x}_i^{(b)}) & \bm{B}_1\\
	\bm{B}_3 & \bm{B}_2
	\end{pmatrix},$$ where $\bm{B}_i,~i=1,2,3$ are defined as
	$$\bm{B}_1={\rm Cov}(\bm{x}_i^{(b)},\bm{x}_i^\ast)\in 
 \mathbb{R}^{bp\times(i-b-1)p},~\bm{B}_2={\rm Cov}(\bm{x}_i^\ast,\bm{x}_i^\ast)\in\mathbb{R}^{(i-b-1)p\times(i-b-1)p},~\bm{B}_3=\bm{B}_1^\top,$$ with $\bm{x}_i^\ast=(\bm{x}_{i-b-1}^\top,...,\bm{x}_1^\top)^\top
 \in\mathbb{R}^{(i-b-1)p}$. Further define $\widetilde{\bm{\gamma}}_i=(({\bm \gamma_i^{(b)}})^\top,\bm{0})^\top \in \mathbb{R}^{(i-1)p\times p}$ and  $\widetilde{\bm{\Phi}}_i=((\bm{\Phi}_i^{(b)})^\top,\bm{0})^\top\in 
 \mathbb{R}^{(i-1)p\times p}$, then by Yule-Walker equation we have
	\begin{equation}
\bm{\Gamma}_i\widetilde{\bm{\Phi}}_i=
\widetilde{\bm{\gamma}}_i+
	\Delta\gamma_i,\label{yw1}
	\end{equation}
	where $\Delta\gamma_i=(\bm{0},(\bm{B}_3\bm{\Phi}_i^{(b)})^\top)^\top$.
	Now we employ \cref{perturbation} with $\bm{A}=\bm{\Gamma}_i,~\Delta \bm{A}=\bm{0},~\bm{x}=\bm{\Phi}_i,~\Delta \bm{x}=\widetilde{\bm{\Phi}}_i^{(b)}-
	\bm{\Phi}_i,~\bm{v}=\bm{\gamma}_i,\Delta \bm{v}=\widetilde{\bm{\gamma}}_i-\bm{\gamma}_i-\Delta\bm{\gamma}_i$. Since $\lambda_{\min}(\bm{\Gamma}_i)\ge \kappa_1>0$ by , hence it suffices to control the decay rate of $\Vert \Delta \bm{x}\Vert$ by calculating $\Vert \Delta \bm{v}\Vert \Vert \bm{x}\Vert/\Vert \bm{v}\Vert$. First notice that $\Vert \Delta \bm{v}\Vert\le \Vert \widetilde{\bm{\gamma}}_i-\bm{\gamma}_i\Vert+\Vert\Delta\bm{\gamma}_i\Vert$ and by Assumption 5, it is obvious to see 
	\begin{equation}\label{delta_1}
	\Vert\widetilde{\bm{\gamma}}_i-\bm{\gamma}_i\Vert\le Cb^{-\tau+1/2}.
	\end{equation} 
	Next let us focus on the non-zero part of $\Delta\bm{\gamma}_i$, for $1\le j\le i-b-1$, note that the $j$th block of $\bm{B}_3\bm{\Phi}_i^{(b)}$ can be controlled as
	\begin{align*}
	\Vert (\bm{B}_3\bm{\Phi}_i^{(b)})_j\Vert&=
 \left\Vert\sum_{k=1}^b
	{\rm Cov}(\bm{x}_{i-b-j},\bm{x}_{i-k})\bm{\Phi}_{i,k}^{(b)}\right\Vert\le 
	\sum_{k=1}^b\Vert {\rm Cov}(\bm{x}_{i-b-j},\bm{x}_{i-k})\Vert 
	\Vert \bm{\Phi}_{i,k}^{(b)}\Vert\\
	&\le \sum_{k=1}^{b}
		(b+j+1-k)^{-\tau}k^{-\tau+1}(\log k)^{\tau-1}\\
		&\le C(\log b)^{\tau-1}\sum_{k=1}^{b}
		(b+j+2-k)^{-\tau}k^{-\tau+1} \\
		&= \frac{C(\log b)^{\tau-1}}{(b+j+1)^{\tau}}\sum_{k=1}^b
		\frac{(b+j+1-k+k)^{\tau}}{(b+j+1-k)^{\tau}k^{\tau-1}}\\
		&\le \frac{C(\log b)^{\tau-1}}{(b+j+1)^{\tau}}\left(1+
		\sum_{k=1}^b\frac{b+j+1}{(b+j+1-k)^{\tau}}\right)\\
		&\le \frac{C(\log b)^{\tau-1}}{(b+j+1)^{\tau}}\left(1+
		\frac{b+j+1}{(j+1)^{\tau-1}}\right),
	\end{align*}
	where the first inequality uses Assumption 4 of the paper and \eqref{delta_1}, the third inequality follows by the elementary inequality $(a+b)^\tau\le 2^{\tau-1}(a^\tau+b^\tau)$ for $a,b>0$. As a result, we can calculate
	\begin{align}
	\Vert \Delta\bm{\gamma}_i\Vert&\le (\log b)^{\tau-1}
	\left\{\sum_{j=1}^{i-b-1}\left(\frac{1}{(b+j+1)^{2\tau}}+
	\frac{1}{(b+j+1)^{2\tau-2}(j+1)^{2\tau-2}}\right)\right\}^{1/2}\notag \\
	&\le C(\log b)^{\tau-1}\left\{(b+2)^{-2\tau+1}+(b+1)^{-2\tau+2}\right\}^{1/2}=
	\bigO\left((\log b)^{\tau-1}b^{-\tau+1}\right). \label{delta_2}
	\end{align}
	Moreover, by the conclusion of the first part in this proposition and Assumption 4, we have $\Vert \bm{x}\Vert\le C$ and $\Vert \bm{v}\Vert\ge C>0$. Together with \eqref{delta_1} and \eqref{delta_2}, this finishes our proof of the second part. $\hfill \square$

 \noindent
	\textbf{Proof of Proposition 3}.
	The first part refers to the proof of \citep[Lemma 3.1]{Ding20}. For the second part, recall $\bm{\Phi}_i^{(b)}=(\Phi_{i,1}^{(b)\top},
	...,\Phi_{i,b}^{(b)\top})^\top$ and denote $\bm{\Omega}_n(t)=[\bm{\Gamma}_n(t)]^{-1}$, observe that for any fixed $i$ and $j=1,...,b$,
	\begin{equation}\label{A.3}
	\bm{\Phi}_{i,j}^{(b)}-\bm{\Phi}_j(\frac{i}{n})=\bm{E}_j^\top\bm{\Omega}_i^{(b)}[\bm{\gamma}_i^{(b)}-\bm{\gamma}_n(i/n)]+
 \bm{E}_j^\top\bm{\Omega}_i^{(b)}
 [\bm{\Gamma}_n(i/n)-\bm{\Gamma}_i^{(b)}]\bm{\Omega}_n(i/n)
 \bm{\gamma}_n(i/n),
	\end{equation}
	where $\bm{E}_j$ is a $bp\times p$ block matrix with the $j$th block being $\bm{\I}_{p\times p}$ and other blocks being $\bm{0}_{p\times p}$. Further denote $\bm{\gamma}_i^\ast=(\bm{\Gamma}(\frac{i-1}{n},1)^\top,...,\bm{\Gamma}
	(\frac{i-b}{n},b)^\top)^\top$, then for the first part of \cref{A.3}, we have
	\begin{align*}
	\left\Vert\bm{E}_j^\top\bm{\Omega}_i^{(b)}(\bm{\gamma}_i^{(b)}-\bm{\gamma}_n(i/n))
	\right\Vert&\le \left\Vert\bm{E}_j^\top\bm{\Omega}_i^{(b)}
	\right\Vert \left\Vert\bm{\gamma}_i^{(b)}-\bm{\gamma}_n(i/n)\right\Vert\\
		&\le C\left(\left\Vert\bm{\gamma}_i^{(b)}-\bm{\gamma}_i^\ast\right\Vert+
		\left\Vert\bm{\gamma}_i^\ast-\bm{\gamma}_n(i/n)\right\Vert\right)\\
		&\le C\sqrt{\sum_{k=1}^b\sup_{t\in[0,1]}
  \left\Vert\frac{\partial \bm{\Gamma}(t,k)}{\partial t}\right\Vert^2 \frac{k^2}{n^2}}\le \frac{Cb^{3/2}}{n},
	\end{align*}
	where the second inequality uses triangle inequality and the fact 
	$\lambda_{\max}\{\bm{\Omega}_i^{(b)}\}=1/\lambda_{\min}
	\{\bm{\Gamma}_i^{(b)}\}\le 1/\kappa_1\le C$. The second inequality follows by the mean value theorem and the last inequality holds due to the assumption $\sup_{t\in[0,1]}\Vert \partial\bm{\Gamma}(t,j)/\partial t\Vert \le C$. On the other hand, using Gershgorin circle theorem of \cref{g_circle} for block matrix and similar to the proof in the first part, we can show that 
	\begin{align*}
	&\left\Vert\bm{E}_j^\top\bm{\Omega}_i^{(b)}
	[\bm{\Gamma}_n(i/n)-\bm{\Gamma}_i^{(b)}]\bm{\Omega}_n(i/n)
	\bm{\gamma}_n(i/n)\right\Vert\\ \le& 
	\left\Vert\bm{E}_j^\top\bm{\Omega}_i^{(b)}\right\Vert
	\left\Vert\bm{\Gamma}_n(i/n)-\bm{\Gamma}_i^{(b)}\right\Vert
	\left\Vert\bm{\Omega}_n(i/n)
 \bm{\gamma}_n(i/n)\right\Vert
	\le C\frac{b^2}{n}.
	\end{align*}
	By the triangle inequality, we will attain the final result as
	\begin{align*}
	&\left\Vert \bm{\Phi}_{i,j}-\bm{\Phi}_j(\frac{i}{n})\right\Vert \le \left\Vert\bm{\Phi}_{i,j}-\bm{\Phi}_{i,j}^{(b)}\right\Vert+\left\Vert \bm{\Phi}_{i,j}^{(b)}-\bm{\Phi}_j(\frac{i}{n})\right\Vert\\
	\le &C\left(b^{-\tau+1}(\log b)^{\tau-1}+b^2/n\right).
	\end{align*}
	$\hfill \square$

	\noindent
	\textbf{Proof of Proposition 4}. We will utilize the technique from the proof in \citep[Theorem 1]{liu2013probability} and follow the proof strategy of \citep[Theorem 3.7]{Ding20}. The proof contains two main steps: (i). Find the convergence rate of difference between our projection matrix $\bm{S}_n$ and the deterministic matrix $\bm{W}$ using the trick of $m$-dependent sequence where 
	$$\bm{S}_n=\frac{1}{n}\bm{Y}^\top\bm{Y}=\frac{1}{n}\sum_{i=b+1}^n\bm{y}_i\bm{y}_i^\top,~~\bm{W}=\int_0^1\bm{W}^{(b)}(t)\otimes (\bm{v}(t)\bm{v}^\top(t))\dee t,$$ with $\bm{W}^{(b)}(t)$ defined in Assumption 7; (ii). Follow the proof of \citep[Theorem 3.7]{Ding20} to complete our proof. 
	
	Starting with the first step, we will rewrite the projection matrix by the definition $\bm{y}_i=\bm{x}_i^{(b)}\otimes \bm{v}(i/n)$ as 
	$$\bm{S}_n=
	\frac{1}{n}\sum_{i=b+1}^n[\bm{x}_i^{(b)}(\bm{x}_i^{(b)})^\top]\otimes 
	[\bm{v}(i/n)\bm{v}^\top(i/n)]\in \mathbb{R}^{bcp\times bcp},$$
Denote the corresponding $j$-dependent sequence as $$\bm{S}_{n,j}=\frac{1}{n}\sum_{i=b+1}^n
\EE[\bm{x}_i^{(b)}(\bm{x}_i^{(b)})^\top|\eta_{i-j-1},...,\eta_{i-1}]\otimes 
[\bm{v}(i/n)\bm{v}^\top(i/n)],$$ namely $\EE[\bm{x}_i^{(b)}(\bm{x}_i^{(b)})^\top|\eta_{i-j-1},...,\eta_{i-1}]$ and $\EE[\bm{x}_{i'}^{(b)}(\bm{x}_{i'}^{(b)})^\top|\eta_{i'-j-1},...,\eta_{i'-1}]$ are independent if $|i-i'|>j$. Then we can decompose the difference into three parts: 
\begin{align}\label{decomp_3}
&\bm{S}_n-\bm{W}\notag \\
=&(\bm{S}_n-\bm{S}_{n,n})+\sum_{j=2}^n(\bm{S}_{n,j}-\bm{S}_{n,j-1})+
(\bm{S}_{n,1}-\bm{W})=:{\rm I}+{\rm II}+{\rm III}.
\end{align} 
Note that $[\bm{x}_i^{(b)}(\bm{x}_i^{(b)})^\top]$ and $[\bm{v}(i/n)\bm{v}^\top(i/n)]$ are both positive semi-definite, we can separate the matrix of basis functions and random matrix deviation in I, II and III. To be more specific, we can control the first term I for instance as follows
\begin{align*}
\Vert \bm{S}_n-\bm{S}_{n,n}\Vert &\le \sup_{i}\Vert \bm{v}(i/n)\bm{v}^\top(i/n)\Vert 
\left\Vert\frac{1}{n}
\sum_{i=b+1}^n \left(\bm{x}_i^{(b)}
(\bm{x}_i^{(b)})^\top-\EE[\bm{x}_i^{(b)}
(\bm{x}_i^{(b)})^\top|\eta_{i-j-1},...,\eta_{i-1}]\right)\right\Vert\\
&\le \zeta_c^2 \left|\frac{1}{n}
\sum_{i=b+1}^n \left(\bm{x}_i^{(b)}
(\bm{x}_i^{(b)})^\top-\EE[\bm{x}_i^{(b)}
(\bm{x}_i^{(b)})^\top|\eta_{i-j-1},...,\eta_{i-1}]\right)\right|_F,
\end{align*}
where the second inequality follows by the definition of $\zeta_c$ and the fact $\Vert\bm{A}\Vert\le |\bm{A}|_F$ with $|\cdot|_F$ denoting the Frobenious norm of a matrix. Then, the above bound of the difference comes down to evaluating its entries. Denote $z_{i}^{(kl)}$ as the $(k,l)$-th element of the matrix $\bm{x}_i^{(b)}(\bm{x}_i^{(b)})^\top$ for $k,l=1,...,bp$. Similarly, we can write  $z_{i,j}^{(kl)}$ as the $(k,l)$th element of the matrix $\EE[\bm{x}_i^{(b)}(\bm{x}_i^{(b)})^\top|\eta_{i-j-1},...,\eta_{i-1}]$. In the following, we will control the difference terms in the order of II, I and III. 

\begin{enumerate}[(a)]
	\item \label{step1}Dealing with term II.\\
	 Consider the $(k,l)$th element of ${\rm II}$ without the matrix of basis functions, we have
	$$\Delta_n^{(kl)}:=\sum_{j=1}^n
 \Lambda_{n,j}^{(kl)}:=
	\sum_{j=1}^n\left(\sum_{i=b+1}^n
	(z_{i,j}^{kl}-z_{i,j-1}^{kl})/n\right),~1\le k,l\le bp.$$ 
	Now let $y_{i,j}^{(kl)}=\sum_{h=(i-1)j+b+1}^{(ij+b)\wedge n}(z_{h,j}^{(kl)}-z_{h,j-1}^{(kl)})/n$ where $a\wedge b:=\min(a,b)$ for two real numbers $a$ and $b$. With $l_0=\lfloor \frac{n-b}{j}\rfloor$, we can obtain 
	$$|\Lambda_{n,j}^{(kl)}|=\left|\sum_{i=1}^{l_0} y_{i,j}^{(kl)}\right|=\left|\sum_{i~{\text is~odd}}y_{i,j}^{(kl)}
	+\sum_{i~{\text is~even}}y_{i.j}^{(kl)}\right|.$$
	Observe that $y_{1,j},y_{3,j},...$ are independent and $y_{2,j},y_{4,j},...$ are also independent by the definition of the $j$-dependent sequence. Using \cref{rosenthal} and the triangle inequality, we can obtain
 {\small
 \begin{equation}\label{lambda}
	\Vert\Lambda_{n,j}^{(kl)}\Vert_q\le \frac{14.7q}{\log q}\left\{\left\Vert \sum_{i~{\text is~odd}}y_{i,j}^{(kl)}\right\Vert_2+\left(\sum_{i~{\text is~odd}}\Vert y_{i,j}^{(kl)}\Vert_q^q\right)^{1/q}+
	\left\Vert \sum_{i~{\text is~even}}y_{i,j}^{(kl)}\right\Vert_2
	+\left(\sum_{i~{\text is~even}}\Vert y_{i,j}^{(kl)}\Vert_q^q\right)^{1/q}
 \right\}.
	\end{equation}
 }
	Let $\delta_z(j ,q):=\max_{i,k,l}\Vert z_{i}^{(kl)}-\tilde{z}_{i}^{(kl)}\Vert_q$ where $\tilde{z}_i^{(kl)}$ is the coupling random variable of $z_i^{(kl)}$, then we have $\delta_z(j,q)\le C(j+1)^{-\tau}$ for $j\ge 0$ by Assumption 8 of the paper and \cref{eps_decay}. Since $\{(z_{i,j}^{(kl)}-z_{i,j-1}^{(kl)})/n\}$ is a martingale sequence, then by \cref{mar_concentration}, we can calculate
	\begin{align*}
	&\Vert y_{i,j}^{(kl)}\Vert_q\le \sqrt{q-1}[(b+ij)\wedge n-(i-1)j-b]^{1/2}\delta_z(j,q)/n,\\
	&\Vert y_{i,j}^{(kl)}\Vert_2\le [(b+ij)\wedge n-(i-1)j-b]^{1/2}\delta_z(j,2)/n.	
	\end{align*}
	Thus \cref{lambda} implies that for $1\le j\le n$,
	$$\Vert \Lambda_{n,j}^{(kl)}\Vert_q \le \frac{29.4q}{\log q}\left(\delta_z(j,2)/\sqrt{n}+
	(q-1)^{1/2}n^{1/q}j^{1/2-1/q}\delta_z(j,q)/n\right).$$
	Consequently, it yields $$\Vert \Delta_n^{(kl)}\Vert_q\le \sum_{j=1}^n
	\Vert\Lambda_{n,j}^{(kl)}
 \Vert_q=\bigO(n^{-1/2}),$$ then we have 
	$|\Delta_n^{(kl)}|=
 \bigO_\Pr(n^{-1/2})$.
	
	\item \label{step2}For the first term ${\rm I}$, consider the $(k,l)$th element $z_{i}^{(kl)}-z_{i,n}^{(kl)}$. Using \cref{mar_concentration} and a similar argument in the Step \eqref{step1}, we have
	$$\Vert S_n^{(kl)}-S_{n,n}^{(kl)}\Vert_q\le C\sum_{m=n}^{\infty}
	\delta_z(m,q)/\sqrt{n}=\bigO(n^{-\tau+1/2}),$$ which also implies $|S_n^{(kl)}-S_{n,n}^{(kl)}|=
 \bigO_\Pr(n^{-\tau+1/2})$.
	
	\item \label{step3}At last, we divide III into two parts:
	$$\bm{S}_{n,1}-\tilde{\bm{S}}_n,~\tilde{\bm{S}}_n-\bm{W},$$ where $\tilde{\bm{S}}_n=\frac{1}{n}\sum_{i=b+1}^n\EE[\bm{x}_i^{(b)}
 (\bm{x}_i^{(b)})^\top] \otimes [\bm{v}(i/n)\bm{v}^\top(i/n)]$.
	
	Similar to the previous discussion, it yields that
	$$\Vert \bm{S}_{n,1}-\tilde{\bm{S}}_n\Vert \le 
	\zeta_c^2 \left\Vert \frac{1}{n}\sum_{i=b+1}^n \left(\EE[\bm{x}_i^{(b)}
 (\bm{x}_i^{(b)})^\top|\eta_{i-1}]-
	\EE[\bm{x}_i^{(b)}(\bm{x}_i^{(b)})^\top]\right)\right\Vert .$$ Since $\EE[\bm{x}_i^{(b)}(\bm{x}_i^{(b)})^\top|\eta_i]-
	\EE[\bm{x}_i^{(b)}(\bm{x}_i^{(b)})^\top]$ is independent across $i$, then it will controlled by \cref{lemma_berns} under operator norm. By elementary calculations, we obtain $R\le C/n$ and $\sigma^2\le Cp/n$, then
	\begin{equation*}
	\Pr\left(\left\Vert \frac{1}{n}\sum_{i=b+1}^n \left(\EE[\bm{x}_i^{(b)}(\bm{x}_i^{(b)})^\top
 |\eta_{i-1}]-\EE[\bm{x}_i^{(b)}(\bm{x}_i^{(b)})^\top]\right)\right\Vert	
 \ge t\right)\le
	bp{\rm exp}\left\{-\frac{t^2/2}{\sigma^2+Rt/3}\right\} 
	\end{equation*}
	converges to 0 by choosing $t=\bigO\left(\sqrt{bp\log n/n}\right)$.
	
On the other hand, by the definition of Riemann integral and the Lipschitz continuity of $\gamma_{kl}(t,\cdot)$, we have for $1\le k,l\le bp$ and $1\le j_1,j_2\le b$,
	$$\frac{1}{n}\sum_{i=b+1}^nv_{c_1}(\frac{i}{n})v_{c_2}(\frac{i}{n})
	\EE[x_{i-j_1,k}x_{i-j_2,l}]=\int_0^1\tilde{\gamma}_{kl}(t,j_2-j_1,c_1,c_2)\dee t
	+\bigO\left(\frac{b}{n}\right),$$ where $\tilde{\gamma}_{kl}(t,j_2-j_1,c_1,c_2)=
	v_{c_1}(t)v_{c_2}(t)\gamma_{kl}(t,j_2-j_1)$.
\end{enumerate}

Combining all three difference terms, we conclude that
\begin{align}\label{projection_matrix_diff}
\Vert \bm{S}_n-\bm{W}\Vert&\le \Vert \bm{S}_n-\bm{S}_{n,n}\Vert_F+
\sum_{j=2}^n\Vert \bm{S}_{n,j}-\bm{S}_{n,j-1}\Vert_F+
\Vert\bm{S}_{n,1}-\bm{W}\Vert_F \notag\\
&\le C\left(bp\zeta_c^2n^{-\tau+1/2}
+bp\zeta_c^2n^{-1/2}+
\zeta_c^2\sqrt{\frac{bp\log n}{n}}+
\frac{Cb^2cp}{n}\right)\notag \\
&=\bigO_\Pr\left(\frac{bp\zeta_c^2}{\sqrt{n}}\right)=o_\Pr(1),
\end{align}
where \eqref{projection_matrix_diff} holds by Assumption 10.
Similarly, we have $$
	\left\Vert\bm{W}-\bm{S}_n\right\Vert=o_\Pr(1).$$ As a byproduct, one can obtain that 
	\begin{equation}\label{matrix_appro}
	\lambda_{\max}(\bm{S}_n)\le C,~~\lambda_{\min}(\bm{S}_n)\ge \kappa_2>0,~\kappa_2~\text{is~some constant},
	\end{equation}
	by \cref{lemma_w} and  Assumption 7 when $j=b$.
 
Next, we will prove the consistency of the estimator. First, recall $\widehat{\bm{\beta}}=\left(\frac{\bm{Y}^\top\bm{Y}}{n}\right)^{-1}
	\left(\frac{\bm{Y}^\top\bm{x}}{n}\right)$, and we have
	$$\widehat{\bm{\beta}}=\bm{\beta}+
	\left(\frac{\bm{Y}^\top\bm{Y}}{n}\right)^{-1}
	\left(\frac{\bm{Y}^\top\bm{\epsilon}}{n}\right)+\left(\frac{\bm{Y}^\top\bm{Y}}{n}\right)^{-1}
	\left(\frac{\bm{Y}^\top(\bm{Q}_1+\bm{Q}_2)}{n}\right),$$ where $\bm{Q}_1,\bm{Q}_2$ are defined in Eq. (20) in Section 3.3 of the paper. Note that $\widehat{\bm{\Phi}}_j(t)-\bm{\Phi}_j(t)=
 (\widehat{\bm{\beta}}-\bm{\beta})^\top\bm{E}_j\bm{A}(t)$, where $\bm{E}_j$ is a $bcp\times cp$ matrix with $j$th $cp\times cp$ block being $\bm{I}_{cp}$ and other blocks being $\bm{0}_{cp}$, $\bm{A}(\cdot)$ is defined in Eq. (21). To simply the notation, we denote the standard deviation term in $\widehat{\bm{\Phi}}_j(t)-\bm{\Phi}_j(t)$ as $$\bm{\Phi}_j^\dagger(t):=
 \left(
 \frac{\bm{\epsilon}^\top\bm{Y}}
 {n}\right)\left(\frac{\bm{Y}^\top\bm{Y}}{n}\right)^{-1}\bm{E}_j\bm{A}(t).$$ The remaining term involving the bias of $\widehat{\bm{\Phi}}_j(\frac{i}{n})-\bm{\Phi}_j(\frac{i}{n})$ for $i=b+1,...,n$ and $j=1,...,b$ can be represented as 
 $$\bm{\Delta}_e:=\bm{\Delta}_c
 +\left(\frac{(\bm{Q}_1+
 \bm{Q}_2)^\top\bm{Y}}{n}\right)\left(\frac{\bm{Y}^\top
 \bm{Y}}{n}\right)^{-1}\bm{E}_j
 \bm{A}(\frac{i}{n}),$$ where $\bm{\Delta}_c$ is defined in Eq. (22) in Section 3.3 of the paper. Through elementary calculations, we have 
 \begin{equation}\label{bias}
 \sup_{i}
 \Vert\bm{\Delta}_e\Vert=\bigO_\Pr\left(
 \zeta_c\sqrt{\frac{p}{n}}
 b^{-\tau+2}(\log b)^{\tau-1}+\zeta_c\sqrt{\frac{p}{n}}
 b^{3}/n+\zeta_c\sqrt{bp}c^{-d_2}\right)=\bigO\left(
 \zeta_c\sqrt{bp}c^{-d_2}\right).
 \end{equation}
 Now, we will derive the magnitude of the variance term $\bm{\Phi}_j^\dagger(t)$. By the mean value theorem and Assumption 9(i), for some constant $C_0>0$ and $t,t^\ast\in[0,1]$, we have
	\begin{align}
	&\left\Vert\bm{\Phi}_j^\dagger(t)
 -\bm{\Phi}_j^\dagger(t^\ast)
 \right\Vert \notag \\
 \le &
\left\Vert\left(\frac{
\bm{\epsilon}^\top\bm{Y}}{n}\right)
	\left(\frac{\bm{Y}^\top\bm{Y}}{n}\right)^{-1}\bm{E}_j
	[\bm{A}(t)-\bm{A}(t^\ast)]\right\Vert \label{coef_matrix_diff_opt_1}\\
	\le& C_0n^{\omega_1}c^{\omega_2}|t-t^\ast|\left\Vert\frac{\bm{\epsilon}^\top\bm{Y}}{n}\right\Vert
	\left\Vert\left(\frac{\bm{Y}^\top\bm{Y}}{n}\right)^{-1}\right\Vert.
 \label{coef_matrix_diff_final}
	\end{align}
	By the result of \eqref{matrix_appro}, it is obvious to find out $\Vert\bm{S}_n^{-1}\Vert=\Vert(\bm{Y}^\top\bm{Y}/n)^{-1}\Vert=\bigO_\Pr(1)$. Furthermore, we can deduce $\Vert\bm{\epsilon}^\top\bm{Y}/n\Vert=\bigO_\Pr(\sqrt{bcp/n})$. For some constant $M>0$, denote the event $\mathcal{B}_n$ on which $C_0\Vert\bm{\epsilon}^\top\bm{Y}/n\Vert\Vert(\bm{Y}^\top\bm{Y}/n)^{-1}\Vert\le M$, then it is easy to obtain $\Pr(\mathcal{B}_n^c)=o(1)$. Similar to the discussion of \citep[Eq.s (42) and (43)]{Chen15}, we conclude that on $\mathcal{B}_n$, for some constant $C>0$, there exist some positives $\delta_1,~\delta_2$ such that whenever $|t-t^\ast|\le \delta_1n^{-\delta_2}$, \cref{coef_matrix_diff_final} becomes
	\begin{equation}\label{bound}
	C_0n^{\omega_1}c^{\omega_2}|t-t^\ast|\left\Vert\frac{\bm{\epsilon}^\top\bm{Y}}
	{n}\right\Vert \left\Vert\left(\frac{\bm{Y}^\top\bm{Y}}{n}\right)^{-1}\right\Vert
	\le C_1n^{\omega_1}c^{\omega_2}n^{-\delta_2}
	\sqrt{\frac{bcp}{n}} \le 
	C_2\zeta_c^2\sqrt{\frac{bp\log(n)}{n}},
	\end{equation}
	where the last inequality follows by Assumption 9(ii). Denote $\mathcal{T}_n$ be the smallest subset of $[0,1]$ such that for each fixed $t\in [0,1]$, there exists a sequence $t_n\in \mathcal{T}_n$ such that $|t_n-t|\le \delta_1n^{-\delta_2}$. Let $t_n(t)$ be the nearest sequence of $\mathcal{T}_n$ to $t$. Then by elementary calculations, we conclude that
	\begin{align*}
	&\Pr\left(\sup_{t\in[0,1]}\left\Vert\bm{\Phi}_j^\dagger(t)\right\Vert\ge 8C\zeta_c^2\sqrt{bp\log(n)/n}\right)\\
	\le&\Pr\left(\left\{\sup_{t\in[0,1]}\left\Vert\bm{\Phi}_j^\dagger(t)
	\right\Vert\ge 8C\zeta_c^2\sqrt{bp\log(n)/n}\right\}\cap \mathcal{B}_n\right)+\Pr(\mathcal{B}_n^c)\\
	\le&\Pr\left(\left\{\max_{t_n\in\mathcal{T}_n}
	\left\Vert\bm{\Phi}_j^\dagger(t_n)\right\Vert\ge 4C\zeta_c^2
	\sqrt{bp\log(n)/n}\right\}\cap\mathcal{B}_n\right)\\
	&{}+\Pr\left(\left\{\sup_{t\in[0,1]}\left\Vert\bm{\Phi}_j^\dagger(t)-
	\bm{\Phi}_j^\dagger(t_n)\right\Vert\ge 4C\zeta_c^2\sqrt{bp\log(n)/n}\right\}\cap 
	\mathcal{B}_n\right)+\Pr(\mathcal{B}_n^c)\\
	\le& \Pr\left(\left\{\max_{t_n\in\mathcal{T}_n}
	\left\Vert\bm{\Phi}_j^\dagger(t_n)\right\Vert\ge 4C\zeta_c^2
	\sqrt{bp\log(n)/n}\right\}\cap\mathcal{B}_n\right)+o(1),
	\end{align*}
	where the first inequality uses $\Pr(A)\le \Pr(A\cap B)+\Pr(B^c)$ for two events $A$ and $B$. The second inequality follows by $\Pr(X+Y\ge x)\le \Pr(X\ge x/2)+\Pr(Y\ge x/2)$ for two random variables $X$ and $Y$. The last inequality is by \cref{bound} and the fact $\Pr(\mathcal{B}^c_n)=o(1)$. According to the definition of $\bm{\Phi}(\cdot)$, we can decompose it and use the similar technique to control the above probability. Namely,
	\begin{align}
	&\Pr\left(\left\{\max_{t_n\in\mathcal{T}_n}\left\Vert\bm{\Phi}_j^\dagger(t_n)
	\right\Vert\ge 4C\zeta_c^2\sqrt{bp\log(n)/n}\right\}
	\cap\mathcal{B}_n\right) \notag\\
	\le& \Pr\left(\max_{t_n\in\mathcal{T}_n}\left\Vert
	\frac{\bm{\epsilon}^\top\bm{Y}}{n}
	\left(\bm{S}_n^{-1}-\bm{W}^{-1}\right)\bm{E}_j\bm{A}(t_n)
	\right\Vert\ge C\zeta_c^2\sqrt{bp\log(n)/n}\right) \label{consistency_1}\\
	&{}+ \Pr\left(\max_{t_n\in\mathcal{T}_n}
	\left\Vert\frac{\bm{\epsilon}^\top\bm{Y}}{n}
	\bm{W}^{-1}\bm{E}_j\bm{A}(t_n)
	\right\Vert\ge C\zeta_c^2\sqrt{bp\log(n)/n}\right) \label{consistency_2}
	\end{align}
   Here, it is easy to obtain that the maximum operator norm in \eqref{consistency_2} is bounded by $C\zeta_c\sqrt{bcp/n}$ and can be consequently represented as $o(\zeta_c^2\sqrt{bp\log n/n})$. Therefore, we have the probability in \eqref{consistency_2} is $o(1)$. Lastly, by the fact $\left\Vert\bm{\epsilon}^\top\bm{Y}/n\right\Vert=\bigO_\Pr(\sqrt{bcp/n})$ and the result in \eqref{projection_matrix_diff}, we conclude that \eqref{consistency_1}$\to 0$ as $n\to\infty$. Hence, we have finished the proof. $\hfill \square$
	
	\noindent
	\textbf{Proof of Theorem 4}.
	Recall the theoretical best linear forecast based on VAR$(b)$ process as
	\begin{equation*}
	\widetilde{Y}_{n+1}^{(b)}(u)=\sum_{j=1}^b\bm{\alpha}_f^\top(u)
	\bm{\Phi}_j(1)\bm{x}_{n+1-j}.
	\end{equation*}
	By definitions of ${\rm PE}_{n+1}^{(b)}$ and $\widehat{\rm PE}_{n+1}^{(b)}$, we have
	\begin{align*}
	\widehat{\rm PE}_{n+1}^{(b)}&=Y_{n+1}(u)-Y_{n+1}^{(p)}(u)+{\rm PE}_{n+1}^{(b)}
	+\widetilde{Y}_{n+1}^{(b)}(u)-
 \widehat{Y}_{n+1}^{(b)}(u)\\
	&=Y_{n+1}(u)-Y_{n+1}^{(p)}(u)+
 {\rm PE}_{n+1}^{(b)}+
 \sum_{j=1}^b\bm{\alpha}_f^\top(u)
 \left[\bm{\Phi}_j(1)
	-\widehat{\bm{\Phi}}_{j}(1)\right]\bm{x}_{n+1-j}.
	\end{align*}
	Following the similar arguments in Theorem 3 and \cref{trun_l2}, we can obtain 
	$$\EE|\widehat{\rm PE}_{n+1}^{(b)}|
 _{\mathcal{L}^2}^2-\EE\left|{\rm PE}_{n+1}^{(b)}\right|_{\mathcal{L}^2}^2=\bigO\left(p^{-(2d_1+1)}+
	\frac{bp\zeta_c^4\log(n)}{n}+b\zeta_c^2p^2c^{-2d_2}\right).$$
	Finally, combining with the result in Theorem 3, we conclude
	\begin{align*}
	&\EE\left|{\rm PE}_{n+1}\right|_{\mathcal{L}^2}^2-
	\EE\left|\widehat{\rm PE}_{n+1}^{(b)}\right|
 _{\mathcal{L}^2}^2\\
 =&\EE\left|{\rm PE}_{n+1}\right|_{\mathcal{L}^2}^2-
	\EE\left|{\rm PE}_{n+1}^{(b)}\right|
 _{\mathcal{L}^2}^2+
 \EE\left|{\rm PE}_{n+1}^{(b)}\right|
 _{\mathcal{L}^2}^2-
\EE|\widehat{\rm PE}_{n+1}^{(b)}|_{\mathcal{L}^2}^2\\
	=&\bigO\left(p^{-(d_1+2)}+pb^{-2\tau+3}(\log b)^{2\tau-3}+pb^5/n^2+
 \frac{bp\zeta_c^4\log(n)}{n}+bp\zeta_c^2c^{-2d_2}\right).
	\end{align*}
	$\hfill \square$
	
	\section{Some auxiliary lemmas}\label{add_lemma}
	In this section, we collect some preliminary lemmas which will be used for our technical proofs.
	
	First, let $\bm{A}=(a_{ij})$ be a complex $n \times n$ matrix. For $1\le i\le n$, let $R_i=\sum_{j\neq i}|a_{ij}|$ be the sum of the absolute values of the non-diagonal entries in the $i$th row. Let $D(a_{ii},R_i) \subseteq \mathbb{C}$ be a closed disc centered at $a_{ii}$ with radius $R_i$. Such a disc is called a Gershgorin disc. Next lemma provides a deterministic bound for the spectrum of a square matrix.
	
	\begin{lemma}[Gershgorin circle theorem]\label{g_circle}
		Every eigenvalue of $\bm{A}=(a_{ij})$ lies within at least one of the Gershgorin discs $D(a_{ii},R_i)$, where $R_i=\sum_{j\neq i}|a_{ij}|$.
	\end{lemma}

 The following lemma is the extension of \cref{g_circle} for block matrices.
	\begin{lemma}[Gershgorin circle theorem for block matrices]\label{g_circle_block}
		Suppose $\sigma(\cdot)$ is the spectrum of a matrix. Let $\bm{A}=(\bm{A}_{ij})\in\mathbb{R}^{dn\times dn}$ where $\bm{A}_{ij}\in\mathbb{R}^{d\times d}$. Then every eigenvalue of $\bm{A}$ lies within at least one of the Gershgorin discs $$G_i:=\sigma(\bm{A}_{ii})\cup \left\{\cup_{k=1}^n D\left(
		\lambda_k(\bm{A}_{ii}),\sum_{j=1,j\neq i}^n\Vert \bm{A}_{ij}\Vert \right)\right\}.$$
	\end{lemma}
     
    Next lemma provides the inequality of matrix sums.
    \begin{lemma}[Weyl's inequality]\label{lemma_w}
    	Let $\bm{A}$ and $\bm{B}$ be Hermitian $n\times n$ matrices with eigenvalues $\lambda_1(\bm{A})\ge \cdots\ge \lambda_n(\bm{A}),~\lambda_1(\bm{B})\ge \cdots\ge \lambda_n(\bm{B})$, respectively. If $1\le k\le i\le n$ and $1\le l\le n-i+1$, then
    	$$\lambda_{i+l-1}(\bm{A})+\lambda_{n-l+1}(\bm{B})\le 
    	\lambda_i(\bm{A}+\bm{B})\le \lambda_{i-k+1}(\bm{A})+\lambda_k(\bm{B}).$$
    	In particular,
    	$$\lambda_i(\bm{A})+\lambda_n(\bm{B})\le \lambda_i(\bm{A}+\bm{B})\le 
    	\lambda_i(\bm{A})+\lambda_1(\bm{B}).$$
    \end{lemma}

	We say that $\bm{A}$ is $m$-banded if
	$$A_{ij}=0,~\text{~if~}|i-j|>m/2.$$
	The following lemma indicates that, under suitable condition, the inverse of a banded matrix can also be approximated by another banded-like matrix. It will be used in the proof of Proposition 2 and can be found in \citep[Proposition 2.2]{Demko84}.
	
	\begin{lemma}\label{lemma_band}
		Let $\bm{A}$ be a positive definite, $m$-banded, bounded and bounded invertible matrix. Let $[a,b]$ be the smallest interval containing the spectrum of $\bm{A}$. Define $r=b/a,~q=(\sqrt{r}-1)/(\sqrt{r}+1)$ and set 
		$C_0=(1+r^{1/2})^2/2ar$ and $\lambda=q^{2/m}$. Then we have
		$$|(\bm{A}^{-1})_{ij}|\le C\lambda^{|i-j|},$$ where
		$C:= C(a,r)=\max\{a^{-1},C_0\}$.	
	\end{lemma}

	The following Lemma provides an upper bound for the error of solutions of perturbed linear system. It will be used in the proof of Proposition 3. Recall that the conditional number of a diagonalizable matrix $\bm{A}$ is defined as $$\kappa(\bm{A})=\frac{\lambda_{\max}(\bm{A})}{\lambda_{\min}(\bm{A})}.$$
	
	\begin{lemma}[Discussed in Section 6.10 of \cite{Franklin00}]\label{perturbation}
		Consider a matrix $\bm{A}$ and vectors $\bm{x}, \bm{v}$ which satisfy the linear system $\bm{Ax}=\bm{v}$. Suppose that we add perturbations on both $\bm{A}$ and $\bm{v}$ such that $(\bm{A}+\Delta\bm{A})(\bm{x}+\Delta\bm{x})=\bm{v}+\Delta\bm{v}$. Assuming that there exists some constant $C>0$, such that $$\frac{\kappa(\bm{A})}{1-\kappa(\bm{A})\Vert \Delta\bm{A}\Vert/\bm{A}}\le C$$ holds, then we have
		$$\frac{\Vert \Delta\bm{x}\Vert}{\Vert\bm{x}\Vert}\le C\left(\frac{\Vert \Delta\bm{A}\Vert}{\Vert\bm{A}\Vert}+\frac{\Vert \Delta\bm{v}\Vert}{\Vert\bm{v}\Vert}\right).$$
		\end{lemma}
	
	In the proof of Proposition 3, we employ the following concentration inequalities.
	\begin{lemma}[Theorem 4.1 and its following Remark of \cite{johnson1985best}]\label{rosenthal}
		A version of the Rosenthal inequality for independent variables $\{X_i\}$:
		$$\left\Vert \sum_{i=1}^n X_i\right\Vert_q\le \frac{14.7p}{\log q}(\mu_{n,2}^{1/2}+\mu_{n,q}^{1/q}),$$ where 
		$\mu_{n,q}=\sum_{i=1}^{n}\EE|X_i|^q$.
	\end{lemma}

\begin{lemma}[Theorem 2.1 of \cite{rio2009moment}]\label{mar_concentration}
	A version of the Burkholder inequality for martingale differences $\{X_i\}$:
	$$\left\Vert \sum_{i=1}^n X_i\right\Vert_q^2
	\le (q-1)\sum_{i=1}^n\Vert X_i\Vert_q^2.$$
\end{lemma}
	
	Lastly, we will mention the Bernstein-type inequality for independent random matrices.
	\begin{lemma}[Theorem 6.1.1 of \cite{Tropp15}]\label{lemma_berns}
		Let $\{\bm{\Xi}_i\}_{i=1}^n$ be a finite sequence of independent random matrices with dimensions $d_1\times d_2$. Assume $\EE(\bm{\Xi}_i)=\bm{0}$ for each $i$, $\max_{1\le i\le n}\Vert\bm{\Xi}_i\Vert\le R_n$ and define 
		$$\sigma_n^2=\max\left\{\left\Vert\sum_{i=1}^n\EE\left(
		\bm{\Xi}_i\bm{\Xi}_i^\top\right)\right\Vert,
		\left\Vert\sum_{i=1}^n\EE\left(\bm{\Xi}_i^\top\bm{\Xi}_i\right)
		\right\Vert\right\},$$ where the norm represents the largest singular value. Then for all $t>0$,
		$$\Pr\left(\left\Vert\sum_{i=1}^n\bm{\Xi}_i\right\Vert\ge t\right)\le (d_1+d_2)\exp\left(\frac{-t^2/2}{\sigma_n^2+R_nt/3}\right).$$
	\end{lemma}

\bibliographystyle{abbrv}
\bibliography{supp}